%% file: paper.tex
\documentclass[12pt,a4paper,twoside,openright,fleqn,english,dutch]{report}    
\usepackage{fancyheadings}
\usepackage{babel}
\usepackage{mystyle}
\usepackage{epsfig}
\begin{document}
\selectlanguage{english}
\input{titlepage}
\input{opdracht}
\tableofcontents
\input{INTRODUCTION}

\input{ITF}
\input{ITFDMRG}
\input{SBMF}
\input{J1J2DMRG}
\input{FNMC}
\input{APPENDIX}

\input{referencelist}
\input{PUBLICATIONS}
\selectlanguage{dutch}
\input{SAMENVATTING}

\input{CV}

\end{document}

%% file: titlepage.tex
\chapter*{}
\selectlanguage{dutch}
\thispagestyle{empty}

\begin{center}
  
  \vspace*{3.0cm}
  
  {\LARGE\bf Density Matrix Renormalisation Group Variants}

  ~\\

  ~\\
  
  {\LARGE\bf for Spin Systems}

\end{center}

\vspace*{4.5cm}

\begin{center}
  
  {\Large Proefschrift}\\ ~\\ {\large ter verkrijging van de graad van
    Doctor}\\ {\large aan de Universiteit Leiden, }\\ {\large op gezag
    van de Rector Magnificus Dr. W. A. Wagenaar, }\\ {\large
    hoogleraar in de faculteit der Sociale Wetenschappen,}\\ {\large
    volgens besluit van het College voor Promoties}\\ {\large te
    verdedigen op woensdag 1 september 1999}\\ {\large te klokke 16:15
    uur} ~\\ ~\\ ~\\ {\large door}\\ ~\\ ~\\ {\Large Maarten Steven
    Lucas du Croo de Jongh}\\ ~\\ {\large geboren te Gorssel op 27
    november 1970}

\end{center}

\newpage

\thispagestyle{empty}

\vspace*{6ex}

\noindent Promotiecommissie:\\[2ex]
\parbox[t]{4cm}{Promotoren:\\
  \\
  Referent:\\
  Overige leden: }~\parbox[t]{11cm}{Prof. Dr. J. M. J. van Leeuwen\\
  Prof. dr. Ir. W. van Saarloos\\
  Prof. dr. H. W. J. Bl\"ote \\
  Dr. P. J. H. Denteneer \\
  Prof. dr. D. Frenkel (AMOLF)\\
  Prof. dr. L. J. de Jongh\\
  Dr. J. Zaanen \\
  }\\

\vfill


Het onderzoek beschreven in dit proefschrift is uitgevoerd als
onderdeel van het wetenschappelijk programma van de Stichting voor
Fundamenteel Onderzoek der materie (FOM) en de Nederlandse Organisatie
voor Wetenschappelijk Onderzoek (NWO).

\selectlanguage{english}
The research described in this thesis has been carried out as part of
the scientific program of the Foundation for Fundamental Research on
Matter (FOM) and the Netherlands Organization for Scientific Research (NWO).

%% file: opdracht.tex
\newpage
\thispagestyle{empty}
\begin{flushright} 
{\it The proof of the pudding is in the eating ... }
\end{flushright}
\newpage
\thispagestyle{empty}

%% file: INTRODUCTION.tex
\chapter*{Introduction} \addcontentsline{toc}{chapter}{Introduction}

Quite a few physical problems give rise to quantum lattice models.
Among these are descriptions of high-$T_c$ superconducting materials,
metals, insulators and magnets. A quantum lattice model is
characterised by quantum objects ---spins, fermions or bosons---
nicely positioned on a regular lattice. We will restrict ourselves to
spins. These are located at the lattice points. The low temperature
behaviour of such a model is strongly influenced by quantum zero point
fluctuations. These usually suppress the classical
order. Sometimes the classical order is even destroyed giving room for
other, non-classical types of ordering.

In this thesis insight in the low temperature behaviour will be
obtained by a numerical study of the ground state at zero temperature.
Two schemes are possible for this purpose and both will be employed.

The first scheme makes a direct estimate of the ground state wave
function. An approximate ground state is built and the properties are
analysed afterwards. The most successful member of this class is the
Density Matrix Renormalisation Group (DMRG). It consists of a
systematic, iterative procedure. At every iteration step the energy is
minimised in a given subspace of the configurational Hilbert space.
This gives a variational approximation to the ground state.
Afterwards, a part of the subspace is enlarged and another part is
truncated. This transformation is tuned to keep an optimal fraction of
the approximation within the altered subspace. To preview one of the
comparisons with the second scheme, the DMRG does not suffer from the
'sign-problem' that hampers many other approaches. Below the
sign-problem will be explained.

White introduced the DMRG in 1992 \cite{white92} and it has proven to
be extremely successful for one-dimensional quantum models
\cite{gehring97,noack93,white93} and two-dimensional classical models
\cite{bursill94,carlon97}. Applications to higher dimensional quantum
models are relatively rare
\cite{ducroo98,liang94,white96,white98,xiang96}. In this thesis we
investigate what can be achieved in two dimensions and find that the
method has substantial more difficulty with the two-dimensional
geometry. We will try to explain this limitation.

The second scheme does not attempt to approximate the ground state but
aims at a direct sampling of the properties instead. From this class
we will employ the Green Function Monte Carlo simulation (GFMC)
\cite{ceperley80,pollock87,trivedi90,runge92}. Dimensionality does not
play such an important role here as it does in the DMRG method. The
essential assumption of the Green Function Monte Carlo simulation,
like that of any Monte Carlo simulation, is that the properties of a
system can be obtained by measuring them in many representative
configurations.  Every measurement $X_\alpha$ is accompanied by a
weight $M_\alpha$ to express its importance and the average over these
measurements, $\sum_\alpha X_\alpha M_\alpha/\sum_\alpha M_\alpha$,
will yield the properties.  Green Function Monte Carlo simulations
suffer from two important shortcomings. The most important one is that
quantum mechanical models that contain either frustration or fermions
require an extension of these weights $M$ to negative values and the
average $\sum_\alpha X_\alpha M_\alpha/\sum_\alpha M_\alpha$ becomes
prone to noise as the individual, positive and negative weights
$M_\alpha$ are exponentially larger that their sum. This complication
is called the sign-problem.  The second limitation is that a priori a
good estimate of the ground state has to exist to help the simulation
distinguish relevant configurations from less relevant ones. With the
introduction of the Fixed Node Monte Carlo (FNMC) the sign-problem can
also be cured by incorporating a very good approximation to the ground
state in the simulation. The extension of the Fixed Node Monte Carlo
to the Green Function Monte Carlo with Stochastic Reconfiguration
(GFMCSR) \cite{sorella98} makes the end result less dependent on the
quality of the approximation.  Still a good approximation remains
vital for the method.

The objective of this thesis is two-fold. First, we study the DMRG
method to understand its capabilities and limitations in two
dimensions. Second, the DMRG will be integrated with a Green Function
Monte Carlo simulation. DMRG provides a systematic approximation to
the ground state based on the energy. The correlation functions are
biased by the implementation of the method. This shortcoming can be
overcome by a Green Function Monte Carlo simulation using the DMRG
wave function as a guiding wave function. This thesis introduces this
new and promising combination of the DMRG and a Green Function Monte
Carlo simulation for the first time.

To achieve these objectives, we will first study a well-known model,
the Ising model in a Transverse Field (ITF) to analyse the DMRG
method.  Afterwards the DMRG is applied to an unsolved problem, the
two-dimensional, antiferromagnetic, frustrated Heisenberg model.
Although this model has been attacked by a variety of methods
\cite{ducroo97,einarsson,feiguin,ivanov,sachdev,schulz,singh99,
  sorella98,zhitomirsky} no definite results exist so far for the
quantum phase diagram of this model as Monte Carlo simulations are
hampered by a fundamental problem called the sign-problem.  Strong
indications for the existence of a phase without classical ordering
are found by DMRG calculations and the combined effort of the DMRG and
the Green Function Monte Carlo with Stochastic Reconfiguration. The
next paragraphs describe this brief outline in more detail.

The first two chapters focus on the method itself. Chapter one
introduces the two-dimensional Ising model in a Transverse Field. It
has a direct and clear mapping to a highly anisotropic
three-dimensional Ising model, making it almost a blue-print for a
model with a quantum phase transition. It does not suffer from the
sign-problem and cluster Monte Carlo simulations have yielded high
quality numerical results \cite{bloete} to which we can compare our
results. DMRG can only handle relatively small systems and finite-size
scaling techniques are necessary to extend the results to larger
system sizes.  A large fraction of the first chapter is devoted to
developing these finite-size scaling techniques. Thanks to the power
of the DMRG combined with these scaling relations we can numerically
establish the critical field of the two-dimensional Ising model in a
Transverse Field upto three significant figures.

The main subject of chapter two is the DMRG technique. A new variant
of the method is introduced and afterwards the general properties are
described. It seems that DMRG will need amendments or modifications
for larger two-dimensional systems. In combination with scaling
relations we can however establish the two-dimensional behaviour.

In the following two chapters the DMRG technique is applied to the
frustrated Heisenberg model. The frustration in that model appears by
competition of nearest-neighbour and next nearest-neighbour
interactions. an the consequences for the phase diagram are unclear.
The essential question is whether a phase with no classical equivalent
exists for intermediate range of frustration.  Chapter three is
dedicated to the description of the model and the introduction of the
spin stiffness $\rho_s$. The spin stiffness is an excellent indicator
of long-range order and should reveal whether an intermediate phase
exists. It is studied using a mean field approximation in the
Schwinger-boson representation. The reason to resort to a relatively
complex mean field approximation is that the ground state of the
frustrated Heisenberg model is rotationally invariant. The mean fields
can thus not be simply the spin expectation values ( $\langle
\vec{\cal S} \rangle = {\bf 0}$). The Schwinger-Boson Mean Field
approximation (SBMF) inserts the mean fields in the interactions of
neighbouring spins, $\vec{\cal S}_i \cdot \vec{\cal S}_j$. This not
only overcomes the complication of the rotational invariance, but it
also extends the mean fields to local correlations.  It yields a
rotationally invariant ground state but the spin length $\frac{1}{2}$
cannot be strictly conserved.

The spin stiffness in this approximation does not reveal an
intermediate phase, but it serves very well for an analysis of the
finite-size scaling behaviour. This helps us to extend the numerical
results of the next chapter to an infinitely large system.

Chapter four outlines the numerical calculation including the
technique to obtain the spin stiffness, guided by the results of
chapter three.  Like many other methods, The DMRG do not give a
definitive answer on the existence of an intermediate phase, but it
provides clear information on infinitely long strips of widths upto
eight sites.

The final chapter combines the DMRG and the Green Function Monte Carlo
with Stochastic Reconfiguration. The frustrated Heisenberg model
belongs to the class where Monte Carlo simulations suffer from the
sign-problem.  As mentioned above a good guidance is essential for the
Green Function Monte Carlo with Stochastic Reconfiguration.  For a
long time finding an proper approximation to the ground state has been
the bottleneck of all Green Function Monte Carlo variants as a large
amount of research time had to be spent on designing it. The DMRG can
provide such an approximate ground state for many different models,
including the frustrated Heisenberg model. In this chapter it is
outlined how to combine both methods and the phase diagram of a
$10\times10$ system is studied. This combination of the DMRG and a
Green Function Monte Carlo simulation is new and promising. Further
extensions, along the lines of forward-walking schemes
\cite{calandra98}, may even be able to obtain accurate values for the
spin-spin correlations.

This thesis will hopefully provide a good understanding of the
intricacies of the DMRG method. The last chapter resolves a long
standing problem of the Green Function Monte Carlo and the last three
chapters give indications of the intermediate phase of the frustrated
Heisenberg model although no definite statements can be made.\\

\noindent {\large \bf Acknowledgement}\\

Although this thesis only bears the name of one author, there are many
people I had the pleasure of sharing ideas with. Let me mention prof.
dr. Steve White, who I visited at University of California in Irvine,
and prof. dr. Daniel Aalberts, who invited me to Williams College in
Massachusetts. Jeroen Doumen performed some of the computations
described in the first chapter.

%% file: ITF.tex
\chapter{Ising model in a Transverse field \label{chap:ITF}}

\section{Introduction}

Since the beginning of the 1960s, the Ising model in a Transverse
field (ITF) has been studied. In first instance, this quantum
mechanical model was employed to describe the order-disorder
transition in some double-well ferroelectric systems like $KH_2PO_4$
crystals. This interest has survived to the present day, but the scope
has widened.

A decade later the renormalisation group and with it the notion of
universality was introduced. The Ising model often served as a test
ground and consequently it was scrutinised. The $d$ dimensional ITF
can be mapped onto a $d+1$ dimensional Ising model. This relation
makes it an excellent vehicle to introduce the concepts of phase
transitions in the realm of quantum mechanics.

Sachdev, Read and others \cite{sachdev97,chakrabarti} have used the
ITF for the same role as the Ising model has played in the context of
classical critical phenomena: a blue-print of phase transitions and
universality. Maybe the most important difference lies in the fact
that it is not the temperature $T$ that induces a phase transition,
but a coupling constant $H$ that can drastically alter the properties
of the system.  With the disappearance of the temperature, $T=0$, it
is the ground state that exhibits this quantum phase transition.

On the outset our intention is to investigate the density matrix
renormalisation group (DMRG). The ITF is chosen as a 'toy-model' both
because of its rich behaviour and its simple description.

In this chapter, the model will be explained. The exact results in one
dimension (a chain) will be reviewed and subsequently a large effort is
made to uncover the finite-size scaling behaviour for the two-dimensional case. Given the restriction of the DMRG, which will be
discussed in the next chapter, it is worthwhile to scale the
length of the system to infinity first, after which the finite
width can be scaled away. The results, table
\ref{tab:critical_properties}, clearly support such a two-stage process.

A good review of the ITF has been published by Chakrabarti et al.
\cite{chakrabarti}, relieving us of the duty to go into great detail.
The numerical treatment is left to the next chapter.

\section{The Model}

Consider a two dimensional square lattice with length $L$ and width
$W$. The lattice is periodic in both directions and each lattice site
contains a spin-$\frac{1}{2}$. The Hamiltonian is given by
\begin{equation}
  {\cal H}_{ITF} = \sum_{l=1}^{L} \sum_{w=1}^{W} \left \{ -4{\cal
    S}_{l,w}^{x} ({\cal S}_{l+1,w}^{x} + {\cal S}_{l,w+1}^{x} ) + 2H
  {\cal S}_{l,w}^z \right \} \label{eq:ITFdef}
\end{equation}
where the ${\cal S}^{\alpha}_{l,w}$ are the usual spin-$\frac{1}{2}$
matrices satisfying the commutation relations
\begin{equation}
  [{\cal S}_{l,w}^{\alpha},{\cal S}_{l',w'}^{\beta}] = i
  \delta_{l,l'} \delta_{w,w'} \epsilon_{\alpha \beta \gamma} {\cal
    S}_{l,w}^{\gamma} ~~,~~ \alpha,\beta,\gamma=x,y,z
\end{equation}
Clearly the energy scale of this Hamiltonian is set to unity. We have
chosen the convention of spin-$\frac{1}{2}$, in contrast with the
Pauli-matrices frequently used in this field. The factor of 4 and 2
are inserted in the definition (\ref{eq:ITFdef}) to make our results
directly comparable to the main part of the literature.

Let us summarise the main symmetry properties of this model.  The
Hamiltonian with field $H$ can be transformed into one with $-H$ by a
spin rotation round the ${\cal S}^x$-axis over 180 degrees. This is a
unitary transformation so we have the freedom to choose $H>0$.  The
model is translation and reflection symmetric in both directions.

An important symmetry that will be extensively used, is the spin-reversal operator ${\cal S}$. It is associated with a rotation over
180 degrees round the ${\cal S}^z$ axis;
\[
{\cal S}^\dagger {\cal S}^x {\cal S} = - {\cal S}^x ~,~ {\cal
  S}^\dagger {\cal S}^y {\cal S} = - {\cal S}^y ~,~ {\cal S}^\dagger
{\cal S}^z {\cal S} = {\cal S}^z.
\]
and can be expressed as
\begin{equation}
{\cal S} = \exp\left [ i \pi \left  (\sum_{l,w} {\cal S}_{l,w}^z +
  \frac{LW}{2} \right ) \right]. \label{eq:defspinreversal}
\end{equation}
The offset of $LW/2$ allows us to associate the quantum number ${\cal
  S}=1$ with the ground state $|\psi_0 \rangle$ for different system
sizes. ( In short: ${\cal S} | \psi_0 \rangle = |\psi_0 \rangle$). One
can state that the spin reversal operator samples the number of
up-spins and returns whether it is even, ${\cal S}=1$, or odd, ${\cal
  S}=-1$.

If $H \rightarrow 0$, we end up with a simple, classical,
two-dimensional Ising model. The ground state is degenerate; all spins
point forwards or backwards in the ${\cal S}^x$-direction. The
associated phase is named the classically ordered phase. The two
classical solutions, forward and backward pointing spins, can be
superposed in a quantum mechanical sense. In this manner states can be
obtained that are either even (${\cal S}=+1$) in spin-reversal or odd
(${\cal S}=-1$). The statement made before about the ground state,
${\cal S} |\psi_0 \rangle=|\psi_0 \rangle$ still holds for one of
them, but the other ground state $|\psi_1 \rangle$ lies in the odd
subspace, ${\cal S} | \psi_1 \rangle = - | \psi_1 \rangle$.

In the other extreme, $H \rightarrow \infty $, the degeneracy is
lifted.  The model essentially describes free spins in an external
field.  The ground state is unique and has all spins pointing down in
the ${\cal S}^z$-direction. This is the reference state for the
quantum disordered phase and again has the quantum number ${\cal
  S}=+1$.  The lowest excitation differs from the ground state by the
reversal of one spin.  So it belongs to the class ${\cal S}=-1$. We
will extensively study the energy gap $\Delta$ between the lowest
excitation (in ${\cal S}=-1$) and the ground state (in ${\cal S}=+1$).
$\Delta = E_{1}-E_{0}$.

\begin{figure}
  \centering \epsfxsize=10cm \epsffile{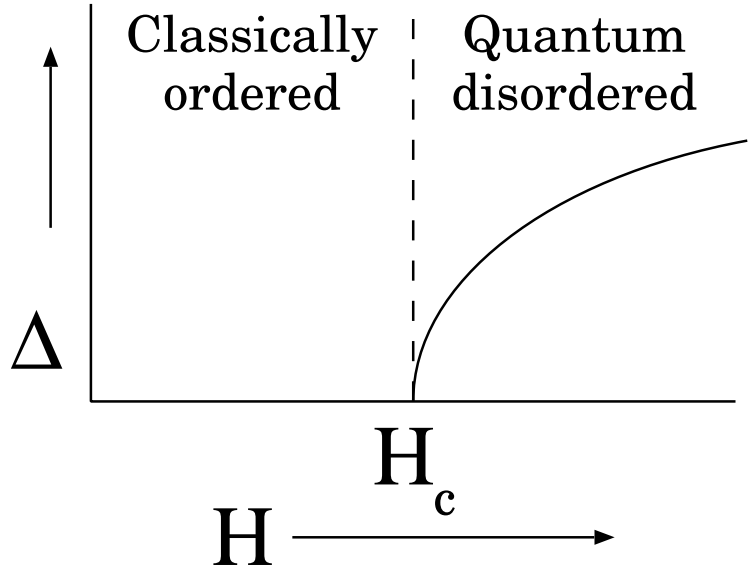}
\caption{\label{fig:ITFphasediagram} The phase diagram for the
  ITF and the energy gap $\Delta$ associated with it.
  At the critical field $H_c$ a phase transition occurs and the gap
  $\Delta$ opens up. Further explanation is given in the text.}
\end{figure}

There is a phase transition between the classically ordered and the
quantum disordered state. A clear signature of this phase transition
is the appearance of the gap $\Delta$, which occurs for a critical
value $H=H_c$. In figure \ref{fig:ITFphasediagram} all these
properties are summarised in a graphical representation.  As we will
show later the relation between the gap $\Delta$ and the field can be
made more explicit by
\begin{equation}
\Delta \sim \left |H-H_c \right |^\nu. \label{eq:2dnu}
\end{equation}
for $H>H_c$. Below the critical field, $H<H_c$, the ground state becomes
degenerate. The gap $\Delta$ should then be defined as the energy
difference between the ground state and the first excitation {\it
  within} the even subspace for equation (\ref{eq:2dnu}) to hold. In the rest of the chapter we will not redefine the gap but only consider $H>H_c$.

It will take another section to prove that the critical exponent
$\nu$ is identical to the critical exponent of the three-dimensional
classical Ising model. After that we focus on extracting both the
critical field $H_c$ and the critical exponent $\nu$ for the
two-dimensional ITF.

\section{Connection with the 3-dimensional Ising model}

The ITF has an intimate connection with a highly anisotropic Ising
model in one dimension higher. Although we will only show this
explicitly for the 2-dimensional ITF, the argument holds in general.

As starting point we take a highly anisotropic, ferromagnetic Ising
model in three dimensions. The classical spins $s^z=\pm 1$ interact
according to the Hamiltonian
\begin{equation}
-\beta {\cal H}_{clas} = K_{\perp} \sum_{(ij)} s_i^z s_j^z + K_{||}
\sum_{(ij)} s_i^z s_j^z,
\end{equation}
on a cubic lattice at an inverse temperature $\beta$. The first term
contains the interactions between nearest neighbours (denoted by
$(ij)$) in the plane perpendicular to the anisotropy axis. The second
term contains those along the anisotropy direction. The coupling
constants $K_{\perp}$ and $K_{||}$ give rise to this anisotropy. We
link them to the field in equation (\ref{eq:ITFdef}) where the energy
scale was set to unity,
\begin{equation}
 K_{\perp}= \varepsilon ~~,~~ e^{-2  K_{||}}= \varepsilon H
~~,~~ \varepsilon \ll 1, \label{eq:paramrelation}
\end{equation}
$K_{\perp}$ is a small, positive parameter and $K_{||}$ is a large
one.

The partition function can be calculated by the transfer matrix
method. We choose the transfer direction along the anisotropy axis.
The Hamiltonian ${\cal H}_{clas}$ can be split in interactions between
slices perpendicular to the anisotropy direction.  For each pair of
adjacent slices we then obtain a local, transfer Hamiltonian ${\cal
  H}_{trans}$ for all $N$ slices. The connections with the ITF will be
found in the partition sum,
\[
Z=\mbox{Tr} e^{-\beta {\cal H}_{clas}} = \mbox{Tr} \prod_{n=1}^{N}
e^{-\beta{\cal H}_{trans}}.
\]
The connection to the quantum mechanical world is made by introducing
summations over complete bases $|\alpha_n \rangle$ for the slices in
this product;
\[
Z=\sum_{|\alpha_n \rangle} \prod_{n=1}^N \langle \alpha_n | e^{-\beta
  {\cal H}_{trans}} | \alpha_{n+1} \rangle,
\]
with periodic boundary conditions, $|\alpha_{N+1} \rangle =
|\alpha_1\rangle$.  The parameters $K_{\perp}$ and $K_{||}$ were
chosen so anisotropic that we may approximate the exponentials;
\begin{eqnarray}
\langle \alpha_n | \exp\left (K_{||} \sum_{(i,j)} s_i^z s_j^z \right ) | \alpha_{n+1} \rangle & =& e^{ K_{||}} \langle \alpha_n |1+2e^{-2  K_{||}} \sum_{l,w} {\cal S}_{l,w}^x | \alpha_{n+1} \rangle, \nonumber \\
\langle \alpha_n | \exp \left( K_{\perp} \sum_{(i,j)} s_i^z s_j^z \right ) | \alpha_{n+1} \rangle& =&
\langle \alpha_n |1+ 4K_{\perp} \sum_{l,w} {\cal S}_{l,w}^z ({\cal S}_{l+1,w}^{z} + {\cal S}_{l,w+1}^{z} ) | \alpha_{n+1} \rangle \nonumber
\end{eqnarray}
All of the above is combined in the resulting partition function
\[
Z e^{- N K_{||}} = \mbox{Tr} \prod_{n=1}^N \left \{1+\varepsilon \left
    [ \sum_{l=1}^{L} \sum_{w=1}^{W} \left ( 4{\cal S}_{l,w}^{z} ({\cal
        S}_{l+1,w}^{z} + {\cal S}_{l,w+1}^{z} ) + 2H {\cal S}_{l,w}^x
    \right ) \right ] \right \}
\]
The right side of the equation can easily be mapped onto the
definition of the ITF in equation (\ref{eq:ITFdef}) by some simple,
unitary transformations: we exchange the ${\cal S}^z$- and ${\cal
  S}^x$-direction and afterwards rotate the spins over $\pi$ round
the ${\cal S}^x$ axis (using the unitary operator $\exp(i \pi \sum_j
{\cal S}^x_j)$). The trace is invariant under these unitary
transformation, so the outcome is
\begin{equation}
Z e^{- N K_{||}} = \mbox{Tr} e^{-N\varepsilon {\cal H}_{ITF}}.
\end{equation}
Careful analysis reveals that this relation holds up to order ${\cal
  O}(\varepsilon^2)$.  Along with this connection come quite a few
others; the correlation function $\xi$, formally defined by the
covariance of Ising spins located at location $(l,w,1)$ and $(l,w,n)$;
\[
\langle s_{l,w,1}^z s_{l,w,n}^z \rangle - \langle s_{l,w,1}^z\rangle
\langle s_{l,w,n}^z\rangle \sim e^{-n/\xi},
\]
can be related to the gap $\Delta$ in the ITF by inserting complete
bases in the same fashion as above. The result is
\begin{equation}
\xi^{-1} = \ln \frac{e^{-\varepsilon E_1}}{e^{-\varepsilon E_0}} = \varepsilon ( E_1-E_0)= \varepsilon \Delta. \label{eq:nastyvareps}
\end{equation}
As mentioned before in the paragraph after equation (\ref{eq:2dnu})
this relation only holds for $H>H_c$.  Below the critical field the
gap has to be redefined.

The reduced temperature $t=(T-T_c)/T_c$ in the classical model is
replaced by the reduced field $h=(H-H_c)/H_c$. The relation for
critical exponent $\nu$ thus also finds its equivalent;
\[
\frac{1}{\xi} \sim t^{z\nu} \rightarrow \Delta \sim h^{z \nu}.
\]
In this expression we use that for the 3D Ising model $z=1$

\section{Exact solutions for the ITF chain}

The properties of the two-dimensional system can only be obtained
by use of finite-size scaling. In doing so, the results for the
open ITF chain will provide a crucial reference. Here we briefly
review those. The reason not to follow Pfeuty \cite{pfeuty} is
that he considers an infinitely long chain, whereas here it is
essential that the chain is both finite and open. Given the beauty
of exact solutions, we can not resist in presenting also the
results for the finite periodic chain.

The essential ingredient to the exact solution of the ITF chain is a
Jordan-Wigner transformation \cite{pfeuty} to spin-less fermions
$c_j$,
\[
{\cal S}_l^+ = \prod_{j=1}^{l-1} e^{i c_j^\dagger c_j} c_j.
\]
In terms of these operators the Hamiltonian reads
\begin{equation}
{\cal H} = -HL + 2H \sum_{j=1}^L c_j^{\dagger} c_j - J\sum_{j=1}^{L-1}
\left ( c_j^{\dagger} - c_j \right) \left ( c_{j+1}^{\dagger} + c_{j+1} \right ) + {\cal H}_{per}.
\label{eq:JWHam}
\end{equation}
The term ${\cal H}_{per}$ governs the interactions between the first
and the last site. If the chain is open, these do not exist, ${\cal
  H}_{per}=0$. For a periodic chain
\begin{equation}
{\cal H}_{per} = J\left ( c_L^{\dagger} - c_L \right ) \left ( c_0^{\dagger} + c_0
\right ) {\cal S}. \label{eq:hper}
\end{equation}
This equation reintroduces another observable, the spin-reversal
operator ${\cal S}$, defined as
\[
{\cal S} = \exp i \pi \sum_{j=1}^L c_j^{\dagger} c_j.
\]
The current definition is identical to the original one in equation
(\ref{eq:defspinreversal}).  It is a conserved quantity as the only
change in the number of fermions occurs in the second term of the
Hamiltonian (\ref{eq:JWHam}) by pair creation or annihilation. As an
empty chain corresponds to all spins pointing downwards, the ground
state has to be in the even, ${\cal S}=1$, subspace. Moreover these
pairs of operators in the Hamiltonian ${\cal H}$ make it already clear
that it can be diagonalised via a Bogoliubov transformation. For the
periodic chain we can go even further and derive exact expressions for
the excitation spectrum. The open-chain properties require the
numerical diagonalisation of a $L \times L$ matrix to get all energy
levels and eigenstates.

\subsection{The periodic chain}

We want to Fourier transform the particle creation and annihilation
operators in the usual manner
\[
c_k=\sum_{l=1}^L e^{ikl} c_l.
\]
The allowed $k$-values depend on the number of fermions present on the
chain; for an odd number , ${\cal S}=-1$, the boundary term ${\cal
  H}_{per}$, equation (\ref{eq:hper}), has the same sign as all other
interactions. The Hamiltonian ${\cal H}$ becomes translational
invariant and we can take the regular $k$-values;
\begin{equation}
k= \frac{2 \pi l}{L} ~~,~~ 0 \le l < L. \label{eq:fouriera}
\end{equation}
For even number of fermions, ${\cal S}=1$, the boundary conditions are
antiperiodic and the set of $k$-values has to be adjusted
accordingly;
\begin{equation}
k= \frac{(2 l+1)\pi }{L} ~~,~~ 0 \le l < L. \label{eq:fourierb}
\end{equation}
The Hamiltonian ${\cal H}$ is now converted to the momentum space
representation and to represent it compactly we define
$p(k)=-2H+2J\cos(k)$ and $q(k)=2J\sin(k)$. It reads
\begin{eqnarray}
{\cal H} &= -HN &- \sum_{0<k<\pi } p(k) \left ( c_k^{\dagger} c_k + c_{-k}^{\dagger} c_{-k} \right )  \nonumber \\
& &+
 i \sum_{0<k<\pi} q(k) \left ( c_k^{\dagger} c_{-k}^{\dagger} + c_{k} c_{-k} \right ) + {\cal H}_{o,\pi}, \nonumber
\end{eqnarray}
where the allowed $k$-values are defined in (\ref{eq:fouriera}) and
(\ref{eq:fourierb}). The wave vector $k=0$ and $k=\pi$ ---treated in
${\cal H}_{0,\pi}$--- are not always allowed. If they are, see table
\ref{tab:0pi}, they play a special role in that they already appear
in a diagonal form;
\[
{\cal H}_{0,\pi} = \delta_{\mbox{0 allowed}}~ p(0) c_0^{\dagger} c_0 +
\delta_{\mbox{$\pi$ allowed}}~ p(\pi) c_\pi^{\dagger} c_\pi.
\]
Apart from these, all the terms have to undergo a Bogoliubov
transformation. Next we provide the main formulas to diagonalise
those: define the canonical transformation
\[
c_k = u(k) \eta_k - i v(k) \eta_{-k}^{\dagger},
\]
with
\[
\begin{array}{rclcrcl}
u(k) &=&\cos \theta_k &,& v(k) &=& \sin \theta_k , \\
\cos 2\theta_k &=& \displaystyle{\frac{p(k)}{\lambda(k)}} &,& \lambda(k) &=& \sqrt{p^2(k) + q^2(k)}.
\end{array}
\]
The final expression for the Hamiltonian now is
\[
{\cal H} = E_{off} + \sum_{0<k< \pi} \lambda(k) \left (
  \eta_k^{\dagger} \eta_k + \eta_{-k}^{\dagger} \eta_{-k} \right) +
{\cal H}_{0,\pi},
\]
with
\[
E_{off} = - \frac{1}{2} \sum_k \lambda(k).
\]
It is important to stress that the form of the Hamiltonian ${\cal
  H}$ depends on the number of fermions present. Creating one
excitation by creating a single fermion will change the parity of the
number of particles; the only allowed excitations within the subspace
are built from pairs of fermions, i. e. $\eta_i^{\dagger}
\eta_j^{\dagger}|\psi_0 \rangle$. To obtain an excitation in the other
subspace, we have to start with the lowest energy state {\it within } that
subspace and create pairs of excitations on that.

\begin{table}
  \begin{center}
\begin{tabular}{|c|c|c|}
 \hline
Length &{\cal S}=-1 & {\cal S}=1 \\
\hline
even & - & $k=0,\pi$ \\
\hline
odd & $k=\pi$& $k= 0$ \\
\hline
\end{tabular}
\end{center}
\caption{The $k=0$ and $k=\pi$ values will only occur for certain length
  and number of fermions. \label{tab:0pi}.}
\end{table}

\subsection{The open chain \label{sub:openchain}}

In the case of the periodic chain we are fortunate that the
translational invariance allows a Fourier transformation. The modes
become almost decoupled afterwards and an exact expression can be
derived by a Bogoliubov transformation.

The open chain is not translation invariant and no further analytical
steps can be taken. The numerical procedure remains straightforward.
Van Hemmen \cite{hemmen80} describes in detail how to derive
both the energy spectrum and all eigenstates of a fermionic system
for a general pair-wise interaction.  Applying the relevant
transformations leads again to a diagonal Hamiltonian containing
excitation occupation numbers offsetted by the ground state energy
$E_0$.

\begin{figure}
  \centering \epsfxsize=10cm \epsffile{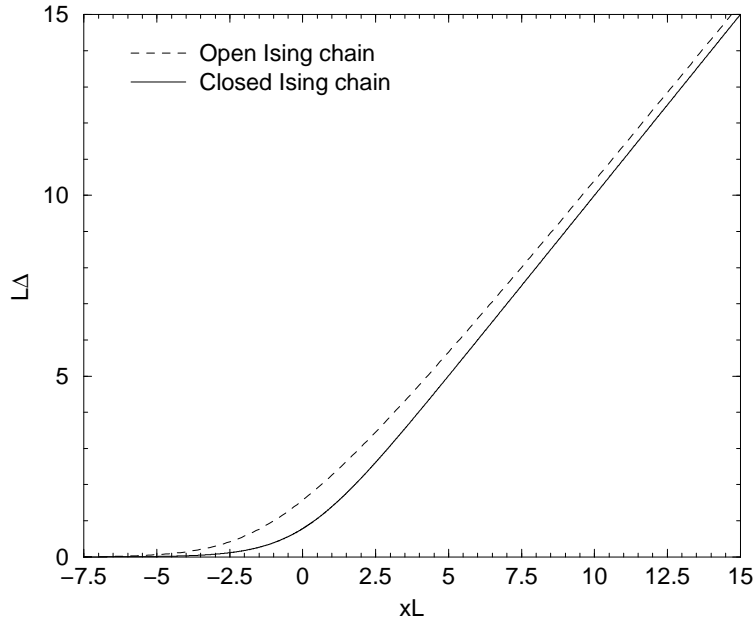}
\caption{ The behaviour of the energy gap $\Delta$ as function
  of the combination of $x=H-H_c$ and the length $L$ of the system.
  The only difference between the lines is the boundary conditions.
  \label{fig:openclosedgap} }
\end{figure}

In figure \ref{fig:openclosedgap} the scaled gap $L \Delta$, on
which we will focus from now on, is depicted for both the open and the
closed chain. They show similar behaviour, although in the vicinity of
the phase transition their ratio becomes quite large.

\section{Finite-size scaling}

We want to establish properties of the 2D ITF through numerical means.
In general it is not possible to obtain direct quantitative
information on an infinitely large system. A widely used indirect
route is to calculate the desired quantities for a set of finite
systems and extract their values for the infinite-size limit by
investigating their dependencies on the systems size. This method is
called finite-size scaling. Anticipating that for our numerical
method, the density matrix renormalisation group, the length of the
system imposes less restrictions than the width, we will first develop
the scaling relations to scale the length $L \rightarrow \infty$.
Afterwards we derive the relations for $W \rightarrow \infty$.  A
final subsection is spent on studying the scaling behaviour for fixed
aspect ratio $L/W$ and $L \rightarrow \infty$. The properties of the
2D system are independent of the route taken to derive them, therefore
both approaches should yield the same results in the limit of infinite
system size.

Although the Density Matrix Renormalisation Group (DMRG) will not be
introduced until the next chapter, the results are incorporated in the
following scaling analysis. DMRG calculations were performed to obtain
the lowest energies in the even and odd subspace. The difference is
the energy gap $\Delta$.  The parameters of the calculations were
$H=3.00 , \dots ,$ $ 3.10$ in steps of $0.01$ and $L/W=2,\dots,5$ in
steps of $0.5$. For widths $W=7,8$ the largest ratio was $L/W=4$.

\subsection{1-Dimensional, $W$ fixed, $L\rightarrow \infty$ \label{sec:1dscaling}}

There are two reasons to scale the length $L$ to infinity first. The
first one has to do with the open boundary conditions. As we will
discuss in the next chapter, closing the boundaries in the length
direction and making the system translation invariant will severely
hamper the accuracy of the calculation. Open systems give rise to
complex finite-size effects by their lack of translational invariance.
These effects will disappear if we scale the length $L \rightarrow
\infty$ where open and periodic systems become indistinguishable.

The second reason is of a more practical nature. The fact that DMRG
functions so extremely well for quantum chains \cite{white92},
indicates that varying the length $L$ of the system will not have a
large impact on the accuracy. Unfortunately this does not hold for the
width $W$ of the system. In the next chapter, section
(\ref{sec:generallim}), we will show that in order to maintain the
accuracy, the size of the calculation grows exponentially with the
width $W$ of the system. With a large range of lengths and only a few
widths that we can handle, it seems wise to remove all the dependence
on the length first. We will thereby obtain fairly accurate results
for infinitely long strips.

Let us now outline the procedure: once the length $L$ has become
sufficiently large, a system of dimensions $L \times W$, with fixed
width $W$, will start to behave as a one-dimensional system. Such a
system will by arguments of universality resemble an open chain, so we
expect for fixed width $W$, that
\begin{equation}
L\Delta(x,W,L) = A(W) \left \{ f_0(B(W)xL) + \frac{1}{L} f_1(B(W)xL) + \dots \right \}. \label{eq:1dscaling}
\end{equation}
This scaling relation contains quite some new elements that deserve
either an introduction or an explanation:
\begin{itemize}
\item $L \Delta$. Given the connection to the Ising model, $\Delta$
  has to correspond to an inverse length. Rescaling the length will
  thus automatically lead to the appearance of the combination $L
  \Delta$.
\item $f_0(\dots)$ and $f_1(\dots)/L$. These are respectively the
  leading order and the first correction of the finite-size behaviour
  of the open ITF chains. We obtained these functions numerically.
  Knowing that the correction arises from a surface contribution (the
  chain is open), the factor $1/L$ in $f_1(\dots)/L$ is obvious.
\item $xL$. Universal behaviour concerns the system properties in the
  vicinity of the phase transition. Every width has its own critical
  field $H_c(W)$, so $x=H-H_c(W)$. One dimensional critical behaviour
  is accompanied by the critical exponent $\nu=1$. The combination
  thus has to be $xL^\nu=xL$.
\item$A(W)$ and $B(W)$. The theory of universality makes predictions
  on the exponents near a phase transition, not on the overall scales.
  These can vary for the different width strips.
\end{itemize}

It has to be stressed that only the form of the leading term,
$f_0(B(W)xL)$, is fixed by universality. The form of the correction
term, $f_1(B(W)xL)/L$, is an assumption. We just take the simplest
form that also suits the open chain.

The scales $A(W)$, $B(W)$ and the critical fields $H_c(W)$ can be
considered fitting parameters to make relation (\ref{eq:1dscaling})
agree with the data. In the following we will describe how we obtained
these parameters and what the outcome is.

\begin{figure}
  \centering \epsfxsize=10cm \epsffile{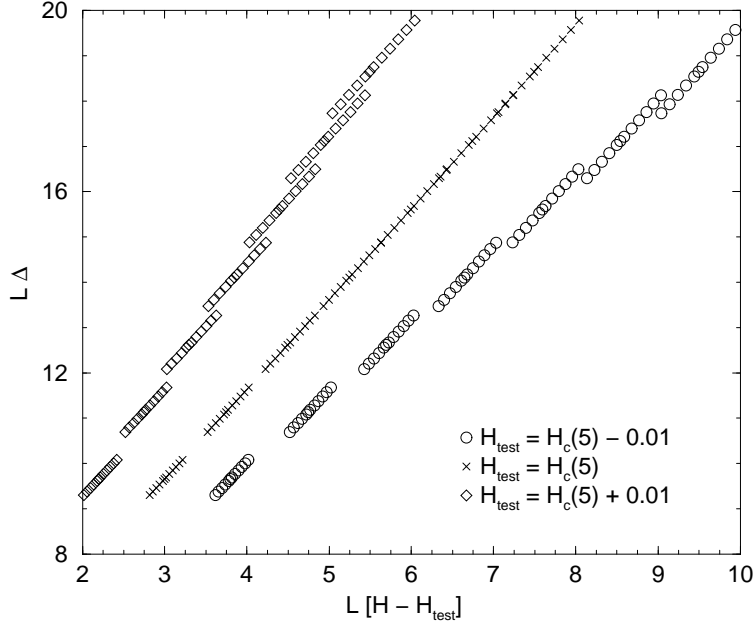}
\caption{The critical field $H_c(W)$ can be obtained by
'smoothing' the curve
  of $L\Delta$ versus $xL$. Depicted is the case for $W=5$ and
  $H_c(W)=2.867$.\label{fig:HcW} }
\end{figure}

The standard fitting procedure is to adjust the parameters such that
all data points are fitted best according to a least-square
functional.  The critical field $H_c(W)$ can also be derived
independently through another, simpler route. In practice this
approach is taken.  If the right $H_c(W)$ is chosen, plotting $L\Delta$
versus $xL$ must give a smooth curve for large enough $L$. From this
feature we will extract $H_c(W)$. An example of this procedure is
given in figure \ref{fig:HcW}.

\begin{figure}
  \centering \epsfxsize=10cm \epsffile{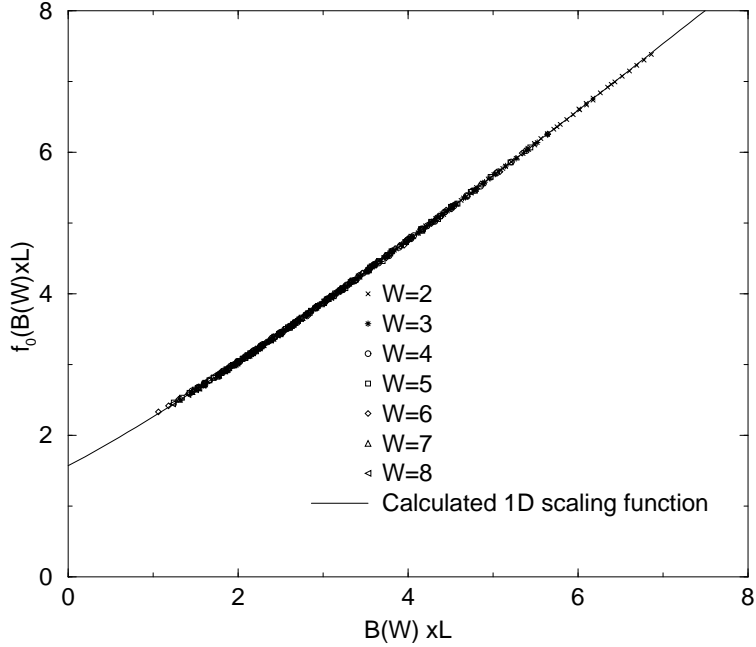}
\caption{The data obtained by DMRG calculations is fitted to scaling relation (\ref{eq:1dscaling}).
  The figure shows scaling collapse nicely. \label{fig:datacollaps} }
\end{figure}

With $H_c(W)$ already found, only $A(W)$ and $B(W)$ have to be
obtained from data collapse. We can fit the data $x,L,\Delta$ to the
formula (\ref{eq:1dscaling}), as both $f_0$ and $f_1$ can be obtained
from the open ITF chain. Figure \ref{fig:datacollaps} reveals that
indeed the data shows nice, universal behaviour with the calculated
$f_0$. The corrections due to $f_1$ are removed from this figure. They
are relatively small.  Table \ref{tab:1dresults} lists the results.

\begin{table}
  \begin{center}
\begin{tabular}{|c|c|c|c|c|}
 \hline
$W$ & $H_c(W)$ &$A(W)$ & $B(W)$ & $A(W)B(W)$ \\
\hline
2 & 2.296 & 2.601 & 0.853 & 2.219\\
3 & 2.646 & 3.096 & 0.829 & 2.567\\
4 & 2.792 & 3.343 & 0.874 & 2.922\\
5 & 2.867 & 3.492 & 0.834 & 3.262\\
6 & 2.912 & 3.586 & 0.998 & 3.579\\
7 & 2.941 & 3.695 & 1.055 & 3.898\\
8 & 2.961 & 3.754 & 1.118 & 4.197\\
\hline
\end{tabular}
\end{center}
\caption{ From the one-dimensional scaling procedure these values
can be extracted. The expression of the gap for the infinite long periodic strip is
 given in equation  (\ref{eq:final1d}) \label{tab:1dresults}.}
\end{table}

The general expression for the gap $\Delta$ at infinite system length
$L=\infty$ can be found by examining the behaviour of $f_0(y)$ and
$f_1(y)$ at large argument $y$ more closely: for large argument $y$
(or field) the ITF Hamiltonian describes free spins in an external
magnetic field with strength $y$. An excitation can be made by
flipping one spin upwards, thus
\[
\lim_{y\rightarrow \infty} \frac{f_0(y)}{y}=1,
\]
with no corrections dependent on size. Therefore the gap $\Delta$ will
become
\begin{equation}
\Delta(x,W) \equiv \lim_{L \rightarrow \infty} \Delta(x,W,L) = A(W)
B(W) [H-H_c(W)]. \label{eq:final1d}
\end{equation}

This relation for the gap $\Delta$ of a finite-width strip and the
numerical value of the quantities appearing in it, table
\ref{tab:1dresults}, are the results of this subsection. In the light
of the gap expression, the trends in the numerics become clearer:

The critical field $H_c(W)$ is monotonically increasing and should
approach the two-dimen{-}sional value,
\[
\lim_{W \rightarrow \infty} H_c(W) = H_c.
\]
The combination $A(W)B(W)$ also shows a trend to the two-dimensional
situation. In two dimensions the critical exponent $\nu < 1$. The
first derivative with respect to the field $H$, which is here
$A(W)B(W)$ and in equation (\ref{eq:2dnu}) it is $\nu
(H-H_c)^{\nu-1}$, should diverge at the phase transition point. So
\[
\lim_{W \rightarrow \infty} A(W)B(W) = \infty.
\]
The trend is in the data but $W=8$ is far to small for this
combination to grow excessively.

\subsection{2-Dimensional, $W \rightarrow \infty$ , $L=\infty$. \label{sec:2dscaling}}

Standard scaling methods as described in \cite{cardy}, can be
implemented now;
\begin{equation}
\Delta(h,W)=\frac{1}{W} g(h W^{1/\nu},u_iW^{y_i}). \label{eq:2dscaling}
\end{equation}
The reduced field $h$ is given by $h=H-H_c$ (remember $H_c \equiv
H_c(\infty)$ ); true two-dimensional scaling behaviour is considered,
in contrast with equation (\ref{eq:1dscaling}). The finite-size
corrections are anticipated by the inclusion of the irrelevant field
$u_i$ and exponent $y_i$. In the previous subsection also an
expression for the gap $\Delta$ was given in equation
(\ref{eq:final1d}). As both, this one and the scaling relation above
must hold, they have to be identical:
\[
A(W)B(W)\left[h+H_c-H_c(W)\right]= \frac{1}{W} g(h
W^{1/\nu},u_iW^{y_i}).
\]
The left side of this identity is linear in $h$. The right part also
contains higher order terms, which we will neglect. Expanding both
sides up to linear order yields:
\begin{eqnarray}
A(W)B(W)W\left [ H_c - H_c(W) \right ] &=& g(0,0) + W^{y_i} u_i \frac{\partial g}{\partial u_i} (0,0), \nonumber \\
A(W)B(W)W &=& W^{1/\nu} \frac{\partial g}{\partial h} (0,0) + W^{1/\nu+y_i} u_i \frac{\partial^2 g}{\partial u_i \partial h} (0,0) . \nonumber
\end{eqnarray}
The derivatives denote differentiations with respect to the first or
the second argument. Given $A(W)$, $B(W)$ and $H_c(W)$ for widths
$W=2,\dots,8$, both the critical properties $H_c$ and $\nu$ and the
fitting parameters $g(0,0)$,  $u_i\partial g(0,0) /\partial u_i$,
 $\partial g(0,0) /\partial h$ and $u_i \partial^2 g(0,0) /$ $\partial u_i
\partial h$ can be obtained.

The results are listed in table \ref{tab:critical_properties}.  The
critical field we obtain is of similar quality as obtain by cluster
Monte Carlo calculation by Bl\"ote \cite{bloete}. The critical
exponent is of substantial less quality. Still, it remains striking to
observe that this quality can already be achieved by considering such
a narrow systems, $W \le 8$ !

\begin{table}
  \begin{center}
\begin{tabular}{|c|c|c|c|}
 \hline
 & Literature & first $L \rightarrow \infty$ & $L=QW \rightarrow \infty$ \\
& &then $W \rightarrow \infty$ & \\
\hline
$H_c$ & 3.0444 &  3.0449 & 3.0439\\
$\nu$ & 0.63029 & 0.61  & 0.62\\
$y_i$ & -0.83 & -1.21 & -1.21$^{(*)}$ \\
\hline
\end{tabular}
\end{center}
\caption[]{Comparison of the results for the critical properties of
 this work with those by Bl\"ote \cite{bloete} \label{tab:critical_properties}.
The first result corresponds to the two step approach: the length is scaled
 to infinity, $L \rightarrow \infty$, afterwards  the width $W \rightarrow \infty$.
The second approach is scaling with fixed aspect ratio. $^{(*)}$ the value for
the irrelevant exponent was taken from the anisotropic scaling.}
\end{table}

\subsection{2-Dimensional, $L,W \rightarrow \infty$ with fixed aspect ratio $L/W$.}

The aspect ratio $Q=L/W$ is fixed and the second step
(\ref{eq:2dscaling}) in the previous approach can be implemented
immediately;
\[
\Delta(h,W,L)=\frac{1}{L} f(h L^{1/\nu} , u_i L^{y_i},Q).
\]
As we consider fields close to the critical field, $|h| \ll 1$, and
the corrections to scaling are expected to be relatively small, this
is expanded up to linear order:
\begin{eqnarray}
L \Delta &=& f(0,0,Q) + L^{y_i} u_i \frac{\partial f}{\partial u_i}
(0,0,Q) \nonumber \\
& &+ h L^{1/\nu} \left (\frac{\partial f}{\partial h} (0,0,Q) +
  L^{y_i} u_i \frac{\partial^2 f}{\partial u_i \partial h}
  (0,0,Q)\right ). \nonumber \\
\end{eqnarray}
The critical properties $H_c$, $\nu$ and $y_i$ are not dependent on the
aspect ratio $Q$. All others, $f(0,0,Q)$, $u_i \partial
f(0,0,Q)/\partial u_i$, $\partial f(0,0,Q)/ \partial h$ and $u_i
\partial^2 f(0,0,Q)/ \partial u_i \partial h$, do. The parameters
were extracted by a least square fit to all of the data available. The
systems considered had a range of properties; widths $4 \le W \le 8$,
$Q=2,2.5,3,3.5,4,4.5,5$ (the fractional aspect ratio's were only
considered with even width $W$) and fields $3.00 \le H \le 3.10$. In
total 471 points were fitted to this behaviour.  An optimal fit can be
made for a large range of parameters. We therefore had to fix
$y_i=-1.21$; the value found in the anisotropic scaling. If the least
square fit is made unrestricted, unreasonable critical exponents will
result. The outcome is listed in table
\ref{tab:critical_properties}.

\section{Conclusion}

In this chapter we derived the critical field $H_c$ and exponent $\nu$
of the two-dimensional ITF. Two routes were taken: one was to scale
the length of the system to infinity and afterwards the width. The
other route followed the usual path of scaling; the linear dimension
was rescaling in both directions.

The two-step scaling approach was prompted by the strengths and
weaknesses of the DMRG. Table \ref{tab:critical_properties} clearly
indicates that this makes sense; the accuracy of the critical field
$H_c$ may be considered unexpected in view of narrow systems
considered. The critical exponents are not too impressive.

Although the essential ingredient in these calculations is the DMRG,
no time was spent on the procedure or the properties. The next
chapter contains that half of the research. The complication that {\it
  was} mentioned, strong limitations on the width of the system, is a
very general feature of the method. Later on, in chapter
\ref{chap:J1J2DMRG} where the frustrated Heisenberg is investigated,
the same problem will reappear.

Apart from two constants, $H_c$ and $\nu$, this chapter has given us
an excellent playground to test the DMRG and a first taste of
finite-size scaling in quantum magnetism. This will be of use for the
frustrated Heisenberg model.

%% file: ITFDMRG.tex
\chapter{Critical properties of the ITF through DMRG calculations \label{chap:ITFDMRG}}

\section{Introduction in the Density Matrix Renormalisation Group}

In this chapter the density matrix renormalisation group (DMRG) is
introduced and studied. This method was originally proposed by White
\cite{white92} in 1992 to resolve some of the problems that the real
space renormalisation group (RSRG) suffers from; in contrast with its
successful treatment of the Kondo problem, the RSRG gives ---relatively---
poor results for strongly interacting quantum chains.

It is well established by now, that the DMRG can handle this class of
lattice problems extremely well, making it the method of choice for
one-dimensional quantum lattice systems. Already in the initial papers
\cite{white92} the promise became apparent. The treatment of the
spin-1 Heisenberg chain \cite{white93} demonstrated the capabilities
of the method. The accuracy was unprecedented.

The method is variational and systematic. If the
computation is scaled up sufficiently, the approximation the ground
state that is made, has to be accurate. The surprise lies in the fact
that this situation can easily be achieved with a minor computational
effort. Some theoretical investigation on the grounds for this
tremendous accuracy has been made \cite{ostlund95}, but no clear
understanding exists at the present. At the same time the applications
have started to diversify. Three main trends can be observed.

The first one is usage of the method in two-dimensional classical
problems. Bursill et al. \cite{bursill94} followed the standard route
of a transfer matrix description to transform a two-dimensional
classical model into a one-dimensional quantum system. Carlon and
Drzewi\'n\-ski \cite{carlon97} build on this extension to settle some of
the outstanding issues in that field.

The second diversification is in the direction of chemical
molecules. Historically, physicists and chemists alike have tried
to develop simple models of chemical compounds that allowed a
theoretical treatment. A good example is the Su-Schrieffer-Heeger
Hamiltonian to model the valance electrons in a polymer. It
simplifies the individual atoms to lattice points, whereby a
one-dimensional quantum lattice model is formed. For a proper
treatment the Coulomb interaction has to remain long ranged and
should not be restricted to individual sites. Despite this
long-range interaction, it seems that the DMRG can still achieve
good accuracy although the issue is not completely decided yet
\cite{bendazzoli99}.

The other direction, upgrading the method to deal with more realistic
molecular models, has also been tried. White and Martin \cite{white99}
have applied the DMRG to the orbital description of water. They
concluded that the accuracy compares favourably with many other
numerical methods.

The third extension is also our main interest, the application to
wider systems. Noack et al. \cite{noack94} focused on the two-leg
Hubbard ladder, while White and Scalapino \cite{white98}
investigate the t-J model on a $16 \times 8$ lattice. We
\cite{ducroo98} were specifically interested in the limitations of
the DMRG on wide systems. It seems that DMRG is not well suited to
handle two-dimensional quantum systems, although with tricks and
brute-force computer power reasonable results can be achieved. In
this chapter we
 explain what the difficulties are.

As a testing ground we use the ITF. The ITF constitutes of a well
understood and simple case of a two dimensional model with a
quantum phase transition. The performance of the DMRG is expected
to decrease in the vicinity of a phase transition, making the ITF
our model of choice. Moreover cluster Monte Carlo calculations
\cite{bloete} have resulted in accurate numerical values for its
critical properties to which we can relate.

The aim of the method is to find an approximation to the ground state
wave function. This is done by 'bootstrapping'; the approximation to
the ground state is improved iteratively. To represent the ground
state a basis in the Hilbert space is necessary. This basis is
systematically improved by the DMRG.  Typically, the route taken is to
start with a basis for one site and iterative enlarge the Hilbert
space by adding sites. Without further ado the size of the basis would
then grow exponential like the Hilbert space at each addition. To
avoid this, the most relevant basis states are selected and all others
are removed.

The route we follow in this chapter is first to introduce and discuss
the density matrix which is the key ingredient of the method. Next the
geometrical properties of the systems we study are listed. Of the two
procedures we implemented, one has extensively been described in the
literature \cite{white92,gehring97}. The other one was introduced by
us \cite{ducroo98} and will be reviewed here. With the method in
place, a link with the RSRG is made. After that, the performance and
the flexibility of the two procedures is compared. The original
proposal by White is the more flexible of the two whereas our
implementation has the potential to be the fastest.

The remainder contains two less related sections, one on the actual
results for the ITF model and one on implementation issues.

\section{The Density Matrix \label{sec:DM}}

\begin{figure}
  \centering \epsfxsize=15cm \epsffile{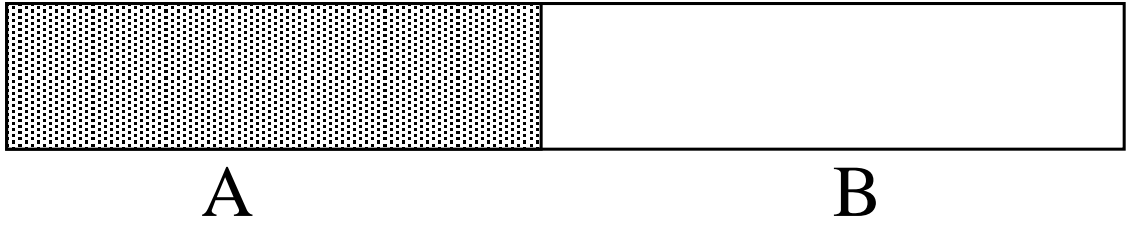}
\caption{The system is divided in a left part $A$ and a right
  part $B$, both containing a number of sites. With both these parts
  an incomplete basis is associated. In the future, a generalisation
  will often be used; the system is split in three or four parts by
  treating either one or both of the intermediate sites (last site of
  $A$ and first site of $B$) separately.
  \label{fig:basicsplitup}}
\end{figure}

Suppose we have a state $|\phi_0 \rangle$. This can be the ground
state, but for the moment we leave that open.  In the DMRG we want
to find a basis to represent this $ |\phi _{0}\rangle $ as well as
possible.  For the entire system, this statement is trivial; the
only required basis vector is the state itself. If on the other
hand the system is divided in a part $A$ and $B$, both containing
a number of sites, as depicted in figure \ref{fig:basicsplitup},
we can find the most relevant fraction of the basis in part $A$.
Our aim is to truncate the basis to this relevant part and remove
the rest of the basis.  A restriction to these basis states should
allow an optimal representation $ |\tilde{\phi }_{o}\rangle$ of
the ground state $ |\phi _{0}\rangle $. By optimal we mean that
the truncation error $P$,
\[
P=\left| |\phi _{0}\rangle -|\tilde{\phi }_{0}\rangle \right| ^{2},
\]
is minimal.  We have to select those $m$  basis states in $A$ that
 minimise the truncation error $P$.

Now that the aim is conceptually clear, we will perform the algebra
and obtain the optimal set. We have a basis $ \left\{ |i\rangle
\right\} $ in $ A $ and $ \left\{ |j\rangle \right\} $ in $ B $ to
represent $ |\phi _{0}\rangle $. Note that these bases are not
necessarily complete. The only requirement is that the space contains
$ |\phi _{0}\rangle $,
\[
|\phi _{0}\rangle =\sum _{ij}\phi _{ij}|i\rangle |j\rangle.
\]
We want to select the $ m $ most relevant basis states $ \left\{
  |\alpha \rangle \right\} _{\alpha =1}^{m} $ from part $ A $. They
are contained in the basis $ \left\{ |i\rangle \right\} $ and
orthonormal, thus
\begin{equation}
|\alpha \rangle =\sum _{i}u_{i}^{\alpha }|i\rangle ~ ,~ \langle \alpha
|\alpha '\rangle =\sum _{i}u_{i}^{\alpha ^{*}}u_{i}^{\alpha '}=\delta
_{\alpha ,\alpha '}. \label{eq:orthonormal}
\end{equation}
The best approximation $ |\tilde{\phi }_{0}\rangle $ can be expressed
in these states,
\[
|\tilde{\phi }_{0}\rangle =\sum _{\alpha j}\tilde{\phi }_{\alpha
  j}|\alpha \rangle |j\rangle \,
\]
and the truncation error $P$, which we want to minimise, is a function
of all these parameters,
\begin{equation}
P=\left| |\phi _{0}\rangle -|\tilde{\phi }_{0}\rangle \right|
^{2}=\sum _{ij}\left( \phi _{ij}-\sum _{\alpha }\tilde{\phi }_{\alpha
    j}u_{i}^{\alpha }\right) ^{2}.  \label{eq:firsttruncationerror}
\end{equation}
The truncation error has to be minimised over the parameters, whilst
the orthonormality condition (\ref{eq:orthonormal}) is fulfilled.
Lagrange multipliers $ \mu _{\alpha \alpha '} $ are introduced for
this purpose. The optimum satisfies
\[
\frac{\partial P}{\partial u_{i}^{\alpha }}=\frac{\partial P}{\partial
  \tilde{\phi }_{\alpha j}}=\frac{\partial P}{\partial \mu _{\alpha
    \alpha '}}=0.
\]
We introduce the density matrix $ \rho _{ii'} $ of $ A $,
\[
\rho _{ii'}=\sum _{j}\phi ^{*}_{ij}\phi _{i'j}.
\]
After some elementary algebra the prefactors $\tilde{\phi }_{\alpha
  j}$ and the Lagrange multipliers $\mu _{\alpha \alpha '}$ can be
obtained,
\[
\tilde{\phi }_{\alpha j} = \sum_i \phi_{ij} u_i^\alpha ~~,~~
\mu_{\alpha \alpha'}=0.
\]
The remaining equations read
\[
\sum _{i'}\rho _{ii'}u_{i}^{\alpha }-\sum _{i'i''\alpha
  '}u_{i}^{\alpha '}u_{i'}^{\alpha '^{*}}\rho _{i'i''}u_{i''}^{\alpha
  }=0.
\]
This equation becomes more transparent when we switch to vector
notation,
\[
\left[ {\bf 1}-\sum _{\alpha '}{\bf u}^{\alpha '}\left( {\bf
      u}^{\alpha '}\right) ^{\dagger }\right] \cdot {\bf
  \vec{\vec{\rho}} }\cdot {\bf u}^{\alpha }={\bf 0}.
\]
This equation can be interpreted as a projector acting on the vector
${\bf \vec{\vec{\rho}} }\cdot {\bf u}^{\alpha }$. This projector
removes all components of the vector along one of the vectors ${\bf
  u}^{\alpha''}$ as
\[
\left[ {\bf 1}-\sum _{\alpha '}{\bf u}^{\alpha '}\left( {\bf
      u}^{\alpha '}\right) ^{\dagger }\right] \cdot {\bf u}^{\alpha''
  }={\bf 0}.
\]
The basis spanned by $ \left\{ {\bf u}^{\alpha }\right\} $ has to
contain all vectors $ \left\{ {\bf \vec{\vec{\rho}} }\cdot {\bf
    u}^{\alpha }\right\}$. A set of eigenvectors of $\vec{\vec{\rho}}
$ clearly satisfies this condition.  Set $ \left\{ {\bf u}^{\alpha
    }\right\} $ to be eigenvectors of $ {\vec{\vec{\rho}} } $ with
eigenvalues $ \lambda _{\alpha } $ so $\vec{\vec{\rho}} \cdot {\bf
  u}^{\alpha }=\lambda _{\alpha }{\bf u}^{\alpha } $.  We know that
\[
\lambda ^{\alpha }={\bf u}^{\alpha ^{\dagger }}\cdot {\bf
  \vec{\vec{\rho}} }\cdot {\bf u}^{\alpha }=\sum _{ii'j}u_{i'}^{\alpha
  ^{*}}\phi ^{*}_{i'j}\phi _{ij}u_{i}^{\alpha }=\sum _{j}\left| \sum
  _{i}\phi _{ij}u_{i}^{\alpha }\right| ^{2}\geq 0,
\]
The complete set of eigenvectors ${\bf u}^\alpha$ is orthonormal, so
\begin{eqnarray}
\sum _{\alpha }\lambda _{\alpha } &=& \sum _{ii'j \alpha}u_{i'}^{\alpha ^{*}}\phi ^{*}_{i'j}\phi
_{ij}u_{i}^{\alpha } = \sum _{ii'j}\phi ^{*}_{i'j}\phi
_{ij} \delta_{ii'} \nonumber \\
&=& \sum _{i}\rho _{ii}=\sum _{ij}\phi_{ij}^2=1.
\end{eqnarray}
To decide which eigenvectors to select, we note that if $m$ vectors
are selected with eigenvalues $\lambda_\alpha$, then
\begin{equation}
\label{eq:truncationerror}
P=1-\sum ^{m}_{\alpha =1}\lambda _{\alpha }.
\end{equation}
This relation can be derived by inserting $u_i^\alpha$ and
$\tilde{\phi}_{\alpha j}$ in the definition of the truncation error
(\ref{eq:firsttruncationerror}).

Given (\ref{eq:truncationerror}) it is immediately evident that the
eigenvectors $ {\bf u}^{\alpha } $ of $ {\bf \vec{\vec{\rho}} } $
corresponding to the $ m $ largest eigenvalues $ \lambda _{\alpha } $
have to be selected to build the basis $\{\bf |\alpha \rangle
\}_{\alpha=1}^m$.

It is worth to continue these algebraic manipulations a bit further
and show how the optimal basis in part $ A $ is related to the optimal
basis in part $ B $. This can be derived step by step. For instance
White \cite{white92} came to similar conclusions when investigating
the consequences of a singular value decomposition. Here we simply
present the result and discuss its consequences.

The relevant bases states $ \left\{ |\bar{\alpha }\rangle ^{m}_{\alpha
    =1}\right\} $ in $ B $ are given by
\[
|\bar{\alpha }\rangle =\sum _{j}v_{j}^{\alpha }|j\rangle ,
\]
where the vector ${\bf v}^\alpha$ is related to the vector ${\bf
  u}^\alpha$ through the relations
\[
v_{j}^{\alpha }=\frac{1}{\sqrt{\lambda _{\alpha }}}\sum _{i}\phi
_{ij}^{*}u_{i}^{\alpha ^{*}}\Longleftrightarrow u_{i}^{\alpha
  }=\frac{1}{\sqrt{\lambda _{\alpha }}}\sum _{j}\phi
_{ij}v_{j}^{\alpha }.
\]
These bases states are orthonormal and they are indeed the largest
eigenvectors of the density matrix $ \xi _{jj'}=\sum _{i}\phi
^{*}_{ij}\phi _{ij'} $ as can be seen by insertion;
\[
\sum _{j'}\xi _{jj'}v_{j'}^{\alpha }=\frac{1}{\sqrt{\lambda _{\alpha
      }}}\sum _{ii'j'}\phi _{ij}^{*}\phi _{ij'}\phi
_{i'j'}^{*}u_{i'}^{\alpha^* }=\sqrt{\lambda _{\alpha }}\sum _{i}\phi
_{ij}^{*}u_{i}^{\alpha ^{*}}=\lambda _{\alpha }v_{j}^{\alpha }.
\]
The ground state obtains a very elegant form in these bases,
\begin{equation}
|\phi _{0}\rangle =\sum _{ij}\phi _{ij}|i\rangle |j\rangle =\sum
_{\alpha=1 } ^{\tilde{m}} \sqrt{\lambda _{\alpha }}|\alpha \rangle |\bar{\alpha
  }\rangle . \label{eq:exactrepr}
\end{equation}
To every basis state $| \alpha \rangle$ in part $ A $ there
corresponds exactly one $|\bar{\alpha} \rangle$ in part $ B $. The set
size $\tilde{m}$ is the minimal number of states available in either
part $A$ or $B$. This equation (\ref{eq:exactrepr}) makes (\ref{eq:truncationerror})
trivial; the approximation $|\tilde{\phi}_0 \rangle$ is simply be given by $|\tilde{\phi}_0 \rangle = \sum_{\alpha=1}^{m} \sqrt{\lambda _{\alpha }}|\alpha \rangle |\bar{\alpha}\rangle $.

 Let us elaborate on the equation (\ref{eq:exactrepr}), as there are a
few important consequences of it. If there are as many states in $B$
as we wish to select in $A$, two further assessments can be made.

First, the truncation error vanishes, $P=0$; there is no approximation
made in transforming $|\phi_0 \rangle$ to $|\tilde{\phi}_0 \rangle$,
$|\tilde{\phi}_0 \rangle=|\phi_0 \rangle$.  All properties of the wave
function in the truncated basis are identical to those in the full
basis.

Second, if to every element of the basis in $B$ a set of quantum
numbers can be assigned, then the distribution of the quantum numbers
in part $A$, contained in the set $\{| \alpha \rangle \}$, is fixed.
Every element has the correct quantum numbers to pair up with the
quantum numbers of his partner in $B$ to form the required quantum
numbers of the wave function $| \phi_0 \rangle$.

\begin{figure}
  \centering \epsfxsize=10cm \epsffile{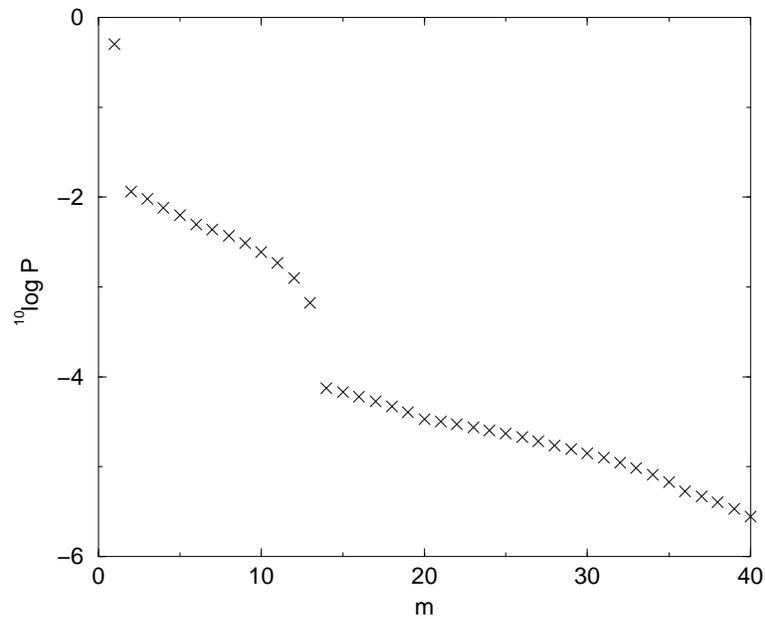}
\caption{The truncation error $P$ as function of the number of
  states kept $ m $. The system is the ITF with periodic boundaries in
  both directions and $ L\times W=6\times 3 $. Part $ A $ contains the
  left $ 3\times 3=9 $ sites. $ H=3.0 $.\label{P_as_function_of_m}}
\end{figure}

In the past few paragraphs we developed a method to distinguish
important states in a part of the system by the density matrix. This
selection criterion can be generalised to incorporate several wave
functions $ |\phi _{\beta }\rangle $. This is not a hypothetical
situation since we will actually target several states in chapter
\ref{chap:J1J2DMRG} .  A weight $ w_{\beta } $ is introduced to
differentiate these states $ |\phi _{\beta }\rangle $ in importance.
We now minimise
\[
P=\sum _{\beta }w_{\beta }\left| |\phi _{\beta }\rangle -|\tilde{\phi
    }_{\beta }\rangle \right| ^{2}.
\]
The underlying framework of minimising $P$ remains linear algebra and
therefore the density matrix becomes the linear superposition of
density matrices for each $ \beta $;
\[
\rho _{ii'}=\sum _{\beta }w_{\beta }\rho _{ii'}^{\beta }.
\]
An approximation scheme based on the density matrix would be useless
if this truncated basis $ \left\{ |\alpha \rangle \right\} _{\alpha
  =1}^{m} $ does not represent $ |\phi _{0}\rangle $ properly. Clearly
the indicator to study is the truncation error $P$.  Figure
(\ref{P_as_function_of_m}) shows that the density matrix can indeed be
an excellent selection criterion as the truncation error $P$ falls off
exponentially with the number of states kept $ m $. In this case only
$ m=13 $ states need to be kept of the $ 2^{9}=512 $ to obtain an
accuracy of $ P \sim 10^{-4} $ in the wave function.

In the following procedure we will make approximations to the ground
state and select the optimal bases to represent them. As these
approximations are not identical to the ground state, some caution is
needed in using the truncation error $P$ as a measure of the accuracy;
As stated before, the truncation error will vanish, $P=0$, if the
environment contains as many states as we want to select. This does
{\it not} mean that $|\tilde \phi_0 \rangle$ is a perfect
representation to the ground state. Instead it means that $|\tilde
\phi_0 \rangle=| \phi_0 \rangle$.

\section{Geometry and symmetries of the ITF}

\begin{figure}
\begin{center}
  \centering \epsfxsize=10cm \epsffile{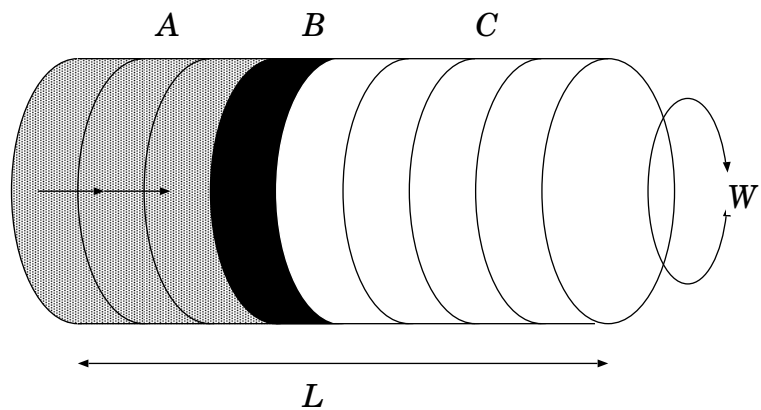}
\caption{The systems we consider are of dimensions $L \times W$.  The system contains three
  parts: a left-hand part $A$ (shaded), an intermediate band $B$
  (black) and a right-hand part $C$ (white).  At every DMRG step part
  $A$ and $B$ are contracted.}
\label{fig:system}
\end{center}
\end{figure}

As a last stop before commencing the description of procedures to
implement the density matrix principle, we have to define the system
geometry and investigate the consequences on the symmetries of the
ITF. As is depicted in figure \ref{fig:system} we consider systems
of sizes $L\times W$. The length $L$ is in general larger than the
width $W$. The system is periodic in the width direction and open in
the length direction. It will be split in a left-hand part $A$ and a
right-hand part $C$, both containing $m$ states. A intermediate band
$B$, containing the complete basis of $2^W$ states, separates them.

The Hamiltonian of such a system contains many symmetries that we can
incorporate in our calculation. The general form of the included
symmetry operators is that they are the direct product of three
components.  Each component acts on one part of the system only. For
example, consider the translation operator ${\cal T}$ in the
width-direction.  This operator is the direct product of three
translations in the individual parts; ${\cal T}={\cal T}_A {\cal T}_B
{\cal T}_C$. The same holds for the reflection ${\cal R}$ in the same
direction,${\cal R}={\cal R}_A {\cal R}_B {\cal R}_C$, and the
spin-reversal operator ${\cal S}=\exp(i\pi \sum_{i,j} {\cal
  S}^z_{i,j}+LW/2)={\cal S}_A {\cal S}_B {\cal S}_C$.

The ground state $| \psi_0 \rangle$ of the system is translational,
reflection and spin-reversal invariant; ${\cal T}| \psi_0
\rangle={\cal R}| \psi_0 \rangle={\cal S}| \psi_0 \rangle=| \psi_0
\rangle$.  For systems of infinite size in the classical ordered
region ($L,W \rightarrow \infty$ and $ H \ll 1$), it will become
degenerate with a state that is spin-reversal anti-symmetric.  In
order to take advantage of the symmetries, the bases of part $A$, $B$
and $C$ are chosen to be eigenvectors of the symmetry operators ${\cal
  T}$ and ${\cal S}$.  ${\cal R}$ is used later on. So if $\{ |a
\rangle \}$, $\{ |b \rangle \}$, $\{ |c \rangle \}$ are the bases of
the individual parts then
\begin{equation}
  {\cal T}_A |a \rangle = e^{ik_a}|a \rangle ~~,~~ {\cal S}_A |a
  \rangle=s_a |a \rangle. \label{eq:symm}
\end{equation}
Similar relations hold for the other two sets. Thus
\[
|\phi_0 \rangle = \sum_{abc} \phi_{abc} |a \rangle |b \rangle |c
\rangle
\]
and application of the symmetry operations together with
(\ref{eq:symm}) yields:
\[
k_a+k_b+k_c = 0 ~\hbox{mod}~ 2\pi ~,~ s_a s_b s_c=1.
\]
It is also possible to set up the program to find the lowest state in
other symmetry classes by forcing other values than $0$ and $1$ in the
equations above.

The Hamiltonian can be written as the sum of Hamiltonians within the
separate parts: ${\cal H}_A,{\cal H}_B$ and ${\cal H}_C$ combined with
interactions between parts: ${\cal H}_{AB},{\cal H}_{BC}$ and ${\cal
  H}_{CA}$; ${\cal H}={\cal H}_A+{\cal H}_B+{\cal H}_C+{\cal
  H}_{AB}+{\cal H}_{BC}+{\cal H}_{CA}$. To show how to implement the
symmetries, we will discuss one element of both types.

First ${\cal H}_A$: it is translational and spin-reversal invariant,
thus
\begin{eqnarray}
  \langle a'| {\cal H}_A |a \rangle &=& \langle a'| {\cal
    T}_A^{-1}{\cal H}_A {\cal T}_A|a \rangle = e^{i(k_a-k_{a'})}
  \langle a'|{\cal H}_A|a \rangle \nonumber \\ &=& \langle a'|{\cal
    S}_A^{-1}{\cal H}_A{\cal S}_A|a \rangle=s_{a'}s_a\langle a'|{\cal
    H}_A|a \rangle \nonumber
\end{eqnarray}
These relations lead to the appearance of delta-functions,
\[
\langle a'| {\cal H}_A |a \rangle = \langle a'| {\cal H}_A |a \rangle
\delta_{s_{a'},s_a} \delta_{k_{a'},k_a}.
\]
It only contains elements within symmetry classes, as one would
expect.

Second ${\cal H}_{AB}$: once again, it is translational and
spin-reversal invariant.  Moreover it can be written as
\begin{eqnarray}
  {\cal H}_{AB} &=&-4\sum_{n=1}^{W} {\cal S}^x_{l,n} {\cal S}^x_{l+1,n}
  \nonumber \\ &=&-4\sum_{n=1}^{W} ({\cal T}_A{\cal T}_B)^{-n+1} {\cal
    S}^x_{l,1} {\cal S}^x_{l+1,1} ({\cal T}_A{\cal T}_B)^{n-1}
  \label{eq:Hab}
\end{eqnarray}
where $l$ is the length of part $A$. ${\cal S}_{i,j}^x$ flips a spin,
so $ {\cal S}_{i,j}^x{\cal S} + {\cal S}{\cal S}_{i,j}^x=0$. Inserting
this and (\ref{eq:symm}) in (\ref{eq:Hab}) gives
\begin{eqnarray}
  \langle a'| \langle b'|{\cal H}_{AB}|a\rangle |b \rangle &=& -4 W
  \langle a'| {\cal S}_{l,1}^x |a \rangle \langle b' | {\cal
    S}_{l+1,1}^x |b \rangle \cdot \nonumber \\ & &\delta_{k_{a'}+k_{b'},k_a+k_b} \cdot
    \delta_{s_{a'},-s_a} \delta_{s_{b'},-s_b}. \nonumber
\end{eqnarray}
This substantially reduces the computational effort. Finally: the
reflection operator ${\cal R}$ is used to make matrix elements like $
\langle a'| {\cal S}_{l,1}^x |a \rangle$ real. In fact we could have
used this last symmetry ${\cal R}$ more, but it only reduces the
effort by a factor of 4 while making the program far more complex.

\section{Procedures \label{sec:bandmethod}}

In the section on the density matrix, a selection criterion was
developed. Now this criterion will be incorporated in a procedure. We
have employed two distinct procedures: the first one was originally
proposed by White \cite{white92}, and trivially extended to two
dimensions by Liang and Pang \cite{liang94}. In this scheme, part $A$,
figure \ref{fig:basicsplitup}, is enlarged iteratively at the
expense of part $B$. Each step one site is moved from part $B$ to part
$A$. The basis on part $A$ will thus incorporate more and more sites,
while the total system size remains the same. Excellent introductions
already exist to which we refer \cite{white92,gehring97}. The second
procedure we introduced ourselves in \cite{ducroo98}. This will be
reviewed next.

When the split-up of figure \ref{fig:system} is made, it is tempting
to use the 1D DMRG method directly: a site is replaced by a band. The
ground state $| \phi_0 \rangle$ of the entire system ABC is calculated
and the optimal basis for block AB is selected through the density
matrix.  However, one runs in severe difficulties. In the section on
the density matrix it was shown that when we select as many states
from part $A$ combined with $B$ as there are in part $C$, the
outcoming basis has to have the appropriate quantum numbers be to
combined with part $C$. So if we also would transform part $C$ in the
density matrix basis, we can write
\[
|\phi_0 \rangle = \sum_{\alpha} \sqrt{\lambda_\alpha} |\alpha \rangle
|\bar{\alpha} \rangle.
\]
We know that ${\cal T} |\phi_0 \rangle ={\cal S}|\phi_0
\rangle=|\phi_0\rangle$, thus
\[
{\cal T}_C | \bar{\alpha} \rangle = e^{ik_{\bar{\alpha}}}
|\bar{\alpha} \rangle ~,~ {\cal S}_C | \bar{\alpha} \rangle =
s_{\bar{\alpha}} | \bar{\alpha} \rangle.
\]
This will immediately dictate the quantum number of the newly built
$|\alpha \rangle$;
\[
{\cal T}_A {\cal T}_B | \alpha \rangle = e^{-ik_{\bar{\alpha}}}
|\bar{\alpha} \rangle ~,~ {\cal S}_A {\cal S}_B| \bar{\alpha} \rangle
= s_{\bar{\alpha}} | \bar{\alpha} \rangle.
\]
Thus $s_\alpha=s_{\bar{\alpha}}$ and
$k_{\alpha}=-k_{\bar{\alpha}}$. The distribution over the symmetry
classes in part $C$ forces the selected states in block $AB$ to be
in ``conjugate''-classes. To overcome this problem, we need to
increase the number of states in part $C$. In that case we can
really make a selection and it allows us to change between
symmetry classes.

In the 1D procedure the solution is to add one extra site to the
environment. The number of states in the environment is then doubled.
In our set-up this would correspond to adding an extra band between B
and C. This is computational far too expensive.

We now introduce variants on White's infinite-size and finite-size
algorithms \cite{white92} that increase the number of states in the
part C.

\begin{figure}
\begin{center}
  \centering \epsfxsize=15cm \epsffile{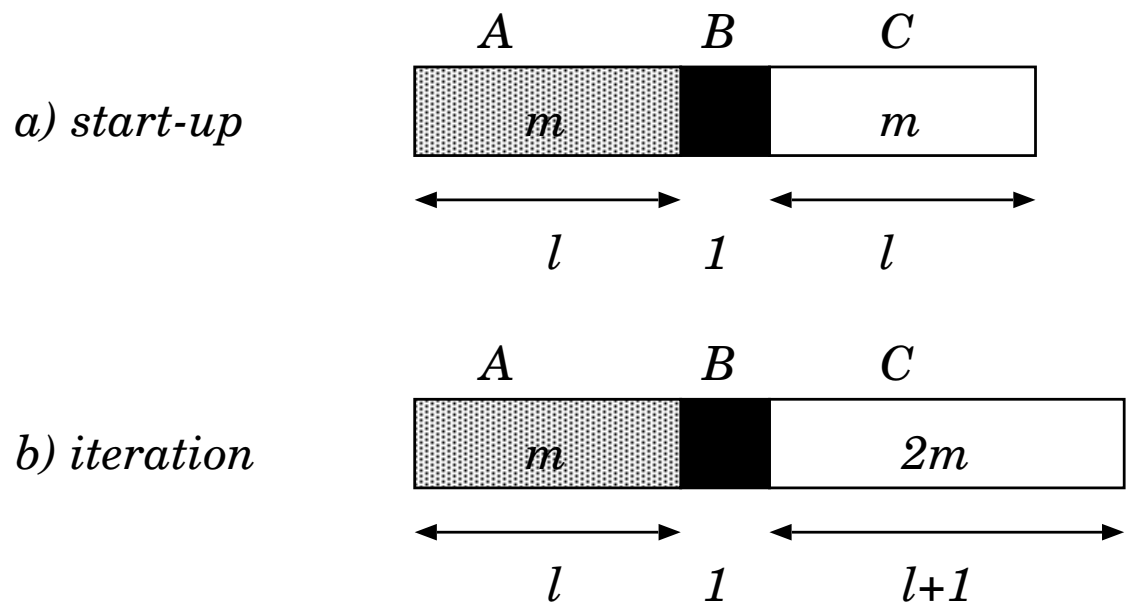}
\caption{A inductive step in the infinite-size procedure consists of
  a start-up to obtain an initial approximation for states in a system
  of length $l+1$ and iterative calculations to make the basis
  converge. The numbers in the rectangle are the number of states in
  the parts. The intermediate band B always contains the complete
  basis of $2^W$ states.}
\label{fig:infinite}
\end{center}
\end{figure}

First we consider our infinite size approach. We only have to describe
one step in the process as it is an inductive method. We have a basis
of $m$ states for a system of length $l$. We then proceed as follows:
\begin{itemize}
\item We construct the combined system as depicted in figure
  \ref{fig:infinite}-a by taking this basis in part $A$ and $C$
  together with the complete basis in the intermediate band $B$.
  ($L=2l+1$)
\item We calculate the ground state $|\phi_0 \rangle$ and obtain $m$
  basis states for a system of length $l+1$ by orthonormalising $\{
  |\beta_c \rangle \}$,
\[
| \beta_c \rangle = \sum_{a,b} \phi_{abc} |a \rangle |b \rangle,
\]
and span the space we previously denoted by $\{ | \alpha \rangle
\}_{\alpha=1}^{m}$.
\item Suppose that block $AB$ has $f$ symmetry classes. To every
  symmetry class we add $m/f$ basis states constructed {\it randomly}
  from the $m2^W$ states in $A$ and $B$. We end up with $m+f\cdot
  m/f=2m$ basis states for a system of length $l+1$.
\item In part $A$ we now take the $m$ basis states for a system of
  length $l$ and in part $C$ we take the newly constructed $2m$ states
  for length $l+1$. ($L=2l+2$) This yields the configuration in figure
  \ref{fig:infinite}-b.
\item We calculate the ground state $|\phi_0 \rangle$ and obtain $2m$
  basis states for length $l+1$ by orthonormalising $\{ |\beta_c
  \rangle \}$. We replace the basis of part $C$ by this basis and
  repeat this step a couple of times ($\sim 3$).
\item We {\it select} from the $2m$ basis states for length $l+1$ $m$
  states on basis of the density matrix.
\end{itemize}
Now we have returned to the original situation with the exception that
$l$ has increased by one. The new ingredient is thus to add $m$ random
states to the basis and iterate until the result has converged.

\begin{figure}
\begin{center}
  \centering \epsfxsize=15cm \epsffile{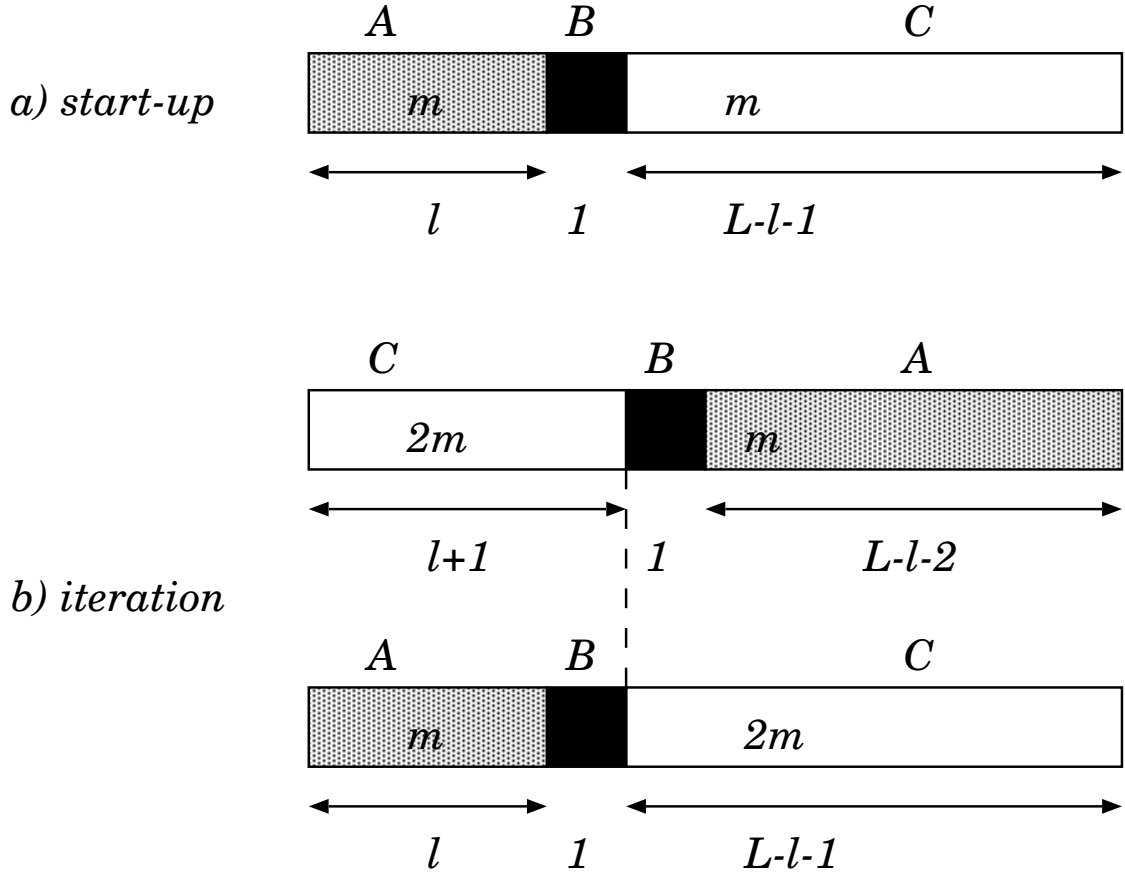}
\caption{A inductive step in the finite-size procedure also consists
  of a start-up to obtain an initial approximation for states in a
  system of length $l+1$. Afterwards we move back and forth between
  lengths $l$ and $l+1$ to make this converge. }
\label{fig:finite}
\end{center}
\end{figure}

In the same line our finite size approach lies. Suppose we have basis
sets of $m$ states for lengths $l$,$L-l-1$ and $L-l-2$, where $L$ is
now fixed and independent of $l$. The iteration step consists of the
following actions:
\begin{itemize}
\item We take the basis for $l$ in part $A$, the basis for $L-l-1$ in
  part $C$ and the complete basis of the band in part $B$. See figure
  \ref{fig:finite}-a.
\item We calculate the ground state $| \phi_0 \rangle $ and obtain a
  basis for length $l+1$ by orthonormalising $\{ |\beta_c \rangle \}$.
\item In the same way as in the infinite-size algorithm we add $m$
  randomly chosen states to this basis.
\item In part $C$ we take the $2m$ basis states for length $l+1$ and
  in part $A$ the $m$ states for length $L-l-2$. This is depicted in
  the first of the two pictures in figure \ref{fig:finite}-b.
\item We calculate the ground state $|\phi_0 \rangle$ and obtain $2m$
  basis states for $L-l-1$.
\item In part $C$ we take the $2m$ basis states for length $L-l-1$ and
  in part $A$ the $m$ states for length $l$; see the second picture in
  figure \ref{fig:finite}-b.
\item We calculate the ground state $|\phi_0 \rangle$ and obtain $2m$
  basis states for $l+1$. These last four steps are repeated a couple
  of times ($\sim 3$)
\item We {\it select} from the $2m$ basis states for length $l+1$ $m$
  states on basis of the density matrix.
\end{itemize}
Once again we have returned to our starting position while increasing
the length $l$ by one. By sweeping through the system we can therefore
systematically improve the basis. This method convergences at a
similar speed as the 1D approach; after 3 sweeps through the system
the final result is achieved.

The approximation scheme to the ground state is both variational and
systematic. The energy of each state in the selected subspace is
higher than the one of the ground state. As this includes the state we
select by minimisation of the energy, our estimate to the ground state
energy is variational.  In section (\ref{sec:calcwave}) we will meet
other examples of variational principles that can be used within the
DMRG scheme.

The systematics come in from the iterations. When a state is selected,
it is truncated and transformed along with the basis. The next
iteration it is used as a starting point for the minimisation routine.
If the density matrix eigenvalues drop off fast enough and the
truncation thus does not severely alter the state, the energy of the
starting point will be almost the same as the outcome of the last
iteration and the during the minimisation this estimate can improve
further.

After several sweeps through the system, the energy will start to
oscillate. This effect is small, but the cause of it gives ground for
further improvement. Suppose we have found a state $|\phi_0 \rangle$
at the previous iteration. This state is truncation to
$|\tilde{\phi}_0 \rangle$. In the truncation a part of the state is
lost, so $|\tilde{\phi}_0 \rangle \neq | \phi_0 \rangle$. The
truncated space of $|\tilde{\phi}_0 \rangle$ lies in the subspace of
$|\phi_0 \rangle$ though. We know that $|\phi_0\rangle$ corresponds to
the lowest energy in that subspace, thus
\[
\frac{\langle \phi_0 | {\cal H} | \phi_0 \rangle}{\langle
  \phi_0|\phi_0\rangle} \le \frac{\langle \tilde{\phi}_0 | {\cal H} |
  \tilde{\phi}_0 \rangle}{\langle
  \tilde{\phi}_0|\tilde{\phi}_0\rangle} .
\]
$|\tilde{\phi}_0 \rangle$ is transformed and thereby embedded in an
other subspace. It is used as a starting point for the next
minimisation. The next estimate $|\bar{\phi}_0 \rangle$ will thus have
a lower energy,
\[
\frac{\langle \tilde{\phi}_0 | {\cal H} | \tilde{\phi}_0
  \rangle}{\langle \tilde{\phi}_0|\tilde{\phi}_0\rangle} \ge
\frac{\langle \bar{\phi}_0 | {\cal H} | \bar{\phi}_0 \rangle}{\langle
  \bar{\phi}_0|\bar{\phi}_0\rangle}.
\]
No strict statement can be made on the relation between the energies
of $|\phi_0 \rangle$ and $|\bar{\phi}_0 \rangle$.  The energy does not
need to decrease monotonically and will start to oscillate.

Once this happens, we can assume that the basis states we use, lie in
the most relevant symmetry classes and it is no longer necessary to
increase at every location in the system the number of states in part
$C$ from $m$ to $2m$.  Leaving out the iterations to increase the
number of states in part $C$, the previous estimate $|\phi_0\rangle$
can still be represented exactly after the basis truncation ,
$|\tilde{\phi}_0 \rangle = | \phi_0 \rangle$. This is explained in the
paragraph after equation (\ref{eq:exactrepr}). The ground state
$|\bar{\phi}_0 \rangle$ of the next step satisfies
\[
\frac{\langle \phi_0 |{\cal H}| \phi_0 \rangle}{ \langle \phi_0
  |\phi_0 \rangle}= \frac{ \langle \tilde{\phi}_0 |{\cal H}|
  \tilde{\phi}_0 \rangle}{ \langle \tilde{\phi}_0 | \tilde{\phi}_0
  \rangle} \ge \frac{\langle \bar{\phi}_0 |{\cal H}| \bar{\phi}_0
  \rangle}{\langle \bar{\phi}_0 |\bar{\phi}_0 \rangle}.
\]
The energy will now decrease monotonically.

\section{Connection with the Renormalisation Group}

The name DMRG has led many people to believe that it is another
implementation of Wilson's renormalisation group. This is incorrect in
at least two respects. First, this method does not contain a semigroup
operation; where in the renormalisation group the Hamiltonian is
mapped onto itself with different parameters, here we change ---read
truncate--- the Hamiltonian all together. The name group is
unfortunate and misleading, but it is well established and to avoid
further confusion, we will stick to the name DMRG.

The second distinction lies in the fundamental difference of basis
selection. The DMRG first calculates the approximate ground state of
the entire system and afterwards makes a selection via the density matrix
of this state. The renormalisation group, on the other hand, does not
consider the properties of the entire system. The basis states are
solely selected on merit of the energy within their own part of the
system.

\begin{figure}
  \centering \epsfxsize=12cm \epsffile{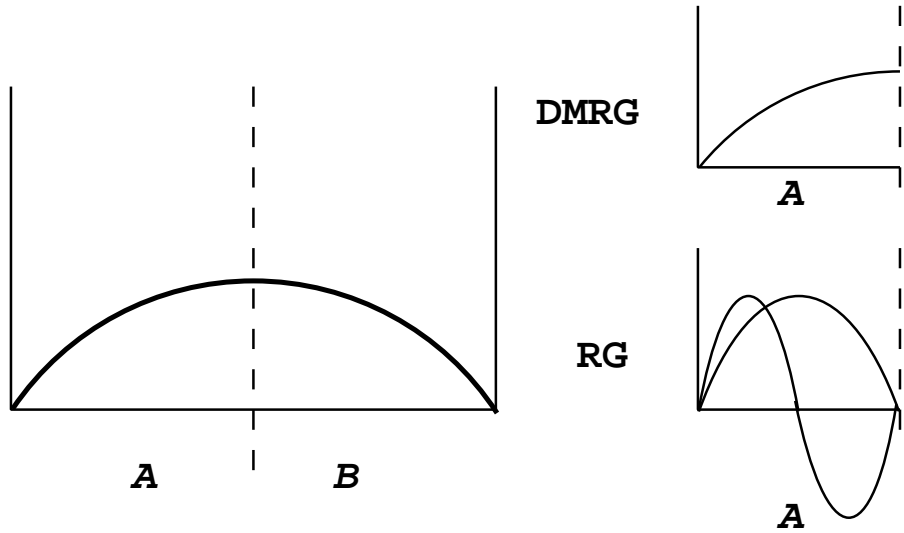}
\caption{\label{fig:particleinbox} On the left the wave function of a particle
  in a box is depicted. At current a continuous version of the DMRG
  does not exist, so one should consider the location in the box a
  very fine grid of discrete points. On the right we see the
  selections of the DMRG, one state suffices, and of the ordinary
  renormalisation group (RG) where many states are needed.}
\end{figure}

The well-known example \cite{white92b,noack93} of a particle in a box
with impenetrable walls illustrates this selection criterion clearly.
By its graphical simplicity figure \ref{fig:particleinbox}
demonstrates that in the DMRG one basis state suffices whereas the
renormalisation group needs many to represent the wave function
properly.

\section{Performance}

In this section a discussion is given on the accuracy that can be
achieved by the DMRG procedures. First we consider issues that affect
both methods, afterwards a comparison between the two methods is made.

\subsection{General limitations. \label{sec:generallim}}

\begin{figure}
\begin{center}
  \centering \epsfxsize=12cm \epsffile{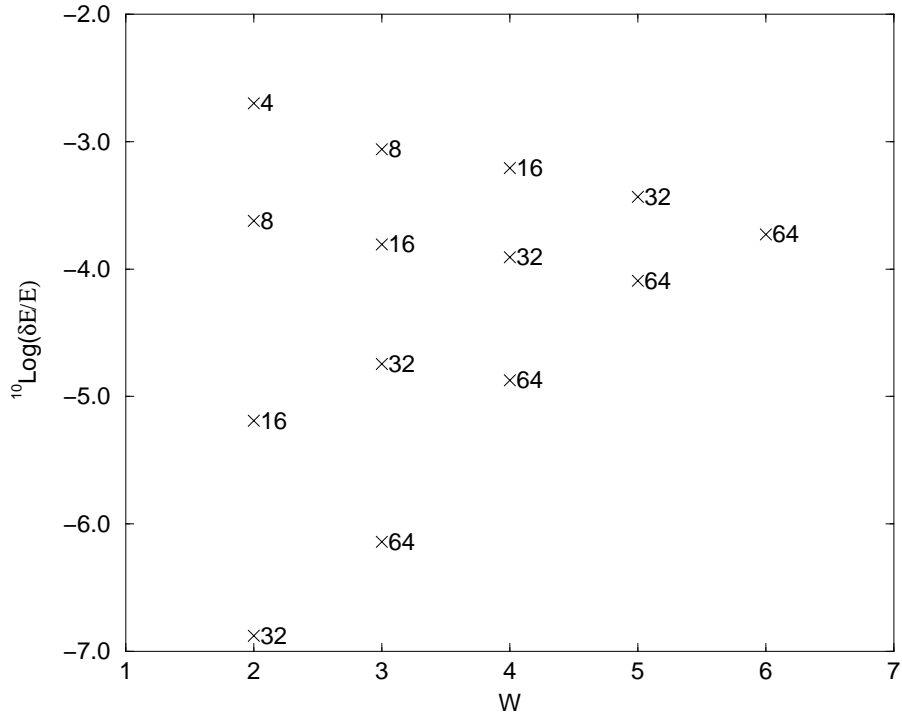}
\caption{The accuracy of the DMRG method for given number of states $m$
  (numbers in graph) as function of the width $W$. $H=3$ and $L=20$.
  The system is periodic in both directions.  The reference value is
  taken from a DMRG calculation with $m=128$.}
\label{fig:error}
\end{center}
\end{figure}

The most striking limitation is that only systems of small width can
be handled. As the width $W$ increases, the number of states $m$ kept
in the procedure has to increase exponentially to maintain the
accuracy \cite{liang94}. In figure \ref{fig:error} this is
exemplified for the ITF. To get a flavour of the background to this
behaviour, we will discuss a pathological example where this statement
can be proven exactly.

\begin{figure}
  \centering \epsfxsize=12cm \epsffile{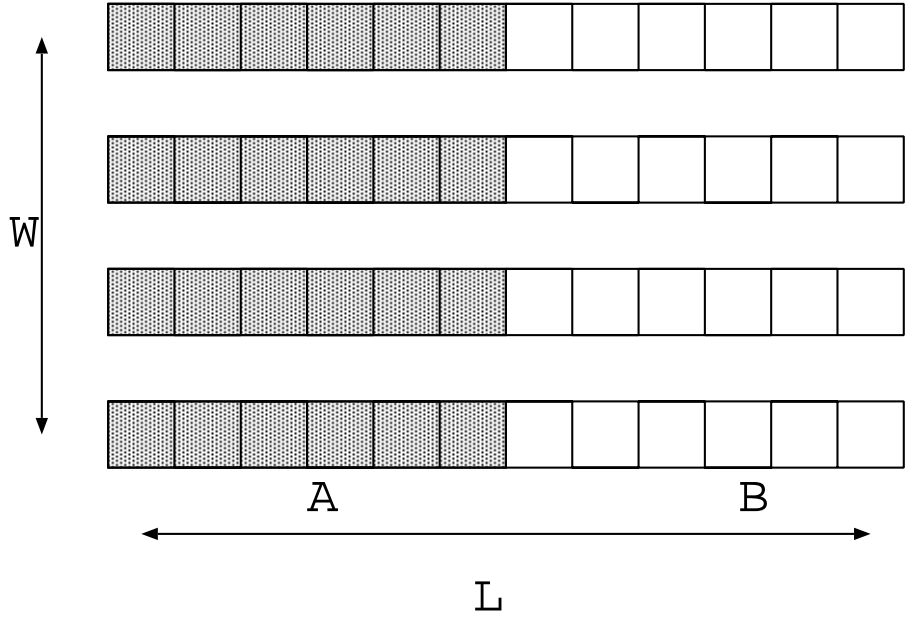}
\caption{\label{fig:chains} The system contains only interaction
  within the chains. Between the chains no interaction exists.  The
  system is split up, and the DMRG procedures grow part $A$ on expense
  of part $B$. Either site by site or row by row.}
\end{figure}

Consider a system containing a set of chains as depicted in figure
\ref{fig:chains}. For each chain a DMRG calculation can be performed
and it is found that $m_0$ states are required to obtain a given
accuracy. To achieve the same accuracy for this system of width $W=4$,
for each chain $m_0$ states have to be preserved as there is no
interaction between the chains. The basis of the entire part $A$ is a
product of bases for the individual chain pieces. Therefore there are
$m=m_0^W$ states necessary for the same accuracy; a clear proof of the
exponential growth of the number of states $m$ with increasing width
$W$.

In the study of the ITF we are fortunate enough to go somewhat further
than this rough-and-ready argument. Far from the phase transition ($H
\ll 1$ or $H \gg 1$) a connection can be made with perturbation
theory.

Consider the quantum disordered phase on a periodic system. Split the
Hamiltonian into a unperturbed part ${\cal H}_0= 2H\sum_{i,j} {\cal
  S}_{i,j}^z$ and a perturbation ${\cal V}=- 4 \sum_{i,j} {\cal
  S}_{i,j}^x ({\cal S}_{i+1,j}^x+ {\cal S}_{i,j+1}^x)$. We split the
periodic, rectangular system of size $L\times W$ again in two parts;
$A$ and $B$ of sizes $l\times W$ and $(L-l)\times W$ where $l$ is an
arbitrary length smaller than $L$.  They both contain $2W$ spins that
border the other part. The unperturbed ground state $|0\rangle $ has
all spins pointing down in the ${\cal S}^z$-direction. It is the
direct product of two equivalent states restricted to $A$ and $B$; $|0
\rangle = |0 \rangle_A |0\rangle_B$. We know that ${\cal H}_0|0
\rangle = -HLW|0\rangle=E_{0}|0 \rangle$.  Perturbation theory yields
\begin{equation}
  |\phi_0 \rangle = |0\rangle + \frac{1}{E_{0}-{\cal H}_0} {\cal V} |0
  \rangle + {\cal O}\left (\frac {1}{H^2} \right )
\end{equation}
The perturbation flips a pair of neighbouring spins. This pair can
be in a single part or it can cross the border between both parts.
In the latter case the spins are adjacent across the boundary
between part $A$ and $B$. Define $\{ |a \rangle_A \}$ to be the
set of states with the flipped pair in part $A$. Analogous for $\{
| b \rangle_B \}$. Moreover let $\{ |n \rangle_A \}$ be the set
with one spin flipped on the $n$th boundary site with $B$ and
define in an equivalent manner $\{ | n \rangle_B \}$. The
perturbation expansion can now be rewritten
\begin{eqnarray}
  | \phi_0 \rangle &=& | 0 \rangle_A |0\rangle_B + \frac{1}{2H} \left
  ( \sum_a |a \rangle_A |0 \rangle_B + \sum_b |0 \rangle_A |b
  \rangle_B +\sum_n |n\rangle_A |n \rangle_B \right) \nonumber \\
&&+{\cal O} \left(
  \frac{1}{H^2} \right ) \nonumber \\ &=& \left ( |0 \rangle_A +
  \frac{1}{2H} \sum_a |a \rangle_A \right) \left ( |0 \rangle_B +
  \frac{1}{2H} \sum_b |b \rangle_B \right ) + \frac{1}{2H} \sum_n
  |n\rangle_A |n \rangle_B \nonumber \\
& & + {\cal O} \left(\frac{1}{H^2} \right ).
  \label{eq:pert}
\end{eqnarray}
As $H \gg 1$, it is necessary to reproduce {\it all} these terms for
an accuracy which is equivalent to the first order perturbation
theory. The minimal number of states needed in part $A$ is therefore
$1$ for the first term in (\ref{eq:pert}) plus $2W$ for all the
boundary terms. We have confirmed this prediction explicitly in both
the small and large $H$ limit ( $H=1/50,50$).

The same line of reasoning also holds for the second and higher order
perturbation terms. We expect for an error comparable to the $n$th
order perturbation theory that $m \sim W^n$, $\delta E \sim (1/H)^n$.
This is always an upper bound for number of states $m$ needed, $m<W^n$
for a given accuracy $\delta E \sim (1/H)^n $. Only when the different
orders in perturbation theory become distinguishable in size ---the
limit of large $H$--- the equivalence holds. Through combinatorics
even the prefactors can be calculated.

Both arguments above indicate that it is of the utmost importance to
limit the interaction between the parts, although the statements are
not identical. If there exists a representation in which the
interaction between the parts is still large, but where the
perturbation expansion needs only few states, the DMRG easily obtains
highly accurate results within that representation.

The question what basis to start from is still open. Xiang
\cite{xiang96} chooses to treat the Hubbard model in momentum space.
This representation introduces an extra quantum number that can be
conserved: the momentum. As this leads to a relatively strong
restriction on those basis states of the different parts that can
combine, the number of states kept $m$ could be increased substantial.
Moreover we expect to have a better starting point for the ground
state by this conservation. On the other hand, from the view of
perturbation theory it is unclear whether the orders in perturbation
theory can be reproduced easily as the interaction between the parts
becomes much more complex and much more extended.

In the case of fermions it is easy to change the representation of
the particle creation and annihilation operators from real space
to for instance momentum space. It is also possible to switch to
the mean field quasi particle representation in which the ground
state contains no quasi particles. This is done in the BCS theory
by means of a Bogoliubov transformation. For a DMRG calculation
such a representation may also be not ideal despite the fact that
the mean field ground state can be represented by exactly one
basis vector (with no quasi particles present). It could be a
source of bias in the final result.

Other attempts have been made to avoid the issue of limited width
all together by smoothing the boundary conditions and thereby
reducing the boundary effects \cite{vekic93}.

\begin{figure}
  \centering \epsfxsize=12cm \epsffile{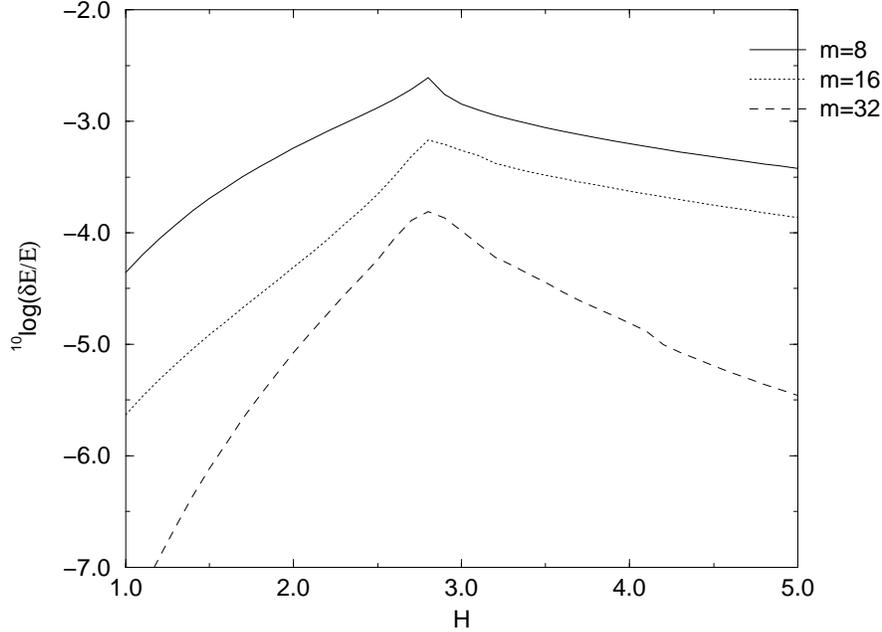}
\caption{The accuracy of the DMRG method for different number of states $m$
  (numbers in graph) as function of the field $H$. The system is
  periodic in both directions with dimensions $W=4$ and $L=20$.  The
  reference value is taken from a DMRG calculation with $m=64$. The critical field of the two-dimensional system is $H_c=3.044$}
\label{fig:errorHz}
\end{figure}

Another restriction is related to phase transitions. A clear
indication of the vicinity of a phase transition is divergence of the
correlation length. At the phase transition algebraic behaviour of the
correlation functions is expected. This is conflicting with the manner
in which the ground state is built by DMRG. Successive basis rotations
and truncations where at every step a new site is included in the
basis clearly favour exponentially decaying correlation functions
\cite{ostlund95}. For the ITF the accuracy of the DMRG at the phase
transition indeed decreases although it remains unclear whether this
is a general feature of phase transitions. Figure \ref{fig:errorHz}
illustrates this.

Several basis rotations and truncations applied to operators cause
them to deteriorate. For the intermediate band or site the basis is
complete and the representation of the operators on these sites is
perfect, but for sites further away from the part boundaries, some
basis truncation have been applied and the representation of the
operators has lost the connection with the states truncated from the
basis. A consequence of this is that the correlation functions for
sites that are simultaneously or successively included in a part by
the procedure are of higher quality than those between sites that lie
further apart.

\subsection{Comparison of both methods.}

In section \ref{sec:bandmethod} a version of the DMRG was introduced
that added bands to a part. White's original proposal was to add sites
to a part.  These two procedures, site and band, can be compared on
two grounds:
\begin{itemize}
\item Computational effort with given accuracy.
\item Flexibility of the procedure.
\end{itemize}
Let us address these criteria in the same order.

A straightforward comparison of the two computer programs favours the
site method heavily. Unfortunately this says more about the software
than about the quality of the method. A fair approach contains two
steps: first the accuracy as function of the number of states kept $m$
is related. Afterwards the computational effort of both procedures as
function of the number of states $m$ is estimated. We have done both
for the ITF.

\begin{figure}
  \centering \epsfxsize=12cm \epsffile{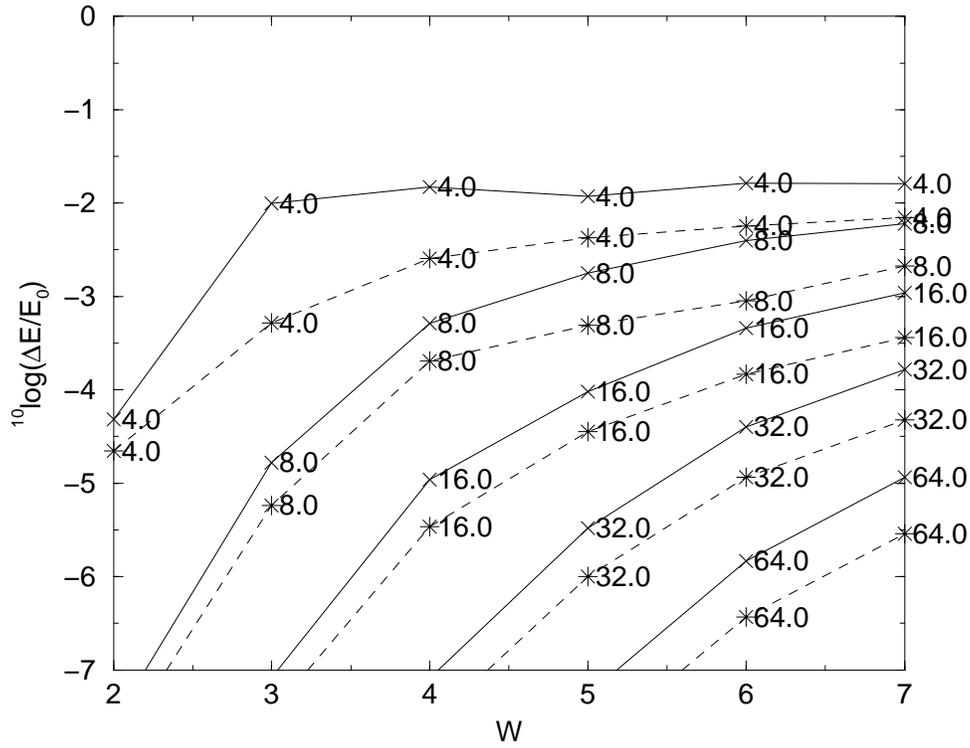}
\caption{The accuracy
  of the site-procedure (asterisk '*') is compared to that of the
  band-procedure (cross $\times$.) for various system widths $W$ and
  number of states $m$ (numbers in the graph). The transverse field
  $H=3$ and the system length $L=20$. The geometry of a cylinder is
  used; Periodic boundary conditions in the short width-direction and
  open ones in the long length-direction.}
\label{fig:sitevsband}
\end{figure}

Figure \ref{fig:sitevsband} depicts the first step. The band method
clearly needs fewer states for the same accuracy.

The bottle neck in the computational effort is finding the ground
state $|\phi_0 \rangle$ in a given subspace. This is done by applying
the Hamiltonian in the order of 5 times to the wave vector. In the
site procedure this costs $m^3 \cdot W/2^2$ operators per projection.
The factor $1/2^2$ follows from conservation of the spin reversal
operator ${\cal S}$. To sweep through the system there are thus
$t_{site}=5 \cdot m^3 \cdot L \cdot W^2/2^2$ operations needed. The
band method has a far larger space in part $B$ but it is possible to
use translational invariance in the width direction and fewer steps
have to be taken to sweep through the system. Per step it costs $m^3
\cdot 2^W/W$ operations ($m \ge 2^W$). With the moving back and
forward at one location ($3$ successive diagonalisations), we get
$t_{band}=3 \cdot 5 \cdot m^3 \cdot L \cdot 2^W/W$ operations for one
complete sweep. Up to width $W \sim 8$ these two estimates are
similar,
\[
\frac{t_{site}}{t_{band}} = \frac{5 m^3 L W^2/2^2}{15 m^3 L 2^W/W}=
\frac{1}{3} \frac{W^3}{2^{W+2}},
\]
with the band method achieving higher accuracy with the same number of
states $m$. After that the site method clearly becomes faster.

To summarise the comparison of the computational performance: as
long as the systems do not become too wide ($W \le 8$ ) , both
methods are comparable in efficiency.  The band procedure is
slightly preferential as it could be faster and the ground state
is more symmetric. The difference in calculation time we observed
is mainly the result of the quality of the software and
specifically the use of the BLAS \cite{BLAS} routines. The
difference in calculative performance is not conclusive.

Next we compare the flexibility. The computational performance is in
practical situations irrelevant as a large fraction of the effort
spent in research is dedicated to probing properties of models by
changing parameters and such. This requires a degree of flexibility in
the procedure. By this criterion the site procedure is clearly
preferential as both the geometry and interactions can easily be
changed.  The band procedure obtains its better calculative
performance from the translation invariance of the ground state. This
fixes both the geometry and the interactions to support translational
invariance. The only advantage of the band method in the terms of
flexibility is that by strictly conserving the translational symmetry,
also states with difference momentum can be targeted.

The overall conclusion is that if one wants to test a model the site
procedure is preferable, whereas if maximum accuracy is required the
band procedure should be used.

\section{Some considerations on the implementation}

Choosing the right software language and tools is very much dependent
on both the hardware and software available. As the progress in these
is rather swift, remarks in this section will be relatively soon
outdated. Still it can be of tremendous use having an idea of what the
implementation could look like. We state our comments going from more
conceptual to language specific ones.

In implementing the Hamiltonian, operators of the form ${\cal S}_1
\cdot {\cal S}_2$ have to be applied to a wave function $|\phi
\rangle$. It is recommended to apply the members of this pair
successively, ${\cal S}_1 \cdot ({\cal S}_2 | \phi \rangle )$, not
simultaneously, $({\cal S}_1 \cdot {\cal S}_2 ) | \phi \rangle$.
Algebraically these two multiplications are identical, but
computational the second one is far more expensive than the first.

Many sophisticated matrix manipulation libraries exist and these can
be used. For that purpose the intermediate sites have to be merged
with the larger parts; in the site procedure, the left intermediate
site can be merged with the left part and the right intermediate site
with the right part. In this fashion we obtain two bases, one for the
left part, $\{ |i \rangle \}$, and one for the right part, $\{ | j
\rangle \}$. Now a wave function $|\phi\rangle$ can be expressed by a
matrix $\Phi$,
\[
|\phi\rangle = \sum_{i,j} \Phi_{i,j} |i \rangle |j \rangle.
\]

The use of this representation touched immediately at another
implementation matter; applying operators to this wave function now
transforms into matrix multiplications. There are quite a few libraries
available that are highly optimised to perform this task of matrix
multiplication, most notably BLAS \cite{BLAS}.

To end our list of remarks, the programming language that we used is
C++. There are two reasons to do this:
\begin{itemize}
\item At current, it is one of the most widely used languages. This
  ensures the availability of tools, libraries and sophisticated
  compilers.
\item C++ allows object oriented programming. This requires a
  different way of analysing a problem than the well-establish
  procedural approach, but it makes the code more transparent and
  reusable. Especially in combination with packages like STL
  \footnote{standard template library, part of the C++ library} this
  provides a powerful programming environment.
\end{itemize}
Naturally C++ is not unique in either of these properties. FORTRAN90
also fits at least the second argument, but is less widely used.

Object oriented code in C++ is slower than FORTRAN or FORTRAN90, but
in a DMRG implementation by far the most time is spend on finding the
ground state. As this is code-wise only a fraction of the entire
program, this can easily be implemented in a procedural fashion using
fast libraries (BLAS).

\section{Results for the ITF}

The scaling analysis of the previous chapter is based on DMRG
calculations. The ground state of the ITF model lies in the space with
spin reversal quantum number ${\cal S}=1$. The first excitation
appears when one spin is flipped, so ${\cal S}=-1$ for that state.  We
calculate in both spaces the ground state energy $E_0^{{\cal S}= \pm
  1}$ and the gap $\Delta$ is given by
\[
\Delta = E_0^{{\cal S}=-1}-E_0^{{\cal S}=1}.
\]
The gap is thus the difference between two large numbers. This causes
the relative error to increase by two orders of magnitude and the
highest achievable accuracy is needed.

Experimentally it has been observed \cite{white96} that the truncation
error $P$ becomes proportional to the error in the energy. This is not
surprising as the truncation error $P$ is a direct indicator of
quality of the basis truncation.  The estimate of the energy is
improved by extrapolating to zero truncation error, $P=0$.

As we mentioned earlier in section (\ref{sec:generallim}), the
accuracy is highly dependent on the length of the boundary between the
different parts. In a periodic system, each part is coupled on both
sides to the rest of the system, so then the boundary is twice as long
as in a open system. With the accuracy already so much under pressure,
we cannot afford to double the length of the boundary.

Unfortunately it is not possible to study system of larger width than $W=8$ with
the DMRG. This means that the DMRG method has  strong competition from  
 cluster Monte Carlo algorithms \cite{bloete}. It has to be
stressed though, that in cases
where the Monte Carlo methods fail as a consequence of the sign-problem, DMRG can provide an excellent replacement.

%% file: SBMF.tex
\chapter{Spin stiffness and finite-size scaling of  the frustrated Heisenberg Model \label{chap:SBMF}}

\section{Introduction}

The interest in the frustrated Heisenberg model arose with the
discovery of high $T_c$-super{-}conductivity. The ground state of the
undoped compound shows long-range antiferromagnetic order, which can
be well described by the Heisenberg model. However introduction of a
small number of holes destroys the long range order and triggers
superconductivity. In a way doping and frustration have similar
effects on the long range magnetic order. The underlying similarity is
not really understood but efforts are made to connect them, see for
instance the book by Auerbach \cite{auerbach}. To improve the
understanding of these effects, it is of interest to study a
frustrated Heisenberg model.

In more recent years it has become a research topic of its own. It may
serve as an example of a system exhibiting a quantum phase transition
going from N\'{e}el order to collinear order with increasing
frustration.  Hopes are that new and so far unknown phases will be
discovered in the intermediate regime.

A link can been made between this frustrated Heisenberg model and the
material $CaV_4O_9$. If the spins are positioned on a rectangular
lattice with every fifth site unoccupied, the model may describe
$CaV_4O_9$ \cite{taniguchi}.

Our intention here is to employ the spin stiffness as a means of
measuring the order present in the system. In this chapter, the
spin stiffness will be calculated in the Schwinger-Boson Mean Field
(SBMF) theory \cite{auerbach,feiguin,mila}. This mean field theory
introduces the mean fields in the interactions between spins and
allows for the ground state to be rotationally invariant. Afterwards
the spin length is no longer conserved to be $\frac{1}{2}$ and we will
discuss the consequences in some detail at the end of section \ref{sec:SBMF}.

The spin stiffness in the SBMF serves as a guideline for the effects
of finite-size scaling, which is an indispensable tool for the
numerical calculations using the DMRG method. The next chapters
contain the numerical parts of our investigations; in chapter \ref{chap:J1J2DMRG}
the DMRG method is combined with finite-size scaling to obtain the spin stiffness of the 
two-dimensional system. In chapter \ref{chap:FNMC} the correlation function are studied by
a combination of the DMRG and  Green Function Monte Carlo simulations.

\section{The frustrated Heisenberg Model}

This model describes interacting quantum spins on distinct lattice sites. The
spins $\vec{\cal S}_j$ have length $\frac{1}{2}$. Hence $|\vec {\cal
  S}_j |^2 =\frac{3}{4}$.  The Hamiltonian is given by
\[
{\cal H}= J_1 \sum_{\langle ij\rangle} \vec{\cal S}_i \cdot \vec{\cal
  S}_j + J_2 \sum_{[ij]} \vec{\cal S}_i \cdot \vec{\cal S}_j.
\]
It incorporates interactions between nearest-neighbour pairs $\langle
ij \rangle$ with strength $J_1$ and next-nearest-neighbour pairs
$[ij]$ with strength $J_2$. We will focus on a square lattice in two
dimensions with $L \times W=N$ sites. At the end of this chapter we
will consider systems where $L \neq W$.  The next-nearest-neighbouring
sites are connected through the diagonals of the lattice. The lattice
is bipartite; it can be split into equivalent sublattices A and B. If
site $j$ is in sublattice A then all its nearest-neighbours are in
sublattice B.  The function $(-1)^j$ will therefore be $1$ if $j \in
A$ and $-1$ if $j \in B$.

Both $J_1$ and $J_2$ are taken to be non-negative. If $J_2/J_1$ is
small the dominating $J_1$ term describes the usual antiferromagnetic
interaction and the ground state will be N\'{e}el ordered. For instance
spin-up on sublattice A and spin-down on sublattice B or $(-1)^j
\langle {\cal S}_j^z \rangle = m_s$.  For the opposite case, $J_2/J_1$
large, the system decomposes in two N\'{e}el ordered sublattices
which, however, have the same quantisation axis.  This is the
so-called collinear ordering. In this ordering alternating strips of
up and down-spins will occur. Suppose the strips are oriented along the
$x$-axis, then $(-1)^{y} \langle {\cal S}_{j}^z \rangle =m_s$ where
the $j$-th spin in on the $y$-th row.

Clearly these couplings frustrate each other on a square lattice when they are comparable
in size. It is beforehand unclear what will happen then. Whether a
different phase exists in the region between the N\'{e}el ordering and
the collinear ordering remains uncertain. In the literature there have
been speculations ranging from dimer - to disordered phases
\cite{einarsson,schulz,zhitomirsky,sachdev}. In all of these cases it
is conjectured that the intermediate phase has no long-range order.
The correlation length thus becomes finite and a gap in the energy
spectrum opens up. This energy gap is a good indicator for the
existence of an intermediate phase. Unfortunately, it is very hard to
achieve high enough accuracy in the numerics as it involves the
subtraction of two ${\cal O}(N)$ numbers (the ground state energy is
subtracted from the energy of the first excitation) to yield a ${\cal
  O}(1)$ number, the gap.

Another indicator for long-range order is the spin stiffness. This
quantity potentially allows for higher numerical accuracy. In the current
chapter the spin stiffness of both the N\'{e}el and the
collinear phase will be derived in the Schwinger-boson mean field approximation.
Part of the material in the chapter is published in \cite{ducroo97}.

\section{The spin stiffness \label{sec:spin-stiffness}}

To investigate the phase diagram, one could look for order parameters.
However, if the frustration increases, a new and different ordering
might appear with an unknown order parameter. So, many different order
parameters would have to be examined. There exists a far more general
approach. It is known that when a continuous symmetry is broken, a
Goldstone mode appears. This is a 'symmetry-restoring' excitation. For
this kind of spin model, these are called spin waves, as they are wave-like fluctuation in the spin orientations. In \cite{halperin69} it was
shown that for both types of order considered here (N\'{e}el and
collinear) the dispersion relation is linear ($\omega=c |{\bf q}|$)
for low energies. The velocity $c$ of this mode satisfies
\begin{equation}
  c=\sqrt{\frac{\rho_s}{\chi_{\perp}}} \label{eq:goldstone}.
\end{equation}
Here $\rho_s$ stands for the spin stiffness and $\chi_{\perp}$ is the
magnetic susceptibility of the system perpendicular to the orientation
of the ordering. A positive spin stiffness ($\rho_s > 0$) is an
indication that a broken-symmetry is present, although the actual
order parameter might be unknown! Therefore it is an excellent measure
of magnetic order that arises from a broken symmetry. For the Heisenberg model
an order related to a local organisation of spins in the intermediate
range of frustration has been suggested
\cite{einarsson,schulz,zhitomirsky,sachdev}. The global rotational
invariance is not broken, therefore the spin stiffness should be zero,
$\rho_s=0$.  In short: the spin stiffness is an excellent indicator to
distinguish between a broken symmetry ground state ($\rho_s \neq 0$)
and other ground states ($\rho_s=0$).

\subsection{Direct derivation at zero temperature}

Fisher, Barber and Jasnow \cite{fisher73} showed that the spin
stiffness $\rho_s$, that is related to the dynamics of the system, can
be obtained by a static twist in the order parameter. They imposed
boundary conditions to achieve this twist. Instead of following them
in detail, we will directly focus on the order parameter. In
\cite{leeuwen} this route has also been followed and we could reduce
the formula there to zero temperature, but the transition from finite
temperature to zero temperature is subtle and we will instead make an
explicit derivation of $\rho_s$. In this way the concepts involved can
be made more transparent and subtleties concerning the necessity of
periodic boundaries can be addressed more directly. The resulting
expression for the stiffness appears in several places in the
literature, e.g. \cite{einarsson}.

We consider the case of N\'{e}el order as an example.  Afterwards it
will be shown that the expression we found holds in more general
cases. Before we start let us mention that a two-dimensional system
exhibiting long-range order at zero temperature is not conflicting
with the Mermin-Wagner theorem \cite{mermin}. The theorem only forbids
long-range order at finite temperature.

The ground state of an antiferromagnet satisfies
\begin{eqnarray}
\langle 0 | {\cal H} | 0 \rangle & = & E_0, \nonumber \\
\langle 0 | {\cal S}_j^x | 0 \rangle & = & (-1)^j m_s. \label{eq:firstm_s}
\end{eqnarray}
It is invariant under translations over lattice vectors respecting
the bipartite breakup of the lattice.  Later on we will come back
to this feature. The search for the spin stiffness $\rho_s$ can be
formulated as finding an excitation $|\tilde{\bf q} \rangle$ such
that
\begin{eqnarray}
  \langle \tilde{{\bf q}}|{\cal H}|\tilde{{\bf q}} \rangle &=& E_0 +
  \frac{1}{2} N \rho_s | {\bf q}|^2 + \dots, \nonumber \\ \langle \tilde{{\bf q}}|
  S^x_j|\tilde{{\bf q}} \rangle &=& (-1)^j m_s \cos({\bf q}\cdot {\bf
    r}_j),  \nonumber \\ \langle \tilde{{\bf q}}| S^y_j|\tilde{{\bf q}} \rangle
  &=& (-1)^j m_s \sin({\bf q}\cdot {\bf r}_j). \nonumber
\end{eqnarray}
The implementation of a twisted orientation like this is a hard task.
It can be avoided by a similarity transformation. Define the unitary
operator ${\cal U}({\bf q})$ by
\begin{equation}
{\cal U}({\bf q})= \exp \left (i {\bf q} \cdot \sum_{j} {\bf r}_j
  {\cal S}^z_j \right ). \label{eq:unitary}
\end{equation}
It corresponds to a rotation about the ${\cal S}^z$-axis. It is easy
to derive that
\[
{\cal U}({\bf q}) {\cal S}_j^+ {\cal U}^{\dagger}({\bf q})= {\cal
  S}_j^+ e^{i {\bf q} \cdot {\bf r}_j} ~~,~~ {\cal U}({\bf q}) {\cal
  S}_j^- {\cal U}^{\dagger}({\bf q})= {\cal S}_j^- e^{-i {\bf q} \cdot
  {\bf r}_j }.
\]
We will now transfer the twist from the order parameter to the
Hamiltonian with this operator ${\cal U}({\bf q})$. Define
\begin{eqnarray}
  {\cal H}({\bf q})&=&{\cal U}({\bf q}) {\cal H} {\cal U}^{\dagger}({\bf q}) \nonumber \\ &=&
  \phantom{+} \frac{J_1}{2} \sum_{\langle ij \rangle} e^{i {\bf q}
    \cdot ({\bf r}_j - {\bf r}_i)} {\cal S}_j^+ {\cal S}_i^- + e^{-i {\bf q} \cdot
    ({\bf r}_j - {\bf r}_i)} {\cal S}_j^- {\cal S}_i^+ + 2 {\cal S}_j^z {\cal S}_i^z \nonumber \\
  & & + \frac{J_2}{2} \sum_{[ ij ]} e^{i {\bf q} \cdot ({\bf r}_j -
    {\bf r}_i)} {\cal S}_j^+ {\cal S}_i^- + e^{-i {\bf q} \cdot ({\bf r}_j - {\bf
      r}_i)} {\cal S}_j^- {\cal S}_i^+ + 2 {\cal S}_j^z {\cal S}_i^z \label{eq:twistH}
\end{eqnarray}
and $|{\bf q}\rangle = {\cal U}({\bf q})|\tilde{{\bf q}}\rangle$.
Combining the expressions for $|\tilde{\bf q}\rangle$ and ${\cal
  H}({\bf q})$ it is trivial to derive that this new state $|{\bf
  q}\rangle$ satisfies
\begin{eqnarray}
  \langle {\bf q}| {\cal H}({\bf q}) | {\bf q}\rangle &=& E_0+ \frac{1}{2} N
  \rho_s |{\bf q}|^2  + \dots, \nonumber \\ \langle {\bf q}| S^x_j|{\bf q} \rangle &=&
  (-1)^j m_s. \label{eq:boundhom}
\end{eqnarray}
$|{\bf q}\rangle$ clearly is an excitation of ${\cal H}({\bf q})$ as
the ground state is trivially given by ${\cal U}({\bf q}) | 0
\rangle$.  We know that
\[
\lim_{{\bf q } \rightarrow {\bf 0}} |{\bf q} \rangle = |0 \rangle,
\]
which can be read as an invitation to apply perturbation theory. Still
there is a quite subtle issue that must not be overlooked. If we would
apply perturbation theory without further ado, the resulting wave
function would be ${\cal U}({\bf q})|0\rangle$. The order parameter
would also be twisted and  consequently it would not satisfy (\ref{eq:boundhom});
\begin{eqnarray}
\langle 0 | {\cal U}^\dagger({\bf q}) {\cal S}_j^x {\cal U}({\bf q}) | 0 \rangle &=& \phantom{-} (-1)^j m_s  \cos({\bf q} \cdot {\bf r}_j), \nonumber \\
\langle 0 | {\cal U}^\dagger({\bf q}) {\cal S}_j^y {\cal U}({\bf q}) | 0 \rangle &=&
- (-1)^j m_s \sin({\bf q} \cdot {\bf r}_j). \nonumber
\end{eqnarray}
The underlying physical idea is that if we apply a twist with wave
vector ${\bf q}$ by slowly turning up this ${\bf q}$, the system and
the order parameter will follow adiabatically and no energy increase
will occur. This is not what we set out to achieve. We are considering
the case where an integer twist over the entire system exists. The
wave vectors ${\bf q}$ that are allowed then fit the lattice; all
terms $\exp(i {\bf q} \cdot ({\bf r}_i - {\bf r}_j)$ can by simplified
to $\exp(i {\bf q}\cdot {\bf r}_{\delta})$ where ${\bf r}_{\delta}$ is
the smallest connecting lattice vector. Even when ${\bf r}_i $ and
${\bf r}_j$ are adjacent across the periodic boundary of the system
this replacement can be done as ${\bf q} \cdot ({\bf r}_i - {\bf r}_j)
\equiv {\bf q}\cdot {\bf r}_{\delta} \hbox{ mod }2\pi$. To summarise,
we seek the homogeneous state $| {\bf q} \rangle$ given an integer
twist over the lattice.  Respecting this subtlety, $|{\bf q} \rangle$
can be derived.

Define
\begin{eqnarray}
  \vec{j} = \left .\frac{d}{d{\bf q}} {\cal H}({\bf q}) \right |_{{\bf q}=0} &=&
  \phantom{+} \frac{i J_1}{2} \sum_{\langle ij \rangle}  ({\bf r}_j -
  {\bf r}_i)  ({\cal S}_j^+ {\cal S}_i^- - {\cal S}_j^- {\cal S}_i^+ )\nonumber \\ &&+ \frac{i
    J_2}{2} \sum_{[ ij ]}  ({\bf r}_j - {\bf r}_i)( {\cal S}_j^+ {\cal S}_i^- - {\cal S}_j^-
  {\cal S}_i^+) ,\nonumber  \\
\vec{\vec{t}} = -\left .\frac{d^2}{d {\bf q}^2} {\cal H}({\bf q}) \right |_{{\bf
      q}=0} &=& \phantom{+}\frac{J_1}{2} \sum_{\langle ij \rangle}
  ({\bf r}_j - {\bf r}_i) ({\bf r}_j - {\bf r}_i)( {\cal S}_j^+ {\cal S}_i^- + {\cal S}_j^- {\cal S}_i^+) \nonumber \\
  && + \frac{ J_2}{2} \sum_{[ ij ]} ({\bf r}_j - {\bf r}_i)({\bf r}_j - {\bf r}_i) ({\cal S}_j^+
  {\cal S}_i^- + {\cal S}_j^- {\cal S}_i^+). \nonumber
\end{eqnarray}
A note on the terms ${\bf r}_i - {\bf r}_j $ is in place here. The
factors $\exp(i {\bf q} \cdot ({\bf r}_i - {\bf r}_j))$ as appearing
in ${\cal H}({\bf q})$ are periodic in the difference ${\bf r}_i -
{\bf r}_j$. Adding or subtracting a vector $(nL,mW)$ leaves this term
invariant. When taking the derivative it is therefore necessary to
insert the smallest possible argument for ${\bf r}_i - {\bf r}_j$. So
when ${\bf r}_i$ and ${\bf r}_j$ are on opposite sides of the
-periodic- boundary we have to replace them by ${\bf r}_\delta$; the
connecting elementary lattice vector.

The state $|{\bf q} \rangle$ is now given in first order perturbation
theory by
\[
|{\bf q} \rangle = \left (1 + i{\bf q} \cdot \frac{1}{E_0 - {\cal H}}
  \vec{j} + \dots \right) | 0 \rangle.
\]
This state clearly satisfies the condition on the order parameter
(\ref{eq:boundhom}) as both the current-current correlation $j$ and
the kinetic term $t$ are translational invariant. The expression for
the stiffness $\rho_s$ can now readily be obtained;
\begin{eqnarray}
\vec{\vec{\rho}}_s &=&
-\frac{1}{N} \langle 0|\vec{\vec{t}} |0 \rangle + \frac{1}{N}
  \langle 0| \vec{j} \frac{1}{E_0 -{\cal H}}
   \vec{j} | 0\rangle \nonumber \\  &\equiv& \vec{\vec{T}} + \vec{\vec{J}}.
  \label{eq:rhofinal}
\end{eqnarray}
The spin stiffness $\vec{\vec{\rho}}_s$ is a tensor and only when the
system has quadratic symmetry it can be denoted by a scalar.

The ground state $|0 \rangle$ is reflection symmetric whereas the
current $\vec{j}$ is not. Therefore $\langle 0|\vec{j}|0 \rangle$ $ = 0$
and we can safely write the current-current correlation as
\begin{equation}
\vec{\vec{J}}= \frac{1}{N} \langle 0| \vec{j} \frac{1}{E_0 -{\cal H}} \vec{j} | 0\rangle =
\frac{1}{N} \sum_{a} \frac{\langle0|\vec{j}|a \rangle\langle a|\vec{j}|0 \rangle
}{E_0-E_a}. \label{eq:jfinal}
\end{equation}

For expression (\ref{eq:rhofinal}) to hold, it is essential that the
system is translational invariant in the direction of the twist
${\bf q}$. Recapturing the subtlety in the derivation, it can be
readily seen that open boundary condition will not lead to the correct
result. In that geometry we will find $|{\bf q} \rangle = {\cal
  U}({\bf q}) | 0 \rangle$. We can borrow a physical argument from the
realm of superfluidity to put this in a broader context.

Periodic boundary conditions can easily be achieved for a bucket of
helium by bending this bucket round a massive cylinder and connecting
both ends together on the other side of the cylinder. The twist we
apply corresponds to a steady rotation of this cylinder with the
bucket around it.  The superfluid will not respond to this twist
whereas the normal component of the helium will start to rotate along. In the
frame of the cylinder it is the superfluid that flows, whereas the normal helium remains inert.  Consequently, the kinetic energy
increases by the superfluid current.

Open boundaries correspond to helium in a closed bucket.  The walls of
a closed bucket prevent a current from running and a zero increase in
energy will be the result. It is worthwhile mentioning that Pollock
and Ceperley \cite{pollock87} have implemented this idea directly in a
Green Function Monte Carlo calculation. They follow a particle winding
around a periodic cell. If it winds faster round a periodic cell than
expected by Brownian motion, superfluidity exists.

The formula (\ref{eq:rhofinal}) is applicable to systems exhibiting
all kinds of long-range magnetic order. If another type of ordering
was twisted with an order parameter which was also defined locally,
the same route would have led to this general result. The assumption
of antiferromagnetic order can therefore be dropped and we end up with
a fairly general expression for $\vec{\vec{\rho}}_s$ at zero
temperature.  The only restriction is naturally that the order must
break the symmetry; otherwise Goldstone modes or spin waves will not
appear and no twist is possible. Equation (\ref{eq:rhofinal}) is
completely equivalent with the expression found in \cite{ceperley95}
and \cite{einarsson}.

\section{Schwinger-Boson Mean Field Approximation \label{sec:SBMF}}

The behaviour of the spin stiffness in a two-dimensional system can be obtained by
finite-size scaling analysis of relatively small systems. It is then
necessary to establish the size dependence of the properties and in
the current and following sections we will derive these in the SBMF
approximation.

The Schwinger-boson representation is in itself a mere reformulation
of the problem. The only extra condition that has to be imposed is
that the number of bosons per site has to be fixed to 1. Any
operator correctly transformed into the language of Schwinger-bosons
will conserve this property.

Standard mean field theory is based on the assumption that there are
only small fluctuations in the orientation of the spins. In our case
that is a poor approximation; quantum fluctuations play a major role.
In the unfrustrated case, $J_2=0$, Monte Carlo calculations
\cite{trivedi90,sandvik98} have shown that the staggered magnetisation
is about 60\% of the classical value.

A possible improvement of the mean field is to incorporate
correlations between neighbouring spins. The Schwinger-boson mean
field approximation is a first step in this direction. it was originally introduced by Arovas and
Auerbach \cite{auerbach} Here we will
apply this approximation and derive expressions for the energies and
wave functions of all states of the frustrated Heisenberg model.  This
will be done both for N\'{e}el and collinear ordered ground state.

The first step in this method is to represent every spin by two
bosons;
\begin{eqnarray}
  {\cal S}^+&=&a^\dagger b, \nonumber \\ {\cal S}^-&=&a b^\dagger, \nonumber \\  {\cal S}^z
  &=& \frac{1}{2} ( a^\dagger a - b^\dagger b). \label{eq:sbtrans}
\end{eqnarray}
We restrict ourselves to the subspace in which $a^\dagger a +
b^\dagger b = 2S$, where $S$ is the total length of the spin. In this
case $S=\frac{1}{2}$. Two observations can be made:
\begin{itemize}
\item The combinations of bosons in (\ref{eq:sbtrans}) satisfy all
  commutation relations of the original operators.
\item None of the operators ${\cal S}^+$, ${\cal S}^-$ and ${\cal
    S}^z$ will connect this subspace to subspaces of other local spin
  since they all commute with $a^\dagger a + b^\dagger b$. Therefore
  the Hamiltonian will only map states in this subspace onto other
  states in the same subspace.
\end{itemize}
The Schwinger-Boson representation is therefore just a reformulation
of the problem. The transformation in (\ref{eq:sbtrans}) does not
favour any specific orientation of the magnetic order and the mean
field approximation for these bosons will also preserve this symmetry.

Now the mean field approximation can be developed. Many authors have
done this before \cite{auerbach} and we follow the route taken by Mila
et al. \cite{mila} for this particular problem.  As stated before we
will consider two types of ordering: N\'{e}el and collinear.  

We begin
with the N\'{e}el ordering. This is expected to be the preferred
ordering when $J_2 \ll J_1$.

A first step towards the mean field approximation is to rotate the
spins over $\pi$ around the ${\cal S}^z$-axis on one of the two
sublattices;
\begin{eqnarray}
  {\cal S}_i^+ &\rightarrow& (-1)^i {\cal S}_i^+, \nonumber \\ {\cal S}_i^- &\rightarrow&
  (-1)^i {\cal S}_i^-, \label{eq:marshall}\\ {\cal S}_i^z &\rightarrow& \phantom{(-1)^i} {\cal S}_i^z. \nonumber
\end{eqnarray}
This transformations is inspired by the Marshall sign rule
\cite{marshall55}, which states that in this new basis the ground
state of the unfrustrated antiferromagnet has only positive definite
coefficients. It is of course not strictly necessary, but
it avoids the need of complex numbers which otherwise would appear.

For the mean fields only those combinations of creation and
annihilation operators can be taken that are invariant under rotation
round the ${\cal S}^z$-axis. Define
\begin{eqnarray}
  {\cal D}_{ij} &=& a_i a_j^\dagger + b_i b_j ^\dagger, \label{eq:repl-ferro}\\
  {\cal B}_{ij} &=& a_i b_j + b_i a_j. \label{eq:repl-antiferro}
\end{eqnarray}
The Hamiltonian becomes
\[
{\cal H}= -\frac{J_1}{2} \sum_{\langle ij \rangle} ({\cal
  B}_{ij}^\dagger {\cal B}_{ij} - \frac{1}{2}) + \frac{J_2}{2} \sum_{[
  ij ]} ({\cal D}_{ij}^\dagger {\cal D}_{ij} - \frac{3}{2}),
\]
where we have inserted (\ref{eq:repl-ferro}) for the ferromagnetic
oriented pairs of spins and (\ref{eq:repl-antiferro}) for the
antiferromagnetic ones. This expression is related to the fact that if
two spins are aligned, ${\cal B}_{ij} | \uparrow \uparrow \rangle = 0$.
Likewise for singlet combinations, it holds that ${\cal D}_{ij}
(|\uparrow \downarrow \rangle - | \downarrow \uparrow \rangle ) = 0$.
Still, the equivalent of ${\cal D}_{ij}$ could in principle appear on
the nearest neighbour bonds. Rotating the spins on one sublattice
yields
\[
{\cal D}_{ij}^\dagger {\cal D}_{ij} \rightarrow (a_i^\dagger a_j -
b_i^\dagger b_j)(a_i a_j^{\dagger} - b_i b_j^{\dagger}).
\]
The optimal mean field solution ---self consistent with lowest
energy--- turns out to satisfy $\langle (a_i a_j^{\dagger} - b_i
b_j^{\dagger}) \rangle=0$ for nearest neighbours. The same holds for
the equivalent of ${\cal B}_{ij}$ on next-nearest neighbour bonds.
Therefore we will neglect these terms.

\begin{figure}
  {\centering \hfill \epsfxsize=8cm \epsffile{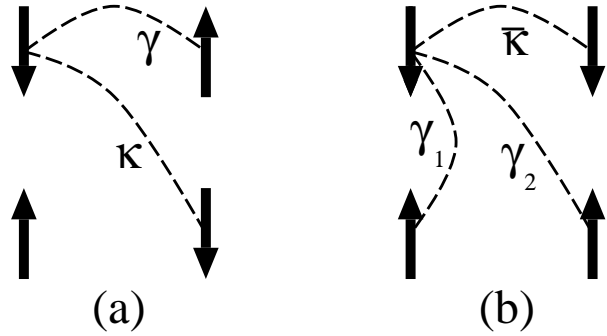} \hfill}
\caption{\label{fig:SBMF-order} The mean fields for the
  N\'{e}el (a) and the collinear order (b).}
\end{figure}

The final step is to replace the products of these two operators
${\cal D}_{ij}$ and ${\cal B}_{ij}$ by their mean field
approximations. Set $\gamma_{ij} = \frac{1}{2}\langle {\cal B}_{ij}
\rangle$ for nearest-neighbour bonds and $\kappa_{ij}=\frac{1}{2}\langle
{\cal D}_{ij} \rangle$ for the next-nearest neighbour bonds. In the regime of N\'eel ordering, we only consider the
translationally invariant solutions, so $\gamma_{ij} \rightarrow \gamma$, $\kappa_{ij} \rightarrow \kappa$. In figure
\ref{fig:SBMF-order} these mean fields are represented.  When the
decoupling is made, the detailed constraint $a^\dagger a + b^\dagger
b=1$ is no longer satisfied as the Hamiltonian will now also map out
of this subspace. Paradoxically, if the constraint were conserved in
detail, the mean fields $\gamma$ and $\kappa$ would become identical
to zero as both alter the number of particles on individual sites. The
constraint $a^\dagger a + b^\dagger b=1$ will be relaxed to only hold
on average, $\langle a^\dagger a + b^\dagger b \rangle =1$. Given the
translational invariance of the system, this can be replaced by a
global version and enforced by a Lagrange multiplier, i.e.
\begin{equation}
{\cal H} \rightarrow {\cal H}_{\rm MF} + \lambda \sum_{i} (a_i^\dagger a_i
+ b_i^\dagger b_i - 1). \label{eq:globalconstraint}
\end{equation}
Introducing the quantities $h_{\bf p}$ and $\Delta_{\bf p}$
\begin{eqnarray}
  h_{\bf p} &=& 4 J_2 \kappa \cos p_x \cos p_y, \label{eq:hq} \\
  \Delta_{\bf p} &=& 2 J_1 \gamma (\cos p_x + \cos p_y), \label{eq:dq}
\end{eqnarray}
the Fourier-transformed Hamiltonian reads
\begin{equation}
  {\cal H}=E_c+\sum_{\bf p} (h_{\bf p}+ \lambda)(a_{\bf p}^\dagger a_{\bf p}
  + b_{\bf p} b_{\bf p}^\dagger) - \Delta_{\bf p}(a_{\bf p}^\dagger
  b_{-\bf p}^\dagger + a_{\bf p} b_{-\bf p}), \label{eq:SBMFH1}
\end{equation}
with
\[
E_c= 2N\left [ J_1 \left ( \frac{1}{4} + 2 \gamma^2 \right ) - J_2
  \left ( \frac{3}{4} + 2 \kappa^2 \right ) - \lambda \right ].
\]
A solution to this Hamiltonian can be found through the Bogoliubov
transformation from $(a_{\bf p},b_{\bf p})$ to $(\alpha_{\bf
  p},\beta_{\bf p})$;
\begin{eqnarray}
  a_{\bf p}&=& \alpha_{\bf p} \cosh \theta_{\bf p} + \beta_{\bf
    p}^\dagger \sinh \theta_{\bf p} ,\nonumber \\ b_{-\bf p}^\dagger
  &=& \alpha_{\bf p} \sinh \theta_{\bf p} + \beta_{\bf p}^\dagger
  \cosh \theta_{\bf p} ,\nonumber \\ \tanh{2 \theta_{\bf p}} &=&
  \frac{\Delta_{\bf p}}{h_{\bf p} + \lambda} . \label{eq:bogoliubov}
\end{eqnarray}
If we define the excitation energies $\omega_{\bf p}$ by
\begin{equation}
  \omega_{\bf p}=\sqrt{(h_{\bf p}+\lambda)^2 - \Delta_{\bf p}^2},
  \label{eq:omegaq}
\end{equation}
the Hamiltonian becomes
\begin{equation}
  {\cal H}=E_c+ \sum_{\bf p} \omega_{\bf p} ( \alpha_{\bf p}^\dagger
  \alpha_{\bf p} + \beta_{\bf p} \beta_{\bf p}^\dagger).
  \label{eq:SBMFH2}
\end{equation}
As in all mean field theories the energy has to be stationary with
respect to $\kappa$, $\gamma$ and $\lambda$. This yields the equations
\begin{eqnarray}
  \kappa &=& \frac{1}{N} \sum_{\bf p} \frac{h_{\bf p}+\lambda}{2
    \omega_{\bf p}} \cos p_x \cos p_y, \label{eq:sumkappa} \\ \gamma
  &=& \frac{1}{N} \sum_{\bf p} \frac{\Delta_{\bf p}}{4 \omega_{\bf p}}
  (\cos p_x + \cos p_y) ,\label{eq:sumgamma} \\ 1 &=& \frac{1}{N}
  \sum_{\bf p} \frac{h_{\bf p}+\lambda}{2 \omega_{\bf p}}.
  \label{eq:sum1}
\end{eqnarray}
As only cosines appears in these equations either as single terms or in the
right combinations, the ${\bf p} = (0,0)$ and the ${\bf p}= ( \pi, \pi
)$ terms are completely equivalent. For future applications it is also
useful to define what will turn out to be the 'condensate' $m_s$
\begin{equation}
  m_s = \frac{h_{\bf 0} + \lambda}{ N \omega_{\bf 0}}.
  \label{eq:condensate}
\end{equation}
This is the sum of the contributions of the ${\bf p} = {\bf 0}$ and
the ${\bf p}= ( \pi, \pi )$ terms in the summations
(\ref{eq:sumkappa}) and (\ref{eq:sum1}). As these are identical, it is
twice the '${\bf p}=0$'-term.  If the system size $N$ becomes
infinite, $N \rightarrow \infty$, $m_s$ equals twice the '${\bf
  p}=0$'-term in (\ref{eq:sumgamma}). It is no coincidence that $m_s$
also appears as the size of the order parameter in equation
(\ref{eq:firstm_s}). By inserting a symmetry breaking term $\eta
\sum_j (-1)^j{\cal S}^x_j$ in the Hamiltonian, it can be shown through a short but subtle calculation that
the equation (\ref{eq:firstm_s}) ---without the staggering $(-1)^j$---
will hold for the ground state of the SBMF Hamiltonian with infinitely
small symmetry breaking field $\eta= {\cal O}(1/N)$. The staggering
$(-1)^j$ has disappeared from the formulae by the transformation (\ref{eq:marshall}).

As long as there is N\'{e}el order the condensate $m_s$ will naturally
be positive. It will obtain a limit value in the case of an infinite
system size $N$; $\lim_{N\rightarrow \infty} m_s=\hbox{const}>0$. This
finishes the discussion of the N\'{e}el ordered ground state.

Next we consider the collinear order.  The route to follow is quite
similar to the one just finished. We can therefore be brief about it.
Introduce the transformation
\begin{eqnarray}
  {\cal S}_{(x,y)}^+ &\rightarrow& (-1)^y {\cal S}_{(x,y)}^+, \nonumber \\
  {\cal S}_{(x,y)}^- &\rightarrow& (-1)^y {\cal S}_{(x,y)}^-, \nonumber \\ {\cal S}_{(x,y)}^z
  &\rightarrow& \phantom{(-1)^y} {\cal S}_{(x,y)}^z. \nonumber
\end{eqnarray}
Again we perform a rotation over $\pi$ around the ${\cal S}^z$-axis, however this
time not on one of the sublattices but on every other row.  Define the
mean fields $\bar{\kappa}=\frac{1}{2}\langle {\cal D}_{i,i+\hat{x}}
\rangle$ over nearest-neighbour bonds in the $x$-direction, $\gamma_1
= \frac{1}{2}\langle {\cal B}_{i,i+\hat{y}} \rangle$ over the
nearest-neighbour bonds in the $y$-direction and $\gamma_2 =
\frac{1}{2} \langle {\cal B}_{i,i \pm \hat{x} \pm \hat{y}} \rangle$
over the next-nearest-neighbour bonds. The quantities $\bar{h}_{\bf
  p}$ and $\bar{\Delta}_{\bf p}$ are now given by
\begin{eqnarray}
  \bar{h}_{\bf p} &=& 2J_1 \bar{\kappa} \cos p_x, \nonumber \\
  \bar{\Delta}_{\bf p} &=& 2 J_1 \gamma_1 \cos p_y + 4 J_2 \gamma_2
  \cos p_x \cos p_y.
\end{eqnarray}
After a Fourier and a Bogoliubov transformation, similar to
(\ref{eq:SBMFH1}) and (\ref{eq:bogoliubov}), we obtain
\[
{\cal H}=\bar{E}_c+ \sum_{\bf p} \bar{\omega}_{\bf p} (
\bar{\alpha}_{\bf p}^\dagger \bar{\alpha}_{\bf p} + \bar{\beta}_{\bf
  p} \bar{\beta}_{\bf p}^\dagger).
\]
This is equivalent to (\ref{eq:SBMFH2}) where $\bar{\omega}_{\bf}$ and
$\bar{E}_c$ are now given by
\begin{eqnarray}
  \bar{\omega}_{\bf p} &=& \sqrt{(\bar{h}_{\bf p}+\bar{\lambda})^2 -
    \bar{\Delta}_{\bf p}^2}, \nonumber \\ \bar{E}_c &=& 2 N \left ( J_1
  (\gamma_1^2 - \bar{\kappa}^2-\frac{1}{4}) + J_2 (\frac{1}{4} + 2 \gamma_2^2) -
  \bar{\lambda} \right).
\end{eqnarray}
The consistency equations for these parameters are
\begin{eqnarray}
  \bar{\kappa} &=& \frac{1}{N} \sum_{\bf p} \frac{\bar{h}_{\bf
      p}+\bar{\lambda}}{2 \bar{\omega}_{\bf p}} \cos p_x
  , \label{eq:barkappa}\\ \gamma_1 &=& \frac{1}{N} \sum_{\bf p}
  \frac{\bar{\Delta}_{\bf p}}{2 \bar{\omega}_{\bf p}} \cos p_y
  , \label{eq:bargamma1}\\ \gamma_2 &=& \frac{1}{N} \sum_{\bf p}
  \frac{\bar{\Delta}_{\bf p}}{2 \bar{\omega}_{\bf p}} \cos p_x \cos
  p_y , \label{eq:bargamma2} \\ 1 &=& \frac{1}{N} \sum_{\bf p}
  \frac{\bar{h}_{\bf p}+\bar{\lambda}}{2 \bar{\omega}_{\bf p}}.
  \label{eq:bar1}
\end{eqnarray}
The 'condensate' $\bar{m}_s$ is defined in a similar manner as before
in (\ref{eq:condensate}), that is
\begin{equation}
\bar{m}_s = \frac{\bar{h}_{\bf 0}+\bar{\lambda}}{ N \bar{\omega}_{\bf
    0}}. \label{eq:condensate2}
\end{equation}
The symmetry of the collinear order differs from that of the N\'eel
order. This condensate contains the contribution of the ${\bf p}={\bf
  0}$ and ${\bf p} = (0,\pi)$ terms in the summations
(\ref{eq:barkappa}) and (\ref{eq:bar1}). These are identical and once
again the condensate $\bar{m}_s$ is twice the '${\bf p}=0$'-term . If
the system size $N \rightarrow \infty$ it also becomes twice the
'${\bf p}=0$'-term of (\ref{eq:bargamma1}) and (\ref{eq:bargamma2}).

The main weakness of the SBMF approach lies in the handling of the
particle constraint $a^\dagger a +b^\dagger b=1$. It is only conserved
on average by use of a Lagrange multiplier, equation
(\ref{eq:globalconstraint}). The ground state will have non-zero
weight in configurations that have either too many or too few bosons
on a specific site. This space is different from the original spin
space and no direct correspondence to the ground state of the
frustrated Heisenberg model exists.

A connection with the original, frustrated Heisenberg model can be
made that involves a restriction of the state and a transformation of
the basis; first the ground state of the SBMF Hamiltonian has to be
restricted to the subspace where the condition $a^\dagger a +
b^\dagger b=1$ holds in detail.  Afterwards, pairs of bosons
($a^\dagger a$ etc.)  have to be replaced by spin operators. The
restricted wave function now lies in the correct space and can serve
as an approximation to the true ground state of the frustrated
Heisenberg model. Wei and Tao \cite{wei94} make this connection for
the unfrustrated case, $J_2=0$. The properties that they extract
from this approximate ground state agree surprisingly well with the
numerical results for the true ground state from various Monte Carlo
calculations.

In this chapter we will not follow their route back to the spin
problem, but remain in the larger, bosonic space. The properties we
obtain should thus be appreciated as such; they are related to those
of the ground state of the frustrated Heisenberg model and should be
seen as indications of the behaviour of the frustrated Heisenberg
model.

The mean field approximation has provided the energies and wave
functions of all states of the Hamiltonian and we can proceed to the
calculation of the spin stiffness $\rho_s$.

\section{Spin stiffness in SBMF \label{sec:schwingerrho_s}}

How can $\rho_s$ be defined properly in the mean field Hamiltonian?
This question might seem trivial, but a closer investigation reveals
that there are two approaches:
\begin{itemize}
\item Start with the original Heisenberg Hamiltonian ${\cal H}$.
  Induce a twist with wavelength ${\bf q}$ (just like has been done in
  section \ref{sec:spin-stiffness}) and afterwards apply the
  appropriate mean field approximation. $\rho_s$ is related to the
  ground state energy of this mean field Hamil\-tonians, i.e.
\[
{\cal H} \rightarrow {\cal H}({\bf q}) \rightarrow {\cal H}_{\rm MF} ({\bf
  q}) ~~, \vec{\vec{\rho}}_s = \frac{1}{N} \frac{d^2 \langle {\cal
    H}_{\rm MF} ({\bf q})\rangle}{d {\bf q}^2}.
\]
\item Swapping the first two steps of the previous approach; first
  apply mean field and then twist, i.e.
\[
{\cal H} \rightarrow {\cal H}_{\rm MF} \rightarrow \tilde{{\cal H}}_{\rm MF}
({\bf q}) ~~, \vec{\vec{\rho}}_s = \frac{1}{N} \frac{d^2 \langle
  \tilde{{\cal H}}_{\rm MF} ({\bf q})\rangle}{d {\bf q}^2}.
\]
\end{itemize}
In the mean field approximation we use here, these two approaches give
the same result. Still, the correct approach is the first one, as
there the energy increase of the original Hamiltonian due to the twist
is approximated. The second approach replaces the original Hamiltonian
by a mean field one and starts to investigate the response of the mean
field Hamiltonian to a twist.

The calculation of the energy $\langle {\cal H}({\bf q}) \rangle$ is a
simple repetition of the approach for ${\bf q}={\bf 0}$. This allows a
fairly direct derivation of the spin stiffness avoiding second order
perturbation theory. Still, the aim of this chapter is to obtain the
finite-size scaling relations for the kinetic term $\vec{\vec{T}}$ and
the current-current correlation $\vec{\vec{J}}$ independently.  These
relations will then be used to extrapolate the data of the DMRG
calculations to infinitely large system sizes. We will therefore
follow the -general- route of second order perturbation theory.

We can use the formulae derived earlier in (\ref{eq:rhofinal}), but
simply taking the mean field ground state to replace the true ground
state in this expression will not suffice; the second term contains an
inversion which we cannot handle in this form. It is necessary to
insert the mean field Hamiltonian ${\cal H}_{\rm MF}$.  In doing so
we extend our SBMF approximations to the entire spectrum.

The SBMF approximation thus has to be performed on the twisted
Hamiltonian, ${\cal H}({\bf q})$, and the first and second derivative
are needed to calculate the stiffness $\rho_s$ in second order
perturbation theory. This in itself is very similar to the description
above for the untwisted situation and left to appendix
\ref{app:SBMFq}. The expressions for the current and kinetic operators
are
\begin{eqnarray}
  \vec{j} &=& \sum_{F} J_{ij} \kappa_{ij}(\vec{\cal  F}_{ij}^\dagger + \vec{\cal  F}_{ij}) -
  \sum_{AF} J_{ij} \gamma_{ij} (\vec{\cal C}_{ij}^\dagger + \vec{\cal C}_{ij}),
  \nonumber \\ \vec{\vec{t}} &=& \frac{1}{2} \sum_{F} J_{ij} \kappa_{ij}
  ({\bf r}_i - {\bf r}_j) ({\bf r}_i - {\bf r}_j)({\cal D}_{ij}^\dagger +
  {\cal D}_{ij}-2 \kappa_{ij}) \nonumber \\ & &- \frac{1}{2} \sum_{AF}
  J_{ij} \gamma_{ij} ({\bf r}_i - {\bf
    r}_j) ({\bf r}_i - {\bf
    r}_j)({\cal B}_{ij}^\dagger + {\cal B}_{ij}-2 \gamma_{ij}). \nonumber
\end{eqnarray}
From this point on, we set ${\bf q} = q ( \cos \phi, \sin \phi )$.  Of
the two terms for $\rho_s$ in (\ref{eq:rhofinal}), $T$ is evaluated
more easily:
\begin{eqnarray}
  \vec{\vec{T}} &=& -\frac{1}{N} \langle 0 |\vec{\vec{t}} |0 \rangle \nonumber \\ &=&
  \frac{1}{N} \sum_{AF} J_{ij} \gamma_{ij}^2
  ({\bf r}_i - {\bf r}_j)  ({\bf r}_i - {\bf r}_j) - \frac{1}{N} \sum_{F} J_{ij}
  \kappa_{ij}^2 ({\bf r}_i - {\bf r}_j)({\bf r}_i - {\bf r}_j). \nonumber
\end{eqnarray}
For the two orderings considered in the last section, this expression
boils down to
\begin{eqnarray}
  T^{\rm neel} &=& J_1 \gamma^2 - 2 J_2 \kappa^2, \label{eq:TYneel}\\
  T^{\rm coll} &=& 2 J_2 \gamma^2_2 + J_1 \left (\gamma_1^2 \sin^2 \phi -
  \bar{\kappa}^2 \cos^2 \phi \right). \label{eq:TYcoll}
\end{eqnarray}
These simple equations hold for all system sizes $N$.

The derivation of $\vec{\vec{J}}$ requires somewhat more effort. First the
matrix element $\langle 0 | \vec{j}|a\rangle$ has to be calculated. As
$|a \rangle$ is an excitation of ${\cal H}_{\rm MF}$, it has to fulfill
the relation
\[
|a \rangle = \alpha_1^\dagger \cdot \dots \cdot \alpha_i^\dagger
\beta_1^\dagger \cdot \dots \cdot \beta_j^\dagger | 0 \rangle.
\]
Of these the only relevant ones are
\[
| a \rangle = \alpha_{{\bf p}_1}^\dagger \beta_{{\bf p}_2}^\dagger | 0
\rangle,
\]
as can be established by applying the Bogoliubov transformations
(\ref{eq:bogoliubov}) to $\vec{\cal C}_{ij}$ and $\vec{\cal F}_{ij}$.
The matrix element itself is a combination of $\langle 0 | \vec{\cal
  F}_{ij} | a \rangle$ and $\langle 0 | \vec{\cal C}_{ij}|a \rangle$.
By means of simple algebra, we obtain
\begin{eqnarray}
  \sum_{i,j=i+{\bf r}_\delta} \langle 0|\vec{\cal F}_{ij}^\dagger+\vec{\cal F}_{ij} | a
  \rangle &=& - 2 \delta_{{\bf p}_1,{\bf p}_2} {\bf
    r}_\delta \sinh 2 \theta_{{\bf p}_1} \sin( {\bf p}_1 \cdot {\bf
    r}_\delta), \nonumber
  \\ \sum_{i,j=i+{\bf r}_\delta} \langle 0| \vec{\cal C}_{ij}^\dagger+\vec{\cal C}_{ij} | a
  \rangle &=& - 2 \delta_{{\bf p}_1,{\bf p}_2}  {\bf
    r}_\delta \cosh 2 \theta_{{\bf p}_1} \sin( {\bf p}_1 \cdot {\bf
    r}_\delta). \nonumber
\end{eqnarray}
The Kronecker-delta forces ${\bf p}_1 = {\bf p}_2 \equiv {\bf p}$,
thus $E_a = E_0 + \omega_{{\bf p}_1}+ \omega_{{\bf p}_2} = E_0 +
2\omega_{\bf p}$.  Inserting all these expressions in the equation for
$J$, (\ref{eq:jfinal}), will give the explicit formula for $J$. The
Hamiltonian in that expression is ${\cal H}_{\rm MF}$. For the
two types of ordering $J$ is:
\begin{eqnarray}
  J^{\rm neel} &=& - \frac{1}{N} \sum_{\bf p} \frac{\sin^2
    p_y}{\omega_{\bf p}^3} \left ( J_1 \gamma (h_{\bf p}+\lambda) - 2 J_2
  \kappa \Delta_{\bf p} \cos p_x \right )^2 \label{eq:JYneel} \\ J^{\rm coll} &=&
  -\frac{1}{N} \sum_{\bf p} \frac{1}{\bar{\omega}_{\bf p}^3} \left (
  \vphantom{\frac{A}{A}} \right . \phantom{-}\cos \phi \sin p_x \left [J_1
  \bar{\kappa} \bar{\Delta}_{\bf p} - 2 J_2 \gamma_2(\bar{h}_{\bf p} +
  \bar{\lambda}) \cos p_y \right ] \nonumber \\ & &\phantom{ -\frac{1}{N} \sum_{\bf p}
    \frac{1}{\bar{\omega}_{\bf p}^3}(} - \sin \phi \sin p_y \left [ J_1
  \gamma_1(\bar{h}_{\bf p} + \bar{\lambda}) + 2 J_2 \gamma_2
  (\bar{h}_{\bf p} +\bar{\lambda})\cos p_x \right ]\left .
  \vphantom{\frac{A}{A}} \right )^2. \nonumber
\end{eqnarray}
In the introduction of the SBMF we already defined the ${\bf p}={\bf
  0}$ term of the summations separately. The condensate is defined in
expression (\ref{eq:condensate}) for the N\'eel order and
(\ref{eq:condensate2}) for the collinear order. Here we also have to
be careful with these terms. For the infinitely large lattice these
equations can be simplified by replacing summations by integrals,
i.e.:
\[
\frac{1}{N} \sum_{\bf p} \dots \rightarrow \frac{1}{(2 \pi)^2}\int d
{\bf p} \dots,
\]
but all divergent terms have to be taken separately, e.g.
\[
1 = \frac{1}{N} \sum_{\bf p} \frac{h_{\bf p}+\lambda}{2 \omega_{\bf
    p}} \rightarrow 1=m_s + \frac{1}{(2 \pi)^2} \int d{\bf p}
\frac{h_{\bf p}+\lambda}{2 \omega_{\bf p}}.
\]
where $m_s$ is defined in equation (\ref{eq:condensate}). Partial
integration over ${\bf p}$ yields:
\begin{eqnarray}
  J^{\rm neel} &=& J_1 \gamma (m_s - \gamma) - 2 J_2 \kappa (m_s - \kappa), \nonumber
  \\ J^{\rm coll} &=& 2 J_2 \gamma_2(\bar{m}_s-\gamma_2) \nonumber \\ & &
  + J_1\left (\gamma_1(\bar{m}_s - \gamma_1) \sin^2 \phi - \bar{\kappa}
  (\bar{m}_s - \bar{\kappa}) \cos^2 \phi \right). \nonumber
\end{eqnarray}
These expressions for $J$ only hold for infinite system size $N$, in
contrast with the expressions for $T$ in (\ref{eq:TYneel}) and
(\ref{eq:TYcoll}).

\begin{figure}
  {\centering \hfill \epsfxsize=8cm \epsffile{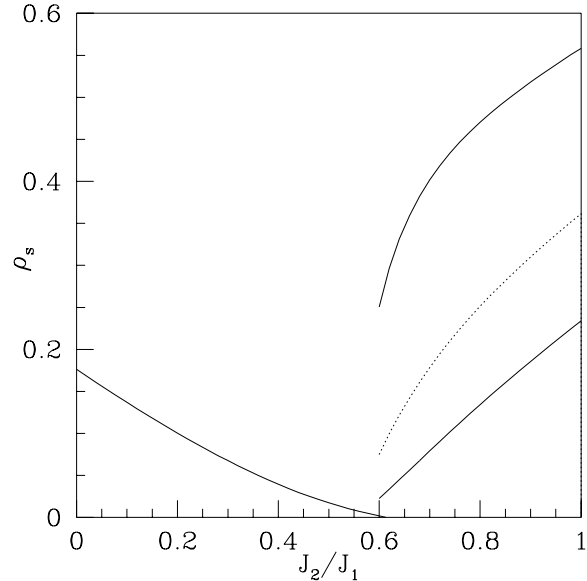} \hfill }
\caption[]{\label{fig:rho_s} The spin stiffness $\rho_s$ in units of $J_1$
  as function of the ratio $J_2/J_1$ (solid lines). For the collinear
  ordering the spin stiffness in the direction of the ferromagnetic
  order (lower solid curve) and in the direction of the
  antiferromagnetic order (upper solid curve) are drawn. The dotted
  line is the result found by Ivanov and Ivanov \cite{ivanov} for the
  collinear ordering.}
\end{figure}

The spin stiffness is given by $\rho_s=T+J$, thus for $N \rightarrow
\infty$
\begin{eqnarray}
  \rho_s^{\rm neel} &=& m_s(J_1 \gamma - 2 J_2 \kappa),
  \label{eq:rhosneel}\\ \vec{\vec{\rho}}_s^{\rm coll} &=& \bar{m}_s \left
  (\begin{array}{cc}2 J_2 \gamma_2 - J_1 \bar{\kappa} & 0 \\ 0 & 2 J_2
  \gamma_2 + J_1 \gamma_1
\end{array} \right ). \label{eq:rhoscoll}
\end{eqnarray}
In figure \ref{fig:rho_s} the numerical results of these formula
are presented. The phase transition is first order and no
intermediate phase exists. This is in contrast with statements in
the literature \cite{einarsson,zhitomirsky}, that suggest an
intermediate phase of dimer or plaquette order. Einarsson and
Schulz \cite{einarsson} studied the spin stiffness on small
clusters and extrapolated those results to the two-dimensional
geometry.  In their results the spin stiffness vanishes in the
region $0.4 \lesssim J_2/J_1 \lesssim 0.6$.

Ivanov and Ivanov \cite{ivanov} have applied a different method to
obtain $\rho_s$. They consider the correlation function $\xi = \langle
\vec{\cal S}_i \cdot \vec{\cal S}_j \rangle$.  Comparison of this
correlation function $\xi$ with the non-linear sigma model where $\xi
\sim \exp(2 \pi \rho_s/T)$ (here $T$ stands for the temperature),
yields
\begin{eqnarray}
  \rho_s^{\rm neel}&=& m_s(J_1\gamma-2J_2\kappa), \nonumber \\
  \rho_s^{\rm coll}&=& \bar{m}_s \sqrt{ (2J_2\gamma_2 - J_1 \bar{\kappa})( 2 J_2
    \gamma_2 + J_1 \gamma_1)}. \nonumber
\end{eqnarray}
Our expression for the collinear ordering is different from theirs.
This is not very surprising as they do not take anisotropy explicitly
into account. Their result is the geometric mean of the two components in (\ref{eq:rhoscoll}).

This section was started with a discussion on the route to be taken;
whether first to twist or apply mean field.  It was stated there that
both would lead to the correct answer. This can now easily be seen. As
$\langle \vec{\cal F}_{ij} \rangle = \langle \vec{\cal C}_{ij} \rangle
= 0$ there is indeed no difference.

\section{Isotropic Scaling}

Often it is impossible to obtain the values of observables in a system
of infinite size ($N \rightarrow \infty$). One approach to overcome
this obstacle is to derive the values for various system sizes $N$ and
afterwards extrapolate to the infinite size. This is for instance done
with data from quantum Monte Carlo calculations and exact
diagonalisation methods. It is necessary to know the size dependence
of the observables to obtain a good approximate for their limiting values.

Recently some controversy has arisen about the size dependence of
$\rho_s$ \cite{feiguin}. It was suggested that the lack of proper
scaling behaviour on small systems for intermediate range of
frustration $0.4 \lesssim J_2/J_1 \lesssim 0.6$ is an indication of
the absence of long-range magnetic order. It is therefore worthwhile
to take a closer look at the scaling behaviour of the various
quantities. This is not a hard task as all formulas in the last
section have an explicit size dependence. We will show that this lack
of scaling behaviour does not only appear for intermediate range of
frustration, but is a general feature of small systems.

We will focus on the N\'{e}el ordering as an example. We want to know
the scaling behaviour of the condensate $m_s$, and the two terms $J$
and $T$ that make up $\rho_s$ ($\rho_s=T+J$). The latter two will turn
out to have different scaling behaviour. The following discussion is
entirely based on the fact that the dispersion relation $\omega_{\bf
  k}$ is a periodic function that is smooth and positive everywhere
except at ${\bf 0}$ and $(\pi,\pi)$ where we see a linear behaviour;
e.g.  $\omega_{\bf k} = c |{\bf k}|$ for $|{\bf k}| \ll 1$
, the spin wave velocity.

First the condensate $m_s$ is considered. It is defined by $m_s(N) =
(h_{\bf 0} + \lambda)/( N \omega_{\bf 0})$ or
\begin{equation}
  \frac{m_s(N)}{2} = 1 - \frac{1}{N} \sum_{{\bf p} \neq 0} \frac{h_{\bf p} +
    \lambda_N}{2 \omega_{\bf p}}. \label{eq:S*scale}
\end{equation}
In the limit $N \rightarrow \infty$ the parameters ($\kappa_N$,
$\gamma_N$, $\lambda_N$) will obtain their limiting value ($\kappa$,
$\gamma$, $\lambda$) and the summation can be replaced by an
integration. Both changes will give rise to corrections. We carefully
investigate the size of these corrections below.

It is known that $\lim_{N \rightarrow \infty}m_s(N) = m_s>0$. This can
also be expressed as $\lim_{N \rightarrow \infty} $
$(h_{\bf 0} +\lambda)/( N \omega_{\bf 0}) =m_s$ or inverting this relation into one
for $\omega_{\bf p}$ at $|{\bf p}| \ll 1$
\begin{equation}
  \omega_{\bf p} = \sqrt{\frac{K_0}{N^2} + c^2 p^2}.
  \label{eq:omegascale}
\end{equation}
The two constants in this formula are given by
\begin{eqnarray}
  K_0 &=& N^2\left[ (4 J_2 \kappa_N + \lambda_N)^2 - (4 J_1
  \gamma_N)^2 \right ] = \left (\frac{4J_2
    \kappa_N+\lambda_N}{m_s(N)} \right )^2, \label{eq:k0isgap} \\
c^2 &=& \frac{1}{2} \left[
  \lambda_N^2 - (4 J_2 \kappa_N)^2 \right]+ {\cal O}(\frac{1}{N^2}). \nonumber
\end{eqnarray}
The suggestive notation $c^2$ anticipates that this is the spin wave
velocity since an antiferromagnet has a linear dispersion relation
$\omega_{\bf p}=c |{\bf p}|$ for low energy.  For finite system size
$N$ the smallest $\bf q$-vector in the summation (\ref{eq:S*scale})
has length $|{\bf p}| = \frac{2 \pi}{L}$ ($L^2=N$).  This means that
the $\bf q$-independent term, $K_0/N^2$, is small compared to the $\bf
q$-dependent term and gives rise to corrections of at least the order
${\cal O}(1/N^2)$ in the summation. We replace $\kappa_N$, $\gamma_N$
and $\lambda_N$ by their limiting values $\kappa$, $\gamma$ and
$\lambda$ and thereby neglect these corrections of order $ {\cal
  O}(1/N^2)$. As the term ${\bf p}=0$ is excluded the summation still
is finite.

The other effect, replacement of the summation by the integration, can
be treated quantitatively due to a lemma by Neuberger and Ziman
\cite{neuberger}.  They consider a function $f({\bf p})$ that is
periodic on the Brillouin zone. If this function $f({\bf p})$
satisfies $\frac{f({\bf p})}{|{\bf p}|} \rightarrow 1$ as $|{\bf p}|
\rightarrow 0$ and $f({\bf p})$ is non-zero and smooth in the rest of
the Brillouin zone, then
\[
\frac{1}{N} \sum_{{\bf p} \neq 0} \frac{1}{f({\bf p})} - \frac{1}{(2
  \pi)^2} \int d {\bf p}\frac{1}{f({\bf p})} =
\frac{\alpha(L^2/W^2)}{\sqrt{N}}
\]
We respect their notation by taking the argument $L^2/W^2$. The
numerical value $\alpha(1)=0.6208$ was computed in \cite{neuberger}.

\begin{figure}
  {\centering \hfill \epsfxsize=8cm \epsffile{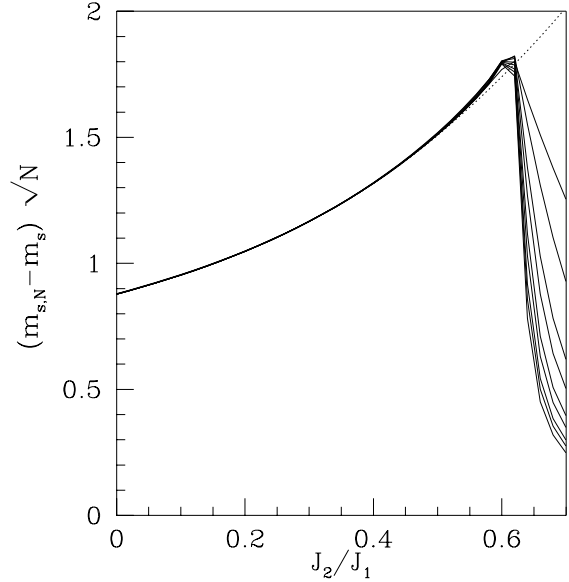} \hfill }
  \caption{\label{fig:S*scale} The numerical scaling behaviour of
    $m_s$ (solid lines) for sizes $N=10^2,20^2,\dots,90^2$ (bottom-up) compared
    with the theoretical curve (dotted line) At $J_2/J_1=0.62$ the
    ground state becomes instable and the discussion on the scaling
    behaviour is no longer applicable.}
\end{figure}

In our situation the limit $|{\bf p}| \rightarrow 0$ is given by
\[
\lim_{{\bf p} \rightarrow 0} \frac{1}{|{\bf p}|} \frac{2 \omega_{\bf
    p}}{h_{\bf p}+\lambda} \sqrt{2} \sqrt{ \frac{\lambda - 4 J_2
    \kappa}{\lambda + 4 J_2 \kappa}}.
\]
The quoted lemma therefore leads to
\[
m_s(N) - m_s = \frac{0.6208}{\sqrt{N}} \sqrt{2} \sqrt{\frac{\lambda +
    4 J_2 \kappa}{\lambda - 4 J_2 \kappa}}.
\]
This result is in excellent agreement with the numerical values we get
when the equations (\ref{eq:sumkappa}-\ref{eq:sum1}) are solved and
the obtained $\kappa_N$, $\gamma_N$ and $\lambda_N$ are inserted in
(\ref{eq:S*scale}). This is depicted in figure \ref{fig:S*scale}. For values of $J_2 > 0.62$
we know from figure \ref{fig:rho_s} that the phase becomes instable and it is no surprise that
the data does not longer fit the scaling relation.

\begin{figure}
  {\centering \hfill \epsfxsize=8cm \epsffile{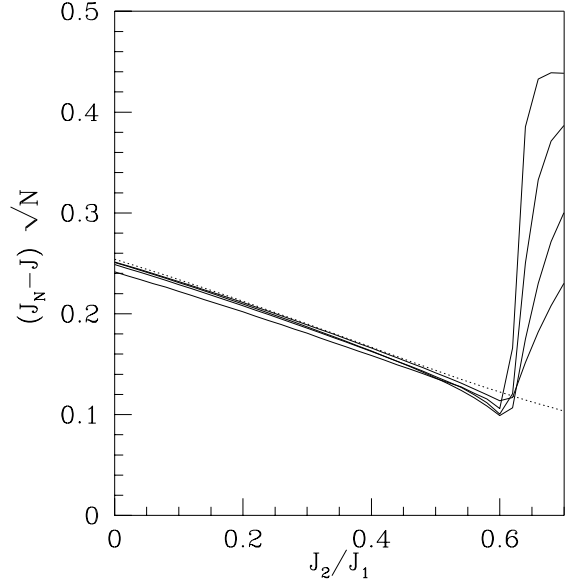} \hfill }
  \caption{\label{fig:JYscale} The numerical scaling behaviour of
    $J$ (solid lines) for sizes $N=10^2,20^2,40^2,100^2$ (bottom-up) compared with
    the theoretical curve (dotted line).}
\end{figure}

The derivation of the scaling behaviour of $J$ proceeds in the same
manner as above. The starting point is (\ref{eq:JYneel}). The result
is
\[
J_N-J = \frac{0.6208}{\sqrt{N}} \frac { \sqrt{\lambda^2 - (4 J_2
    \kappa)^2}}{4 \sqrt{2}}=\frac{0.6208}{\sqrt{N}} \frac{c}{4}.
\]
Again we find excellent agreement with the numerical results in figure
\ref{fig:JYscale} upto $J_2=0.62$, where the N\'eel phase becomes instable.

To obtain the scaling behaviour of $T$ a more involved
reasoning is required.  As is seen in (\ref{eq:TYneel}) it depends both on
$\kappa_N$ and $\gamma_N$. They are part of the set
$(\kappa_N,\gamma_N,\lambda_N)$ of solutions to
(\ref{eq:sumkappa}-\ref{eq:sum1}). As these are mutually dependent
they have to be solved simultaneously. In the formulation of
(\ref{eq:sumkappa}-\ref{eq:sum1}) the divergent '${\bf p}=0$'-terms
are still included. We will rearrange these equations to remove the
poles. Give the equations (\ref{eq:sumkappa}-\ref{eq:sum1}) the
numbers $I$, $II$ and $III$ respectively. For the derivation of the
scaling behaviour we will use the combinations $I-III$, $II-III$ and
$4J_2 \kappa_N I - 4 J_1 \gamma_N II + \lambda_N III $, or (using
(\ref{eq:hq}), (\ref{eq:dq}) and (\ref{eq:omegaq}))
\begin{eqnarray}
  \kappa_N & = & 1 + \frac{1}{N} \sum_{{\bf p}} \frac{h_{\bf p} +
    \lambda_N}{2 \omega_{\bf p}}\left( \cos p_x \cos p_y-1 \right ) ,
  \label{eq:scalingkappa}
  \\ \gamma_N &=& 1 + \frac{1}{N} \sum_{{\bf p}} \frac{1}{2
    \omega_{\bf p}} \left( \frac{\Delta_{\bf p}}{2} (\cos p_x + \cos
  p_y) - (h_{\bf p} +\lambda_N)\right ), \\ \lambda_N &=& 4 J_1
  \gamma_N^2 - 4 J_2 \kappa_N^2 + \frac{1}{2 N} \sum_{{\bf p}}
  \omega_{\bf p}. \label{eq:scalingl}
\end{eqnarray}
It is not possible to extract the scaling behaviour from the equations
completely in analytic form: the exponent of $N$ can be found, but the
prefactor cannot be derived easily. As an example we consider the
equation (\ref{eq:scalingl}).

First we want to replace the set $(\kappa_N,\gamma_N,\lambda_N)$ in
the summation over $\omega_{\bf p}$ by $(\kappa,\gamma,\lambda)$. From
(\ref{eq:omegascale}) we know that $\omega_{\bf p}$ has the general
form
\[
\omega_{\bf p} = \sqrt{\frac{K_0}{N^2} + g_N({\bf p})}.
\]
where $g_N({\bf p})\approx c^2 |{\bf p}|^2$ if $\frac{2 \pi}{L} \leq
|{\bf p}| \ll 1$. On all lattice points except the origin $g_N({\bf
  p}) \gg K_0/N^2$. We have $\omega_{\bf 0} = {\cal O}(1/N)$ or $1/2N
\omega_{\bf 0} = {\cal O}(1/N^2)$. For ${\bf p} \neq 0 $ we can expand
$\omega_{\bf p}$ in $K_0/N^2$;
\[
\omega_{\bf p} = \sqrt{g_N({\bf p})} \left [1+ \frac{K_0}{2 N^2
    g_N({\bf p})} + \dots \right ] = \sqrt{g_N({\bf p})} + {\cal
  O}(\frac{1}{N^{3/2}}).
\]
The corrections in the total summation thus become
\[
\frac{1}{2 N} \sum_{\bf p} \omega_{\bf p} = \frac{1}{2 N} \sum_{\bf p}
\sqrt{g_N({\bf p})} + \frac{1}{2 N} N \cdot {\cal
  O}(\frac{1}{N^{3/2}}).
\]
This mean that replacing the set $(\kappa_N,\gamma_N,\lambda_N)$ by
$(\kappa,\gamma,\lambda)$ leads to errors of the order ${\cal
  O}(1/N^{3/2})$.

As in previous cases the summation has to be replaced by an integral.
Neuberger and Ziman \cite{neuberger} provide a lemma well suited for
this type of summation.  It reads: consider the same $f({\bf p})$ as
before,
\begin{equation}
  \frac{1}{N} \sum_{{\bf p} \neq 0} f({\bf p}) - \frac{1}{(2 \pi)^2}
  \int d {\bf p}f({\bf p}) = -\frac{\beta(L^2/W^2)}{N^{3/2}}
  \label{eq:neuberger}
\end{equation}
The numerical value $\beta(1)=-0.7186$ was computed in
\cite{neuberger}.  Application of this lemma again leads to
corrections of the order ${\cal O}(1/N^{3/2})$. In a similar manner
the other equations for $\kappa$ and $\gamma$ can be treated.

\begin{figure}
  {\centering \hfill \epsfxsize=8cm \epsffile{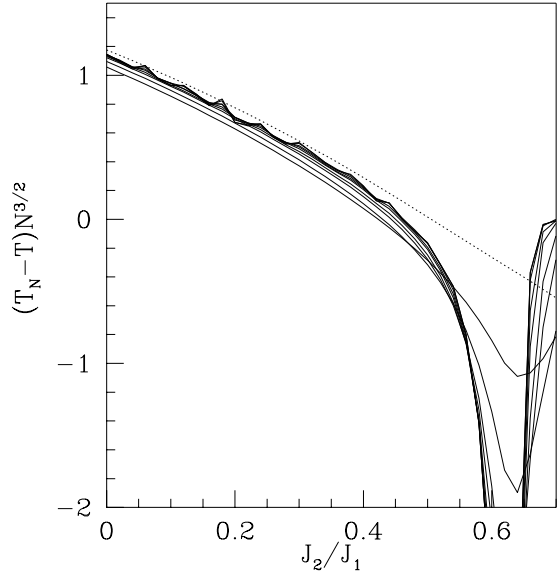} \hfill }
  \caption{\label{fig:TYscale} The numerical scaling behaviour of
    $T$ for square systems of size $N=10^2,20^2,\dots,90^2$. The
    $N^{3/2}$-scale is derived from analytical arguments in the text.
    The dotted line gives the expected behaviour if the size
    dependence of $\kappa $, $\gamma $ and $\lambda$ inside the
    summations (\ref{eq:scalingkappa}-\ref{eq:scalingl}) is
    neglected.}
\end{figure}

We now have two sources of finite-size corrections. On one hand
the dependence of $\kappa$, $\gamma$ and $\lambda$ on the systems
size. We know the exponent of this correction, ${\cal
O}(N^{3/2})$, but not the prefactor. On the other hand we have the
correction arising from the summations. There we not only know the
exponent, ${\cal O}(N^{3/2})$, but also the prefactor in equation
(\ref{eq:neuberger}).  Since both size dependencies are ${\cal
O}(N^{3/2})$ and can therefore not be separated, it is not easy to
obtain the overall prefactor. However if the size dependence of
the parameters $(\kappa_N, \gamma_N, \lambda_N)$ is neglected the
following prefactor is found for $T$:
\begin{equation}
T_N-T = \frac{0.7186}{N^{3/2}} \frac{1}{\sqrt{2}} \sqrt{\frac{\lambda
    +4J_2 \kappa}{\lambda - 4 J_2 \kappa}} \left[ \lambda - 8 J_2
  \kappa \right ]. \label{eq:Tisotropic}
\end{equation}
As can been seen in figure \ref{fig:TYscale} this gives a reasonable
description for not too large $J_2/J_1$. The numerical calculations
show that at $J_2/J_1=0.62$ the scaling behaviour is of order ${\cal
  O}(1/N)$.  In figure \ref{fig:TYscale} one therefore observes
crossing-over behaviour from ${\cal O}(1/N^{3/2})$-scaling to ${\cal
  O}(1/N)$-scaling around this point.

The size dependence of both $T$ and $J$ are now known. The scaling
behaviour of the spin stiffness $\rho_s $ ($=T+J$) is dominated by the
order ${\cal O}(1/\sqrt{N})$-behaviour of $J$.

\section{Scaling for a highly anisotropic geometry}

\begin{figure}
  {\centering \hfill \epsfxsize=8cm \epsffile{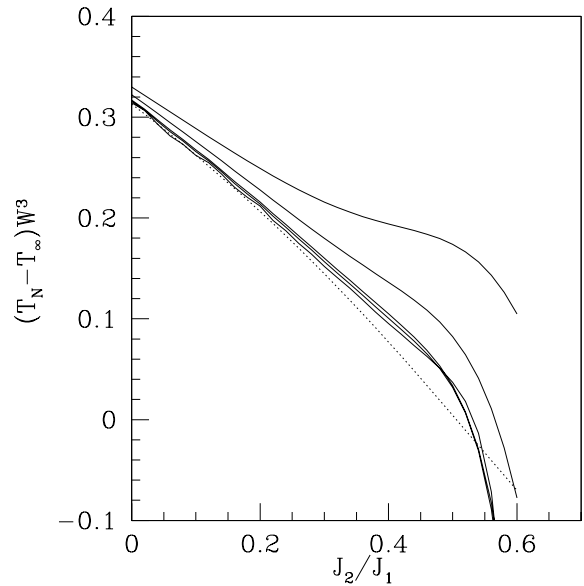} \hfill }
  \caption{\label{fig:Taniso}  The scaling of a system of 'almost'
    infinite length $L=4000$ and finite width $W=6,8,10,12,14,20$
    (top-down). The dotted line is the theoretical curve given in
    equation (\ref{eq:Taniso}).}
\end{figure}

The DMRG method that we will employ in the next chapter, allows us to
study systems of fairly large length $L$ while the width $W$ has to
remain small. This fact can be seen as an invitation to apply a two-step scaling procedure; first the length dependence is removed from
the observables by taking the limit $L \rightarrow \infty$, and
afterwards one takes the limit $W \rightarrow \infty$. For both steps
knowledge of the scaling behaviour is essential. The first step is
fairly trivial and will be discussed in due time. For the second one,
scaling the width $W$ away, some guidance from the SBMF results is
useful. Both the kinetic term $T$ and the energy density $E_0/N$ can
be treated in the same fashion.  The energy density can be simply
obtained by evaluating the Hamiltonian ${\cal H}$ in the SBMF ground
state; the expectation values of the operators can be written in the
mean fields $\kappa$ and $\gamma$. The energy density is given by
\[
\frac{E_0}{N}= -2J_1(2\gamma^2 - \frac{1}{4}) + J_2 (2 \kappa^2 -
\frac{3}{4}).
\]
As we have not spent any time so far on the energy density, we will
start with the scaling behaviour observed in the literature. Neuberger
and Ziman \cite{neuberger} derive
\[
\frac{E_0}{N} - \lim_{N\rightarrow \infty} \frac{E_0}{N} = \frac{2
  c}{N^{3/2}} \beta ( L^2/W^2 ).
\]
The function $\beta$ is the same as defined in (\ref{eq:neuberger}).
We thus seek the behaviour of $\beta$ as $L \rightarrow \infty$.
Neuberger and Ziman's approach \cite{neuberger} can be extended 
to this situation. Still, the algebraic manipulations are quite tedious  \cite{otterlo} and using
\[
\zeta(3)=\sum_{n=1}^\infty \frac{1}{n^3}=1.2021,
\]
 they lead to
\[
\lim_{L\rightarrow \infty} \frac{\beta( L^2/W^2)}{(LW)^{3/2}} =
-\frac{1.2021}{2 \pi W^3}.
\]
As expected all length dependence drops out and only a finite width
dependence remains. When we neglect the dependence of the parameters
$\kappa$, $\gamma$ and $\lambda$ on the system size as before in
(\ref{eq:Tisotropic}), the expressions for the energy density $E_0/N$
and the kinetic term $T$ are
\begin{eqnarray}
\frac{E_0}{N} -  \lim_{N\rightarrow \infty} \frac{E_0}{N} &=& -\frac{1.2021}{\pi W^3} c. \label{eq:E0aniso} \\
T_N-T &=&\frac{1.2021}{2 \sqrt{2} \pi W^3}  \sqrt{\frac{\lambda
    +4J_2 \kappa}{\lambda - 4 J_2 \kappa}} \left[ \lambda - 8 J_2
  \kappa \right ]. \label{eq:Taniso}
\end{eqnarray}
In figure \ref{fig:Taniso} we have numerically checked the second of
these predictions. Just like in the isotropic case, the power is
correct and the prefactor is reasonable.

\begin{figure}
  {\centering \hfill \epsfxsize=8cm \epsffile{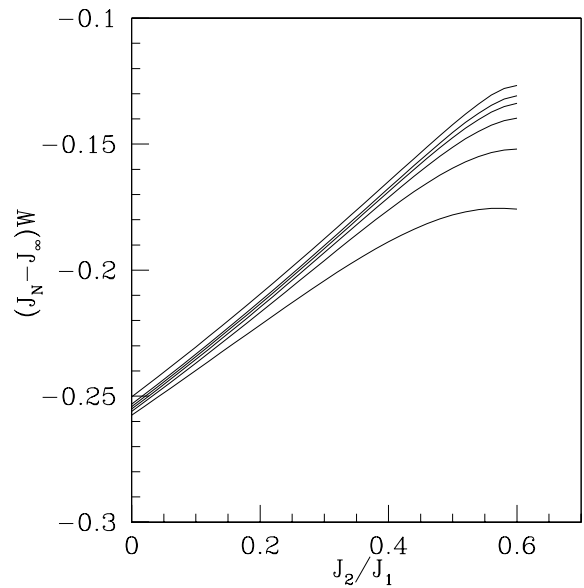} \hfill }
  \caption{\label{fig:Janiso}  The scaling of a system of 'almost'
    infinite length $L=4000$ and finite width $W=6,8,10,12,14,20$
    (bottom-up). }
\end{figure}

Unfortunately a similar limit $L \rightarrow \infty$ can not be taken
in the scaling expression for the current-current correlation $J$. The
reason can be found in the underlying assumptions; we supposed that
the dispersion relation is linear and the correction due to the fact
the $\omega_{\bf k}$ has a gap $\Delta$, as presented in
(\ref{eq:omegascale}) has been neglected. For a highly anisotropic
system, $L \rightarrow \infty$, this assumptions is not allowed. We
can illustrate this by an example: the summation over reciprocal
lattice in the long direction of the system can be replaced by an
integral and the outcome diverges;
\[
\frac{1}{LW} \sum_{k_x,k_y} \frac{1}{|{\bf k}|} = \frac{1}{W}
\sum_{k_y} \frac{1}{2 \pi} \int dk_x \frac{1}{\sqrt{k_x^2+k_y^2}}
\rightarrow \infty,
\]
whereas the infinite size expression stays bounded;
\[
\frac{1}{(2 \pi)^2} \int d {\bf k} \frac{1}{|{\bf k}|} =
\hbox{finite}.
\]
The analytical expression we therefore should derive, involves the gap
$\Delta$. Above we mentioned that it is not possible to obtain the
prefactor of the scaling behaviour of $\gamma, \kappa$ and $\lambda$.
The gap $\Delta$ is directly expressed in these three quantities and
its explicit size dependence therefore becomes just as elusive.
However, we can resort to numerical means. In figure \ref{fig:Janiso}
we establish that the current-current correlation shows a $1/W$
scaling behaviour.

To conclude, the width $W$ has taken over the position of the square
root of the system size , $\sqrt{N}$, in all scaling relations when
the highly anisotropic limit of $L \rightarrow \infty$ is taken.
($\sqrt{N}=\sqrt{LW}$) This is naturally not surprising as the
smallest dimension of the system is $W$ which should set the length
scale.

In the next chapter a brief discussion will be given of the behaviour
of the energy gap $\Delta$. This is the energy difference between the
ground state singlet and the first excitation triplet. For an infinite
large system with spin waves this gap obviously will be zero whereas
for other types of order it might remain finite.  For reasons to be
outlined later, we will consider the member of this triplet with
${\cal S}^z=1$.  In the SBMF approximation the gap $\Delta$ is given
by
\[
\Delta = 2 \omega_0.
\]
The factor 2 can easily be understood if we rewrite the operator
${\cal S}^z$ in the excitation operators $\alpha_{\bf k}$ and
$\beta_{\bf k}$;
\begin{eqnarray}
{\cal S}^z &= &\sum_j {\cal S}^z_j = \frac{1}{2} \sum_j \left (
  a_j^\dagger a_j - b_j^{\dagger} b_j \right ) \nonumber \\
&=& \frac{1}{2} \sum_{\bf
  k} \left ( a_{\bf k}^{\dagger} a_{\bf k} - b_{\bf k}^{\dagger}
  b_{\bf k} \right ) = \frac{1}{2}\sum_{\bf k} \left (
  \alpha_{\bf k}^{\dagger} \alpha_{\bf k} - \beta_{\bf k}^{\dagger}
b_{\bf k} \right). \nonumber
\end{eqnarray}
The ground state contains no excitations so ${\cal S}^z |0 \rangle =
0$.  The ${\cal S}^z=1$-space can be reached by creating {\it two}
excitations; ${\cal S}^z \alpha_{\bf 0} ^{\dagger} \alpha_{\bf 0}
^{\dagger}|0 \rangle = 1 \cdot \alpha_{\bf 0} ^{\dagger} \alpha_{\bf
  0} ^{\dagger}|0 \rangle$. This is not surprising in the view of the
fact that the number of bosons should be conserved on a site.
Applying only one creation operator creates an excitation that does
not satisfy this condition. When two creation operators are applied
to the ground state, their combination contains terms that do satisfy
this condition, for example, it contains terms of the form
$a^{\dagger}_j b_j$.

\begin{figure}
  {\centering \hfill \epsfxsize=8cm \epsffile{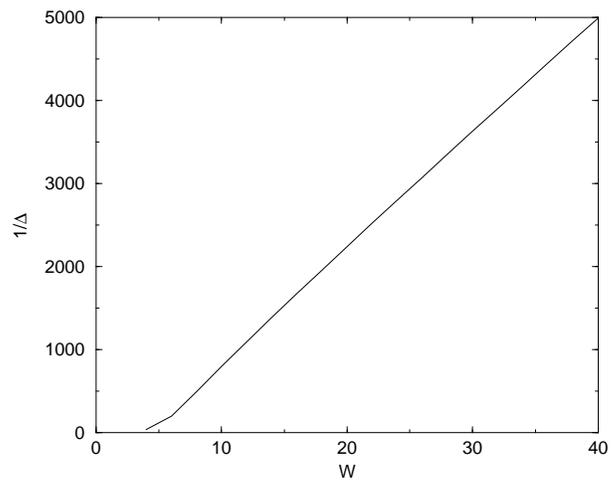} \hfill }
  \caption{\label{fig:finitegap} For almost infinite length, $L=1000$,
    and various widths $W$ we observe that the inverse of the gap
    $\Delta$ is linear proportional to the offsetted width $1/\Delta
    \sim W-W_0$.}
\end{figure}

From (\ref{eq:omegascale}) and (\ref{eq:k0isgap}) we know that the gap
$\Delta=2\omega_{\bf 0} = 2 \sqrt{K_0}/N$ has a very subtle size
dependence as $\kappa$, $\lambda$ and $\gamma$ have an ${\cal
  O}(1/N^{3/2})$ dependence but $K_0/N^2$ has an ${\cal O}(1/N^2)$
dependence. Adding to this the high anisotropy, we can only establish
the finite-size corrections numerically. The best fit with a simple
function is a size dependence $\Delta \sim 1/(W-W_0)$ as can be seen
in figure \ref{fig:finitegap}. This is in line with the idea that the
gap $\Delta$ can be considered proportional to the inverse of the
correlation length. The smallest length scale in the system is $W$
leading to a $1/W$ behaviour.

Unfortunately, the scaling of the gap does not follow the route that previous scaling
relations have followed.  The scaling behaviour of $E_0/N$, $T$ and $J$
in the highly anisotropic limit can be summarised as replacing the
system size $N$ in the formulae for the isotropic geometry by $W^2$.
For the gap $\Delta$ that would lead to a $1/W^2$ dependence whereas
we actually find a $1/(W-W_0)$ dependence.

\section{Conclusions}

In this chapter we have employed the SBMF approximation to get the
approximate phase diagram of the frustrated Heisenberg model. We found
two phases, N\'eel and collinear. The energies of both ground states
suggest a first order phase transition at $J_2/J_1 \approx 0.595$.
There is no evidence for an intermediate phase. SBMF approximations in
line with either dimer-like order always yield higher ground
state energies. As mentioned before there are several articles in the literature where
an intermediate phase is suggested \cite{einarsson, zhitomirsky}.

The second half of the chapter was spent on finite-size scaling. The
scaling behaviour of various quantities ($\rho_s, J,
\dots$) was derived for both square, $L=W$, and a highly anisotropic geometry,
$L \gg W$. These scaling relations will be used in the next chapter to
extrapolate the numerical data to infinite system size.

%% file: J1J2DMRG.tex
\chapter{Density Matrix Renormalisation Group approach to the stiffness \label{chap:J1J2DMRG}}

\section{Expressions for $\rho_s$}

In chapter \ref{chap:SBMF} we studied the spin stiffness in a
Schwinger-boson mean-field approximation. In that case the order has
a distinct orientation and it is clear how to twist it. If we do not
resort to a similar mean-field approximation and consider the
frustrated Heisenberg model on a finite-size lattice, the orientation
of the ground state is not as well defined. This model has a continuous
symmetry; homogeneous rotations in spin space leave the Hamiltonian
invariant. We also know that in a finite system the ground state is
unique. It therefore has to be rotationally invariant. Only in an
infinitely large system spontaneous symmetry breaking can occur and we
can associate a direction with the order of the ground state.  The only
other way to break the symmetry is by enforcing an orientation. This
is for instance the case in a mean-field approximation.

The orientation of the order is thus homogeneously distributed over
the sphere; every orientation in spin space is just as likely.  The
twist we apply has a plane associate with it. Previously we twisted in
the $x-y$ plane. The fraction of the ground state oriented along the
$z$-axis will not be affected by this twist. Leaving out that
fraction, we only twist $2/3$ of the order parameter. To compensate
for that we follow Einarsson and Schulz \cite{einarsson} and introduce
\[
T_{sym}=\frac{3}{2} T ~~,~~ J_{sym}= \frac{3}{2} J,
\]
with $T$ and $J$ given in (\ref{eq:rhofinal}).  The spin stiffness is
given by
\begin{equation}
  \rho_s=T_{sym}+J_{sym}. \label{eq:rhosym}
\end{equation}

A further complication lies in the numerical nature of our approach.
Unfortunately the ground state we calculate, will not be entirely
rotational symmetric. The reason for this touches on the very nature of
spontaneous symmetry breaking. As the Hamiltonian is rotational
invariant, it only takes a small field to orient the ground state.
Basis states in line with this field will prevail. In standard mean-field approximations we use this by introducing external fields to fix
the orientation. Here we go a step further and directly meddle with
the basis. By definition we start our DMRG calculation with an
asymmetric and poor basis. At each step the ground state wave function
will be symmetry broken in the same orientation as the basis. The
following basis truncation will again be asymmetric.  Even if we were
to start off with a symmetric basis, numerical errors would break the
symmetry eventually. It is very difficult to maintain a global
symmetry by means of iterative local basis updates.

To repair this partial symmetry breaking, we would have to incorporate
the rotational symmetry exactly in the procedure.  The DMRG method
allows for certain symmetries to be conserved as was extensively
discussed in chapter \ref{chap:ITFDMRG}. A good example is ${\cal
  S}^z$, which we also conserve in the present calculations.  This is
possible because to every basis state we can assign a quantum number
$s^z$. The ${\cal S}^z$ for the entire system is then the sum of the
$s^z$ of the basis states on the individual parts. It would be nice if
also the total spin ${\cal S}$ could be conserved, but we will argue
that this is not feasible. The ground state $|\psi_0 \rangle$ lies in the
${\cal S}=0$ space. The conditions for this restriction can be easily
derived:
\[
\begin{array}{ccrcl}
{\cal S}^2|\psi_0\rangle=0 &\rightarrow& \sum_{\alpha=x,y,z} {\cal S}^{\alpha^2} | \psi_0 \rangle &=& 0  \\
&\rightarrow& \langle \psi_0| \left({\cal S}^{\alpha} \right )^2| \psi_0 \rangle &=& \left | {\cal S}^\alpha |\psi_0\rangle \right|^2 =0 \\
&\rightarrow& {\cal S}^\alpha | \psi_0\rangle&=&0. \end{array}
\]
These conditions are rephrased to reflect that we work in the
basis that conserves ${\cal S}^z$,
\begin{equation}
{\cal S}^z| \psi_0 \rangle ={\cal S}^+| \psi_0 \rangle ={\cal S}^-| \psi_0 \rangle =0
\label{eq:sconserved}
\end{equation}
A direct consequence of fulfilling (\ref{eq:sconserved}) is a
rotationally invariant ground state, as the global rotations in spin
space over an angle $r$ are given by $\exp(i r {\cal S}^\alpha)$.

Our approximation $|\phi_0 \rangle$ already satisfies the first condition of (\ref{eq:sconserved}), ${\cal S}^z| \phi_0 \rangle
=0$, in a standard implementation of the DMRG.
Conservation of the second and third condition would require for each
basis state in an individual part $A$ of the system $|i \rangle_A$ the
image ${\cal S}^\pm_A |i \rangle_A$.  This would scale up the number
of basis states tremendously and the calculation would become
prohibitively large. We will therefore neglect this symmetry and
evaluate ground state wave functions that are only approximately
rotational invariant.

On the other hand, there is no reason why the symmetry should be {\it
  completely} broken; for narrow systems, the DMRG is accurate enough
to compensate for this symmetry breaking tendency.

In general the final ground state will be somewhere in between a
symmetry broken state and a rotational invariant state. The
expressions for the kinetic term $T$ and the current-current
correlation $J$ can be symmetrised to overcome this orientational
problem;
\begin{eqnarray}
  t_{all} &=& J_1 \sum_{\langle ij \rangle} (\hat{q}
  \cdot(\vec{r}_i-\vec{r}_j))^2 \vec{\cal S}_i \cdot\vec{\cal S}_j
  +J_2 \sum_{[ ij ]}(\hat{q} \cdot (\vec{r}_i-\vec{r}_j))^2
  \vec{\cal S}_i \cdot\vec{\cal S}_j, \nonumber \\ \vec{j}_{all} &=&
  iJ_1 \sum_{\langle ij \rangle} (\hat{q}\cdot
  (\vec{r}_i-\vec{r}_j))\vec{\cal S}_i \times \vec{\cal S}_j
  +iJ_2 \sum_{[ ij ]}(\hat{q}\cdot (\vec{r}_i-\vec{r}_j))
  \vec{\cal S}_i \times \vec{\cal S}_j, \nonumber
\end{eqnarray}
and define
\[
T_{all} = -\frac{1}{N} \langle \phi_0 | t_{all} | \phi_0 \rangle ~~,~~
J_{all}^\alpha = \frac{1}{N} \langle \phi_0 | j_{all}^{\alpha}
\frac{1}{E_0-{\cal H}} j_{all}^{\alpha} | \phi_0 \rangle,
\]
then the stiffness $\rho_s$ is given by
\begin{equation}
  \rho_s = T_{all} + \frac{1}{2} \sum_{\alpha=x,y,z} J_{all}^\alpha. \label{eq:rhoall}
\end{equation}
Three current-current correlation $J_{all}^\alpha$ have to be
calculated for (\ref{eq:rhoall}) whereas only one $J_{sym}$ for
expression (\ref{eq:rhosym}). In the next section we will see that
this is substantial more involved and we prefer to use formula
(\ref{eq:rhosym}) whenever the accuracy permits us to. In practice
this means that for narrow systems we use the symmetric form. For
wider systems the general form is necessary.

\section{Calculating wave functions \label{sec:calcwave}}

The expressions for the stiffness $\rho_s$ in (\ref{eq:rhosym}) and
(\ref{eq:rhoall}), have to be implemented numerically. The first
ingredient is the ground state. A standard implementation of the DMRG
results in a good estimate $|\phi_0 \rangle$ of the ground state. The
kinetic term $T$ can be obtained by a simple measurement on this
wave function $|\phi_0 \rangle$.  However the current-current correlation
$J$ needs a more elaborate approach.

An inversion has to be performed for $J$. We will prove that we can
invert $E_0-{\cal H}$ within the {\it subspace} spanned by the basis
states for the various parts. To derive this we first take a step back
and inspect the method to calculate the ground state. At each iteration
of the DMRG the state $|\phi_0 \rangle$ in the subspace that has a minimal
energy $E_0=\langle \phi_0|{\cal H}|\phi_0\rangle/\langle \phi_0|\phi_0\rangle$, is
selected.  This energy $E_0$ is always larger than the true
ground state energy and we thus have a variational principle.
At every next iteration we can improve upon our estimate by
simple minimising $E_0$ further starting with the -truncated- outcome
of the previous iteration.  This variational principle is crucial for
the method as it enables to distinguish the best approximation to the
ground state from other configurations in the basis.

We can design a similar variational principle for the inversion.
Define $g(x)$,
\begin{equation}
  g(x)=\frac{1}{2} \langle x|{\cal H}-E_0| x \rangle + \langle x |\phi_{\bf j}
  \rangle , \label{eq:mininverse}
\end{equation}
where $|\phi_{\bf j} \rangle = j|\phi_0 \rangle$ and $E_0$ is the best estimate
of the ground state energy known at that point in the procedure
\footnote{Unfortunately this is not always the latest calculated
  energy; it tends to fluctuate. This was already discussed at the end of section \ref{sec:bandmethod}.}. This function has a global minimum
at
\begin{equation}
|x \rangle=|\phi_{\bf i} \rangle = \frac{1}{E_0-{\cal H}} |\phi_{\bf j} \rangle, \label{eq:inverse}
\end{equation}
as the quadratic term, $\frac{1}{2} \langle x |{\cal H} -E_0| x
\rangle$, is positive definite. In the realm of linear algebra $|\phi_{\bf i} \rangle$ is called the correction vector. This function provides us with a
variational principle similar to the one we had before. Moreover the
minimum of the function within a specific {\it subspace} is also given
by (\ref{eq:inverse}) where $|\phi_{\bf i} \rangle$, $|\phi_{\bf j} \rangle$ and
${\cal H}$ are now restricted to that subspace. The inversion within
the subspace is thus the best approximation we can make for the global
minimum.

At every step the subspace changes and we can get closer to the real
inverse $|\phi_{\bf i} \rangle$. If $|\phi_{\bf j} \rangle$ is known with high
accuracy, $1/(E_0-{\cal H})|\phi_{\bf j} \rangle$ can be obtained with
similar accuracy. A nice by-product is that
\begin{equation}
  g(\phi_{\bf i})=\frac{1}{2} \langle \phi_0 | j \frac{1}{E_0-{\cal H}} j |\phi_0 \rangle;
  \label{eq:globalminimum}
\end{equation}
Apart from the prefactor this is essentially the expression for $J$.

The basis has to be tuned to present these wave functions $\{ |\phi_0
\rangle, |\phi_{\bf j} \rangle, |\phi_{\bf i} \rangle \}$ optimally. The reason
for including the first wave function, $|\phi_0 \rangle$, may be evident.
The other two are necessary, since we need expression
(\ref{eq:globalminimum}) accurately. If the basis does not properly
represent $|\phi_{\bf i} \rangle $ or $| \phi_{\bf j} \rangle$, $g(\phi_{\bf i}) = \frac{1}{2}
\langle \phi_{\bf j} | \phi_{\bf i} \rangle$ is incorrect. To adjust the basis
to these wave functions we have to incorporate them in the density
matrix.  Let us briefly outline the reasoning behind that. Define the
truncation error $P$ by
\[
P=\left | |\phi_0 \rangle - |\tilde{\phi}_0 \rangle \right | ^2 + \left | | \phi_{\bf
    j} \rangle - |\tilde{\phi}_{\bf j} \rangle \right | ^2 + \left | | \phi_{\bf
    i} \rangle - |\tilde{\phi}_{\bf i} \rangle \right | ^2.
\]
The tilde denotes the projection of the wave function on the truncated
basis. The truncation error $P$ has to be minimal. A few linear
algebra manipulations leads to the density matrix
\begin{equation}
\rho_{ii'} = \rho^0_{ii'} + \rho^{\bf j}_{ii'} + \rho^{\bf i}_{ii'}. \label{eq:rhoinversion}
\end{equation}
The density matrix is thus the sum of the individual density matrices.
As usual the most important states correspond to the eigenvectors of
this density matrix $\rho$ with the largest eigenvalues.  The
different density matrices could have different weights, but the
effect on the accuracy of the spin stiffness is unknown. We set them
therefore to be equal.

The remaining issue of the last section has now also been answered;
expression (\ref{eq:rhoall}) for $\rho_s$ is more elaborate than
(\ref{eq:rhosym}) as three instead of one inversions have to be
performed.  Moreover all these extra wave functions (four in total; two
extra currents and two extra inverses) have to be included in the
density matrix along the lines of equation (\ref{eq:rhoinversion}).
Naturally the basis will then be less suited for each individual
wave function and an overall loss of accuracy will follow.

\section{Geometry}

The shape of the systems we study is dictated by limitations of the
DMRG. Their width is fairly restricted ( maximally 8 sites wide ) and
periodicity along the length of the system is not feasible.  Earlier
we explained that the spin stiffness can only be measured with the
expressions we derived if the axis along which we twist, denoted by
$\hat{\bf q}$, is periodic. We used two different arguments for
this, that are both valid in their own right; first, the perturbation
theory for an open system will not give the desired state and energy.
Secondly, a similarity with superfluidity exists; a superfluid cannot
flow freely when there is an impenetrable wall in its way.

The system thus has to be periodic in the -narrow- width direction
and open in the length direction; the shape of a cylinder.

The model itself puts some extra constrains on the wrapping of the
lattice around the cylinder. In order to frustrate neither the N\'eel
nor the collinear ordering, the periodicity of 2 lattice sites has to
be satisfied.

\begin{figure}
  \centering \epsfxsize=15cm \epsffile{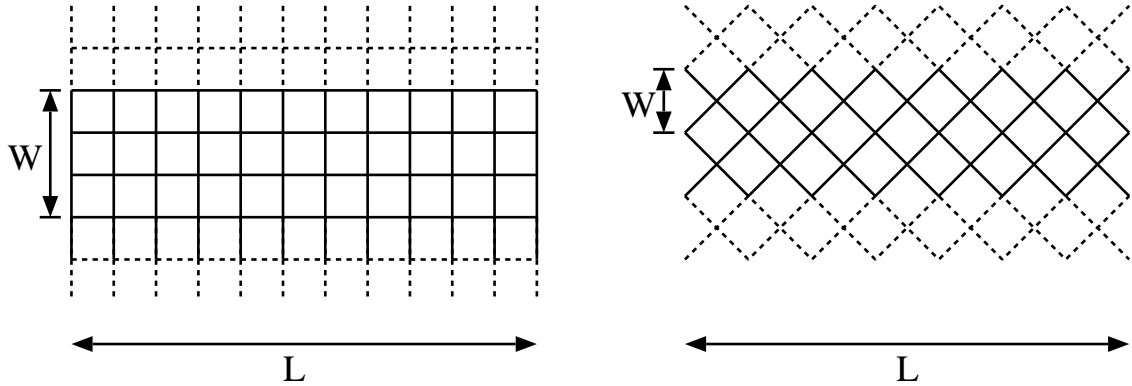}
\caption{\label{fig:squaretilted} The square and the tilted square lattice
  of width $W$ and length $L$. In the square lattice the periodicity
  is along the vertical lattice-axis. In the tilted lattice the
  periodicity is along the (vertical) diagonal of the lattice.
  Periodic images of the nearest neighbour bounds are depicted by
  dashed lines}
\end{figure}

The two lattices depicted in figure \ref{fig:squaretilted} fulfill
both requirements. The width $W$ of the lattice is the number of
sites one passes going round the cylinder. The length on the other
hand is the maximum number of sites one encounters while scanning
along the long direction.

When the next-nearest-neighbour coupling becomes dominant, these
lattices both fall apart in two sublattices. For the square lattice,
these sublattices have an effective width of $W/2$, whereas the tilted
lattice breaks up in sublattices of width $W$. Knowing that the
accuracy of the DMRG rapidly decreases with increasing width of the
system, we expect results of a strongly decreasing accuracy for the
tilted lattices with increasing frustration $J_2$.

The direction $\hat{\bf q}$ in which the stiffness is measured is
different for the square and for the tilted square lattice.  The
square lattice allows a measurement along the axis corresponding to
the width direction of the cylinder. On the tilted lattice the
direction to be taken is the diagonal of the lattice. In short:
\[
\mbox{Square lattice :} ~ \hat{\bf q} = \left ( \begin{array}{c} 0 \\
    1 \end{array} \right ) ~~ , ~~\mbox{Tilted square lattice :} ~
\hat{\bf q} = \frac{1}{\sqrt{2}} \left ( \begin{array}{c} 1 \\ 1
\end{array} \right ).
\]

\section{Scaling}

The DMRG accuracy rapidly decreases with increasing system width $W$.
It becomes therefore necessary to apply finite-size scaling theory to
obtain quantities of the two-dimensional system. Here we implement a
two step scaling where we can use the discussion on anisotropic scaling
in the previous chapter and the scaling analysis of the two-dimensional ITF in sections \ref{sec:1dscaling} and \ref{sec:2dscaling}.

The first step exploits the strength of the DMRG; for a fixed width
$W$ it is numerically not difficult to vary the length $L$
substantially. By doing so, we can extract the dependence of various
quantities on this length $L$ and remove it.  The remaining fraction
corresponds to a system of length $L=\infty$. To obtain this
one-dimensional scaling behaviour we can not employ the SBMF
approximation of the previous chapter. That was based on periodic
boundary conditions whereas the systems considered here have open
boundary conditions in the long direction. Still, the correct exponent
can easily be derived.

While the correlation length $\xi$ is finite, the influence of open
boundaries on both ends only extends over
this length $\xi$. The corrections to the bulk behaviour of all
properties $E_0, T, J$ and $\rho_s$ is then just a surface term;
\begin{eqnarray}
  \frac{E_0(L,W)}{N} - \frac{E_0(\infty,W)}{N} &=& {\cal O} \left(
  \frac{1}{L} \right ), \nonumber \\ T(L,W) - T(\infty,W) &=& {\cal O}
  \left( \frac{1}{L} \right ),\nonumber \\ J(L,W) - J(\infty,W) &=&
    {\cal O} \left( \frac{1}{L} \right ),\nonumber \\ \rho_s(L,W) -
    \rho_s(\infty,W) &=& {\cal O} \left( \frac{1}{L} \right
    ),\nonumber
\end{eqnarray}

For the second step, scaling in the width direction, we fall back on
the expressions derived in the previous chapter. As the length has
become infinite, $L \rightarrow \infty$, it is no longer relevant
whether the corresponding boundary is open or periodic. We can refer
to the results for the periodic case derive before; expressions
(\ref{eq:E0aniso}) and (\ref{eq:Taniso}) contain the ${\cal O}(1/W^3)$
scaling behaviour for the energy density $E_0/N$ and the kinetic term
$T$ respectively. With the help of (\ref{eq:E0aniso}) we can also
extract the spin wave velocity $c$.  Moreover figure \ref{fig:Janiso}
demonstrates the ${\cal O}(1/W)$ scaling behaviour that SBMF yields
for the current-current correlation $J$.

\section{Results}

Reliable finite-size scaling requires a substantial number of data
points. We have considered various system sizes to make -at least- the
first step in the scaling procedure, $L \rightarrow \infty$,
indisputable. For the square geometry (see figure
\ref{fig:squaretilted} widths $W=4,6,8$ were considered and for the
tilted geometry widths $W=2,4$ were studied. In all graphs we set
$J_1=1$.

We have used both DMRG variants, the original one proposed by White
\cite{white92} and our implementation \cite{ducroo98}. For this model
we confirm the statements made in chapter \ref{chap:ITFDMRG}; the
variant of White is much more flexible. On the other hand, our variant
needs $30\%$ fewer states for a similar accuracy in the calculation.
Furthermore the ground state is more symmetric as the translational
symmetry is strictly conserved. A relative small extra gain can be
made by reusing bases; whenever we start a new calculation that
differs from the last one in the size of the frustration $J_2$, the
bases for the various parts for the preceding value of $J_2$ can be
used. This reduces the number of sweeps needed to about three.

\subsection{Scaling to $L=\infty$}

\begin{figure}
  \centering \epsfxsize=10cm \epsffile{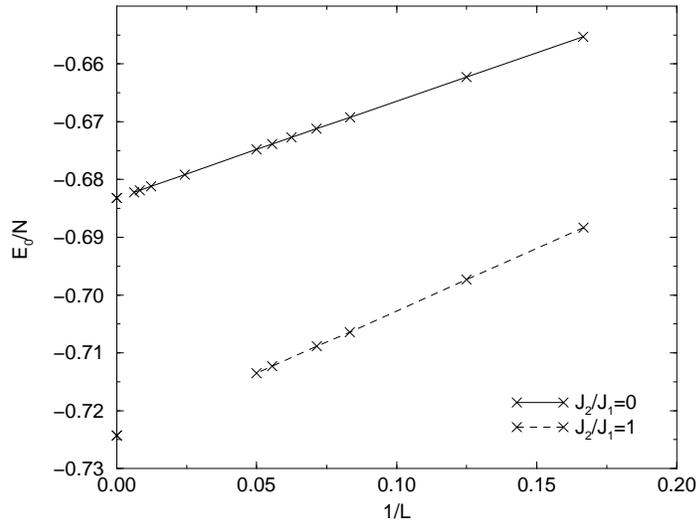}
\caption{\label{fig:E0.W=4} The energy density $E_0/N$ as function of the
  inverse length for width $W=4$ on the square lattice.}
\end{figure}

\begin{figure}
  \centering \epsfxsize=10cm \epsffile{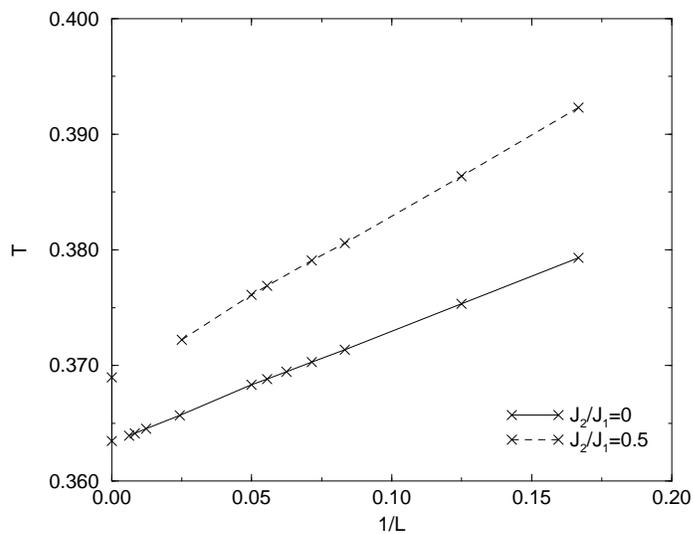}
\caption{\label{fig:T.W=4} The kinetic term $T$ as function of the inverse
  length for width $W=4$ on the square lattice.}
\end{figure}

\begin{figure}
  \centering \epsfxsize=10cm \epsffile{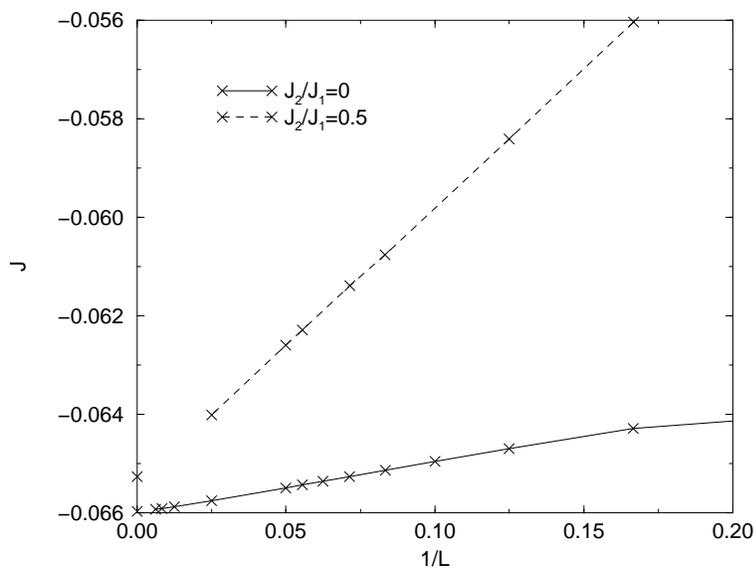}
\caption{\label{fig:J.W=4} The current-current correlation $J$ as function
  of the inverse length for width $W=4$ on the square lattice.}
\end{figure}

\begin{figure}
  \centering \epsfxsize=10cm \epsffile{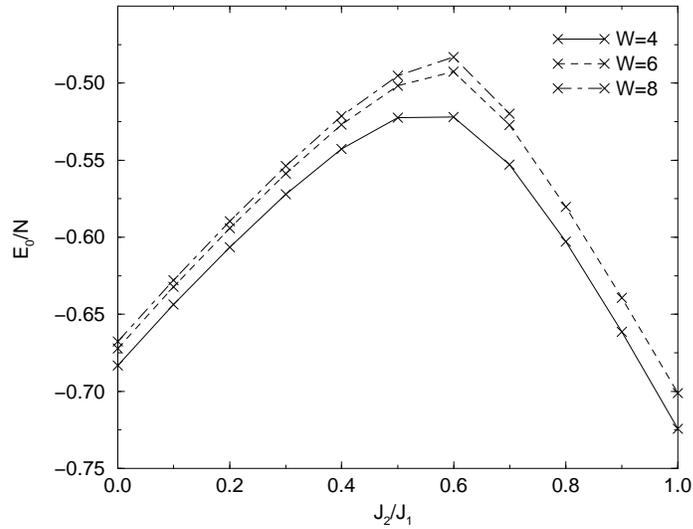}
\caption{\label{fig:E0.W=X} The extrapolated energy density
  $E_0/N$ for widths $W=4,6,8$ on the square lattice. For $W=8$ ratios
  $J_2/J_1 \ge 0.8$ were not considered.}
\end{figure}

\begin{figure}
  \centering \epsfxsize=10cm \epsffile{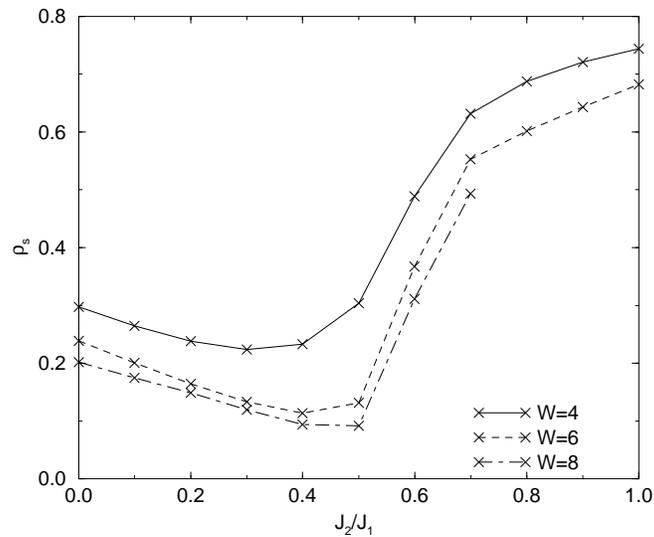}
\caption{\label{fig:rhos.W=X} The extrapolated spin stiffness
  $\rho_s$ for widths $W=4,6,8$ on the square lattice.}
\end{figure}

\begin{figure}
  \centering \epsfxsize=10cm \epsffile{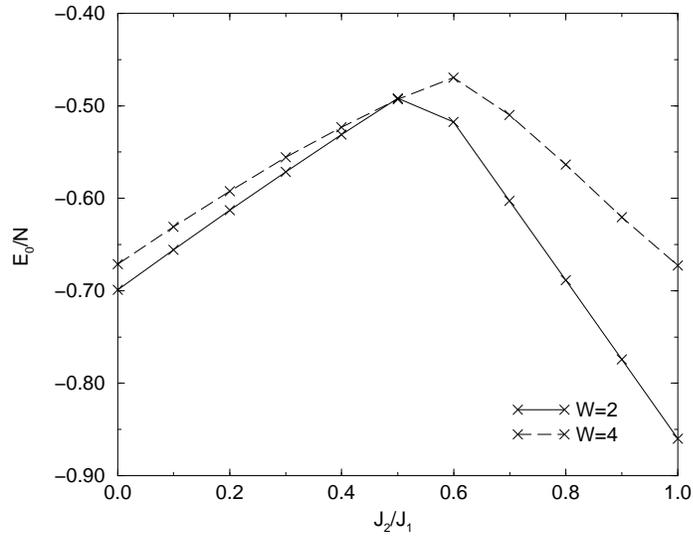}
\caption{\label{fig:E0.W=Xsqrt} The extrapolated energy $E_0/N$
  for widths $W=2,4$ on the tilted square lattice.}
\end{figure}

\begin{figure}
  \centering \epsfxsize=10cm \epsffile{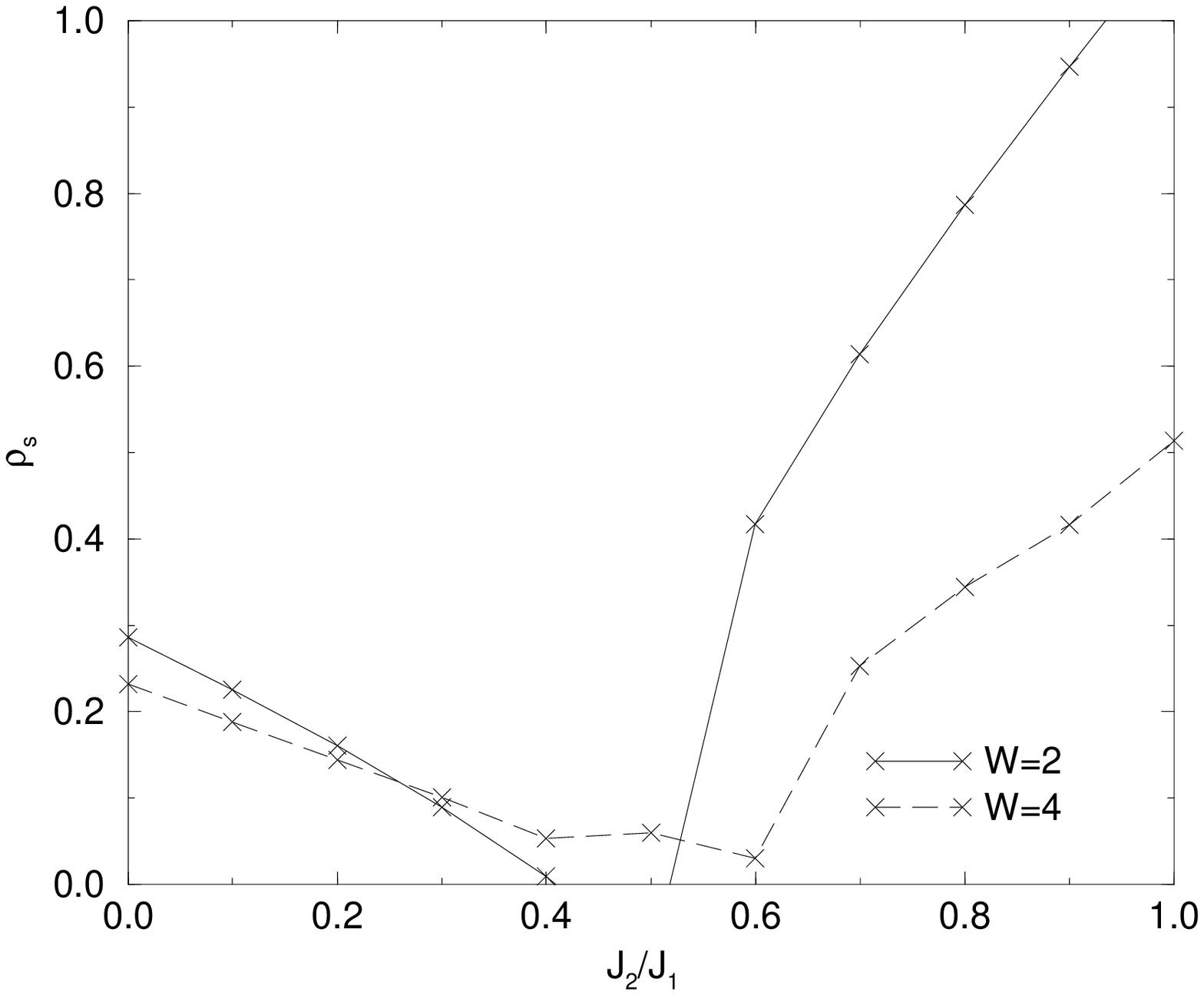}
\caption{\label{fig:rhos.W=Xsqrt} The extrapolated spin stiffness
  $\rho_s$ for widths $W=2,4$ on the tilted square lattice.}
\end{figure}

The width $W$ of the system is fixed and for various lengths $L$ the
properties $E_0/N$, $T$, $J$ and $\rho_s$ are calculated. Usually we
set $L=2W,3W,4W,5W$. In figures \ref{fig:E0.W=4}, \ref{fig:T.W=4}
and \ref{fig:J.W=4} we have depicted this for square lattices with
$W=4$ and two values for $J_2$. Many more lengths $L$ are considered
here as it is computationally fairly easy to achieve enough accuracy
for system sizes up to $L=160$. The scaling behaviour of ${\cal
  O}(1/L)$ is clearly confirmed by the graphs. In the figures is the
extrapolated values for $1/L=0$ are also depicted. The resulting
energy density $E_0/N$ and stiffness $\rho_s$ of the infinitely long
system are collected for various $J_2$ in figures \ref{fig:E0.W=X}
and \ref{fig:rhos.W=X}.  Figures \ref{fig:E0.W=Xsqrt} and
\ref{fig:rhos.W=Xsqrt} contain the equivalent results for the tilted
lattice.

\subsection{Scaling to $W=\infty$}

\begin{figure}
  \centering \epsfxsize=10cm \epsffile{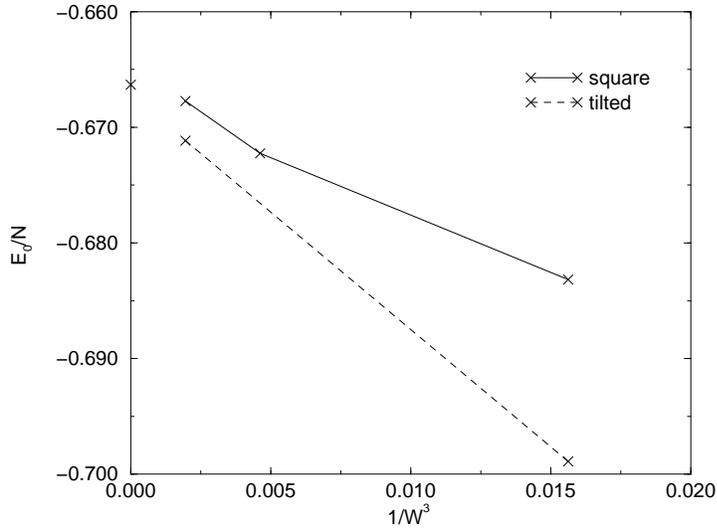}
\caption{\label{fig:E0.W=0} The extrapolated energy density
  $E_0/N$ for square lattices $W=4,6,8$ and tilted lattices $W=2,4$.
  $J_2=0$. The widths of the tilted lattices is multiplied by a factor
  of 2 to get both lines for the square and tilted lattices within the
  same range. The cross $\times$ on the axis denotes the extrapolated
  value.}
\end{figure}

\begin{figure}
  \centering \epsfxsize=10cm \epsffile{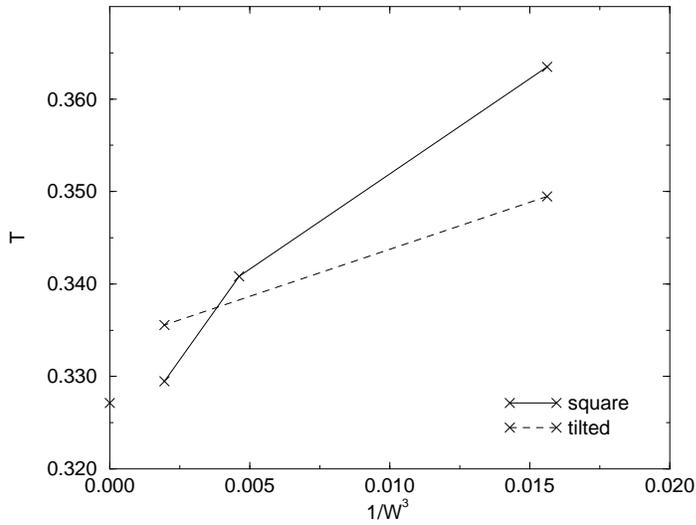}
\caption{\label{fig:T.W=0} The kinetic term $T$ extrapolated. $J_2=0$. The cross $\times$ on the axis denotes the extrapolated
  value.}
\end{figure}

\begin{figure}
  \centering \epsfxsize=10cm \epsffile{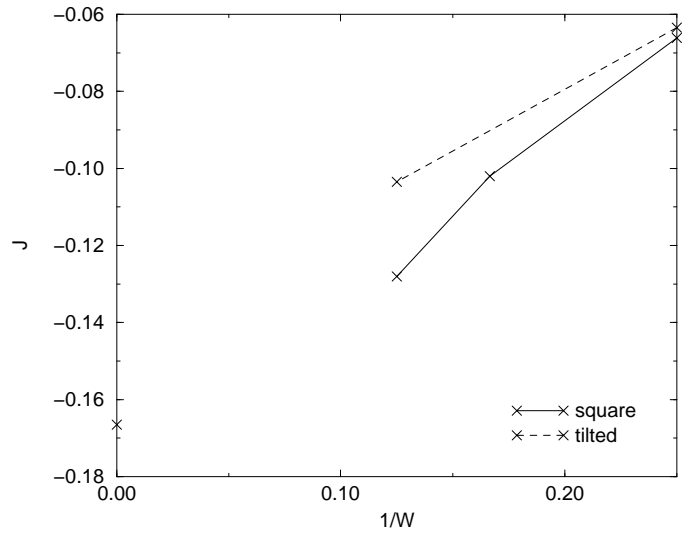}
\caption{\label{fig:J.W=0} The current-current correlation $J$
  extrapolated. $J_2=0$.The cross $\times$ on the axis denotes the extrapolated
  value.}
\end{figure}

In the final, two-dimensional system, the orientation of the lattice,
square or tilted, does no longer matter. The values of all quantities
are equal for both orientations at system size $L \times W = \infty
\times \infty$. However, the prefactors for the finite-size
corrections do not have to be the same. We fit the scaling for both
orientations with the same offset, but with different gradients. In
figures \ref{fig:E0.W=0}, \ref{fig:T.W=0} and \ref{fig:J.W=0} this is
done for $J_2=0$. The resulting extrapolations are also plotted in the
figures. In all these figure we have multiplied the width $W$ of the
tilted lattices by a factor of $2$ so that both the tilted and the
square lattices fall within the same range of $1/W^3$ and $1/W$. All
data can then easily be plotted in the same graph.

\begin{figure}
  \centering \epsfxsize=10cm \epsffile{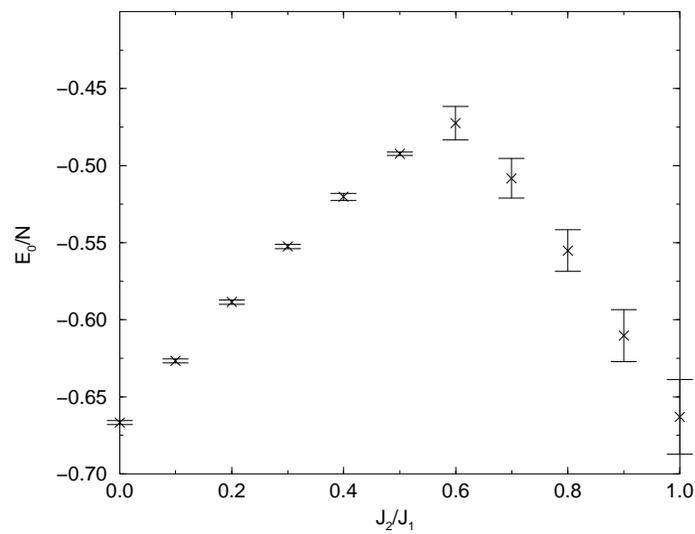}
\caption{\label{fig:E0} The energy density $E_0/N$ for
  a two-dimensional system. This curve is the collection of
  extrapolations done as in \ref{fig:E0.W=0}. The error bars are based
  on fitting the data to the scaling relations.}
\end{figure}

\begin{figure}
  \centering \epsfxsize=10cm \epsffile{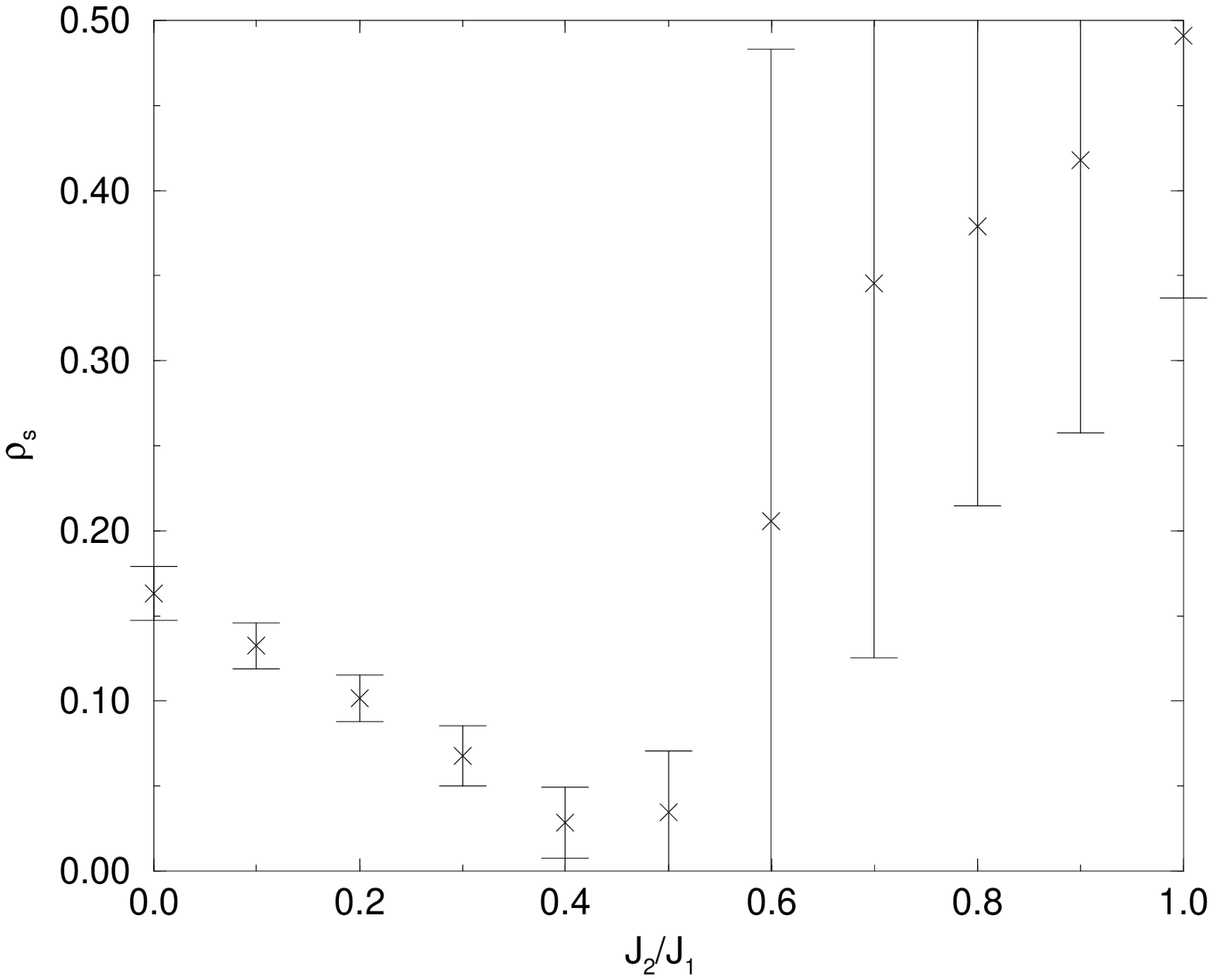}
\caption{\label{fig:rhos} The spin stiffness $\rho_s$ of
  a two-dimensional system. Every point is the sum of the kinetic term
  $T$ and the current-current correlation $J$ obtained by
  extrapolation as in \ref{fig:T.W=0} and \ref{fig:J.W=0}.}
\end{figure}

In figures \ref{fig:E0} and \ref{fig:rhos} we plot the energy density
$E_0/N$ and spin stiffness $\rho_s$ for a two-dimensional system. The
error bars in these graphs are based on fitting the data-points for
$(\infty,W)$ to the assumed scaling relations; the errors in the data
after the first scaling, $L \rightarrow \infty$ are neglected.

For $J_2=0$ there is no sign problem and the literature contains
excellent results with which we compare in table \ref{tab:sandvik}.
The known values for $E_0/N,\rho_s,T$ and $J$ do not contradict with
our estimates, although the differences are up to $6\%$.

\begin{table}[h]
  \begin{center}
\begin{tabular}{|c|rl|rl|}
 \hline
 & & Sandvik \cite{sandvik98} & &This work\\ \hline
$E_0/N$ & -&0.669437(5) & -&0.666(1) \\
$\rho_s$ & & 0.175(2) & & 0.165(10) \\
$T$ & &0.3347185(3) & &0.330(2) \\
$J$ & -&0.160(2) & -&0.165(6) \\
\hline
\end{tabular}
\end{center}
\caption{\label{tab:sandvik} The comparison between our results for $J_2/J_1=0$
  and those by Sandvik \cite{sandvik98}.}
\end{table}

Figure \ref{fig:E0} suggest a first order phase transition as the
gradient of energy curve seems to change drastically around $J_2/J_1
\approx 0.6$. This is the same behaviour as observed in the SBMF
approximation, figure \ref{fig:rho_s}.

The error bars of the stiffness $\rho_s$, figure \ref{fig:rhos},
increase dramatically while sweeping past $J_2/J_1=\frac{1}{2}$. The
reason for this is that the kinetic term $T$ and the current-current
correlations $J$ for the tilted lattices and the square lattices no
longer seem to have the same limit in the two-dimensional case. We
still enforce this and as a consequence the error bars increase
dramatically. Einarsson and Schulz \cite{einarsson} suggest a region
$0.4 \lesssim J_2/J_1 \lesssim 0.6$ where the spin stiffness vanishes,
$\rho_s=0$. This is not in contradiction with our results although we
also can not confirm it.

\section{Other Indicators}

\begin{figure}
  \centering \epsfxsize=10cm \epsffile{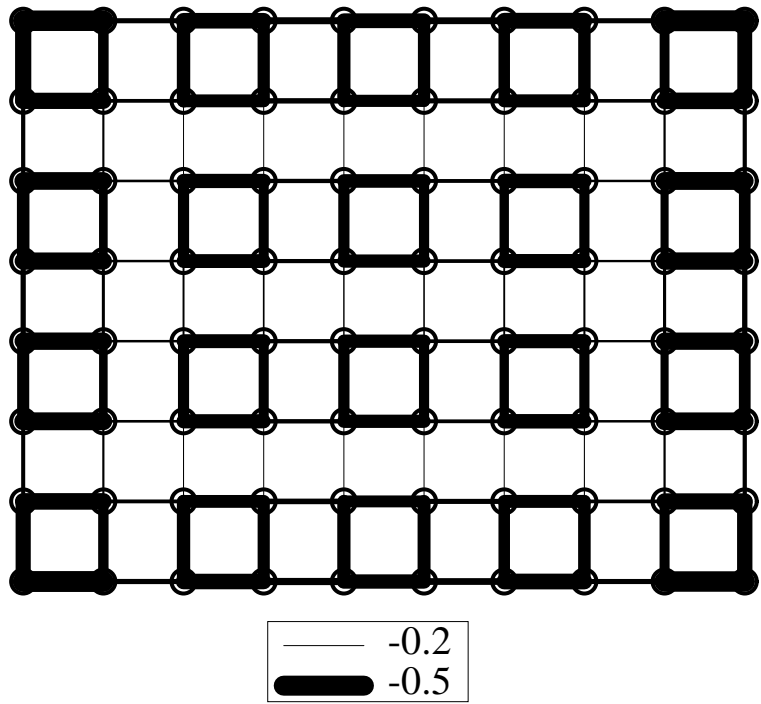}
\caption{\label{fig:scdots} For 10x8 system that has open boundary conditions in both directions, the correlations
  $\langle \vec {\cal S}_i \cdot \vec {\cal S}_j \rangle$ where $i$
  and $j$ are nearest neighbours, are depicted. This is for
  $J_2/J_1=0.5$. A perturbation (\ref{eq:dh_dimer}) is included to
  distinguish between plaquette and dimer order. }
\end{figure}

In the process of calculating the spin stiffness $\rho_s$, it is easy
to generate the spin-spin correlations $\langle \vec {\cal S}_i \cdot
\vec {\cal S}_j \rangle$. Although no finite-size scaling was
performed, the correlations already given a clear hint what to expect
in the intermediate range of frustration. For these correlation
functions there are no restrictions on the boundary conditions as was
the case for the spin stiffness $\rho_s$. To achieve highest possible
accuracy we set the boundary conditions in both directions to be open.
In figure \ref{fig:scdots} the correlations are depicted for
$J_2/J_1=0.5$. To be honest, the Hamiltonian has been modified to be
more conclusive. Let us explain this.

In the literature \cite{zhitomirsky,sachdev} there are many
suggestions for the intermediate phase. Two of these would yields such
a correlation picture, namely dimer and plaquette phases. They distinguish
themselves in a very quantum mechanical manner: the dimer phase
consists of nearest neighbour singlets nicely stacked next to each
other on the lattice and all aligned in the same direction.The
ground state in the plaquette phase basically is a direct product of
two vertical singlets plus two horizontal singlets on {\it each }
plaquette.

Figure \ref{fig:scdots} could correspond to a superposition of two
discrete orientations of the ground state; one oriented in the vertical
direction and one in the horizontal direction.  Inserting a small
perturbation in the Hamiltonian,
\begin{equation}
\delta {\cal H} = \frac{J_1}{10} \sum_i \vec{\cal S}_i \cdot \vec{\cal S}_{i+\hat{x}},
\label{eq:dh_dimer}
\end{equation}
will lift the degeneracy of the dimer orderings, after which only
dimers in the length direction will remain visible, whereas the
plaquette phase would not suffer severely from it. The system depicted
in \ref{fig:scdots} has this perturbation (\ref{eq:dh_dimer})
included. It provides clear evidence in favour of a plaquette phase.

As a further indicator, the spin gap $\Delta$ was also briefly
studied. This is the energy difference between the singlet ground state
and the first excitation in the ${\cal S}^z=1$ space. This excitation
is a member of the triplet of lowest excitations. Both for the N\'eel
and the collinear ordering the gap $\Delta$ should disappear as there
exist spin waves. If there is a phase between these two, the gap
$\Delta$ might open up. The phases suggested in the literature
actually all imply a gap $\Delta$. The advantage of taking a member
outside the ${\cal S}^z=0$ space is that it can be considered the
ground state in its own space and calculated in exactly the same
fashion as the ground state itself with the quantum number ${\cal
  S}^z=1$. Unfortunately the energies of both the ground state and the
excitation grow as $N$ whereas the gap remains of order $J_1$; $\Delta
\sim J_1$. Only for $W=4,6$ we could obtain enough accuracy to scale
away the length (an ${\cal O}(1/L)$ correction). With only these two
points a scaling analysis in the width direction is impossible for the
simple reason that no optimal fitting can be done (which needs at least
three points).

\section{Discussion}

The scaling analysis of the spin stiffness $\rho_s$ we performed, does
not give accurate results. It neither supports nor contradicts the
existence of an intermediate phase. The two successive steps of the
scaling analysis have very different degree of success and we will
discuss them separately;

The first step, scaling $L \rightarrow \infty$, gives reliable values
for the properties of an infinitely long cylinder. There are two
sources of errors in the properties of this cylinder: corrections to
scaling and systematic error in the DMRG due to a insufficient number
of states kept. For the square lattices of width $W=4,6$ and the
tilted lattice of width $W=2$, the scaling corrections are the
dominant source of errors. The least square fits estimate the relative
error to be of the order $10^{-3}$. For the square lattice of width
$W=8$ and the tilted lattice for width $W=4$, the systematic errors of
the DMRG should also contribute. The peculiar behaviour of the
stiffness in the region round $J_2/J_1=0.5$ for the tilted lattice,
figure \ref{fig:rhos.W=Xsqrt}, indicates that this source is there
even determinant for the overall accuracy.

The second part, scaling $W \rightarrow \infty$, is by no means as
successful as the first. Figures \ref{fig:E0.W=0}, \ref{fig:T.W=0} and
\ref{fig:J.W=0} do not justify the scaling behaviour we assume.  Most
likely the scaling behaviour we derived in the last chapter is not
valid for these small system widths $W$. There we also found that for
Schwinger bosons this scaling behaviour set in at the linear dimension
$L=10$.

The literature provides accurate results for the unfrustrated
Heisenberg model, $J_2=0$. Sandvik \cite{sandvik98} is the latest in a
whole line of authors who have performed finite-size scaling on
square, periodic systems. Their results fit the scaling behaviour
nicely and the results are of high quality. While the results are not
contradicting ours, our inaccurate fit is in sharp contrast with
theirs. Naturally we consider an unusual geometry (infinitely long
cylinders), but the origin of this discrepancy is essentially not
understood.

Einarsson, Schulz et al. \cite{einarsson, schulz} performed a similar
analysis to ours. Instead of scaling the length $L \rightarrow \infty$
first, they considered square lattices with periodicity in both
directions. From sizes $4 \times 4, 2\sqrt{5} \times 2\sqrt{5},
4\sqrt{2} \times 4 \sqrt{2}$ and $ 6 \times 6$ they inferred the
properties of the 2D case. They observe reasonable scaling behaviour
in line with the unfrustrated case. That their systems are much
smaller than ours, makes the contrast with our findings even more
striking.

The stiffness thus does not give a definite answer and we switch our
attention to the correlation functions. If there exists an
intermediate phase, the correlations in the ground state clearly hint
at a plaquette phase.  Zhitomirsky and Ueda \cite{zhitomirsky}
suggested before that indeed the plaquette phase is favourable to a
dimer one. Still, careful study of the dependence of the correlation
functions give rise to a few other suspicions: the plaquette
correlations arise far sooner than the stiffness becomes negligible.
Perhaps a super solid phase exists? Moreover similar behaviour is
observed coming from the collinear order, although there dimer (and
not plaquette) correlations are appearing. This even makes rooms for
two intermediate phases. The abrupt change of the energy in figure
\ref{fig:E0} suggested that between two of these phases a first
order phase transition exists. If that is the case, it is most likely
that it will occur between a plaquette phase and a dimer phase
(possibly both with long range magnetic order).

In the next chapter we will study the correlations further to get
insight in possible intermediate phases.

%% file: FNMC.tex
\chapter{Combination of DMRG and Fixed-Node Monte Carlo \label{chap:FNMC}}

\section{Introduction}

In chapter \ref{chap:ITFDMRG} and \ref{chap:J1J2DMRG} it became clear
that the DMRG can achieve phenomenal accuracy for relatively narrow
strips, but the accuracy deteriorates once the strips get wider.
Section \ref{sec:generallim} provided a connection between this
behaviour and perturbation theory. Although this relation only yields
an upper bound on the number of states needed for a certain accuracy,
it is clear that the number of off-diagonal matrix elements in the
Hamiltonian together with their sizes determine the accuracy. If we
still want to extract ground state information for wider systems,
changes or amendments in the method have to be made.

In this chapter we combine it with the Green Function Monte Carlo
(GFMC). GFMC is based on the notion of projecting out the ground
state. Define the operator ${\cal G}=1-\varepsilon {\cal H}$ with
$\varepsilon \ll 1$ \cite{ceperley80,hetherington84}. The ground state
$|\psi_0 \rangle$ can be found starting with a state $|\phi \rangle$
and letting it relax,
\begin{equation}
|\psi_0 \rangle \sim {\cal G}^n | \phi \rangle~~,~~ n
\varepsilon \gg 1 ~,~ \varepsilon \ll 1. \label{eq:greenproject}
\end{equation}
When $\varepsilon$ is small enough, not the largest but the smallest
eigenvector will be projected out by this projection.  The Hilbert
space is very large ${\cal O}(2^N)$, making it impossible to apply
these matrix operations in detail. We want to perform a representative
sampling of these matrix multiplications. With configurations $R$,
built from individual up and down spins $\sigma=\uparrow,\downarrow$ ,
$| R \rangle=|\sigma_1 \dots \sigma_N \rangle$, we can construct paths
$\vec{R}=R_0,\dots,R_n$. Insertion of complete basis sets $\sum_R |R
\rangle \langle R|$ between the individual projectors ${\cal G}$ in
the previous equation, can be read as a summation over paths;
\begin{equation}
|\psi_0 \rangle \sim \sum_{\vec{R}} |R_n \rangle \left [ \prod_{i=1}^n
  \langle R_i| {\cal G}| R_{i-1} \rangle \right ] \langle R_0|\phi
\rangle. \label{eq:paths}
\end{equation}
A path $\vec{R}$ derives its name from the fact that it only
contributes to the ground state $|\psi_0 \rangle$ if $\langle R_i |
{\cal G} | R_{i-1}\rangle \neq 0$ for all successive configurations;
only specific paths through phase space can be followed. The general
assumption of a Monte Carlo simulation and the GFMC in particular is
that accurate properties can still be obtained when only a few of
these paths $\vec{R}$ are semi-randomly selected to represent equation
(\ref{eq:paths}). As the Hilbert space is very large, ${\cal O}(2^N)$,
even the most extensive GFMC simulations can be considered to contain
only relative few paths. In practice we always generate 6000 paths.

We will apply the GFMC to the frustrated Heisenberg Hamiltonian.
Successive configurations $R_{i-1}$ and $R_i$ can differ at most in
the orientation of two, nearby spins as the Hamiltonian only contains
local spin-pair interactions. The transition strength $\langle R_i |
{\cal G}| R_{i-1} \rangle$ is set by the Hamiltonian so that locally a
good equilibrium is reached. As a consequence the local correlation
functions are of good quality.  The GFMC uses the off-diagonal matrix
elements of the Hamiltonian in an essential way to systematically
probe the Hilbert space.  Both these aspect, high quality of local
correlation functions and the intrinsic use of off-diagonal matrix
elements, touch on weaknesses of the DMRG; first, as mentioned before
the off-diagonal matrix elements severely limit the accuracy of the
DMRG. Second, the correlation functions are biased through the
sequence in which the sites of the lattice are incorporated in the
basis. When the sites of a column are added successively to the basis,
the correlations between the columns are underestimated.

The GFMC allows for a systematic bias towards specific paths $\vec{R}$
without influencing the expectation values. This is done by means of a
guiding wave function $\langle R |\phi_G\rangle$. It becomes a measure
of importance for a configuration $R $ and thereby of a path.  This
guiding state $|\phi_G \rangle$ embodies both the greatest strength of
the method and its main weakness: if a large amount of information on
the ground state $|\psi_0 \rangle$ is incorporated in the guiding
state $|\phi_G \rangle$, the results will improve drastically. On the
other hand, without a proper guiding wave function no reasonable
results can be obtained.

Even more emphasis is put on a good guiding wave function when
handling models with frustration or fermions.  These are typical cases
exhibiting the 'sign-problem'. A good guiding state $|\phi_G \rangle$
can maybe not solve this sign-problem, but is can suppress it to such
an extent that it does not influence the extracted ground state
properties.

With the strengths and limitations of both methods in mind, it seems a
logical solution to combine them; DMRG can make an initial guess
$|\phi_0\rangle$ to the ground state $|\psi_0 \rangle$. Although this
is a systematic approximation, the local correlation functions bear a
clear signature of the method. They depend on the mapping from the
two-dimensional system to a one-dimensional chain that is necessary to
apply the DMRG (site version). The guess $|\phi_0\rangle$ can improve
the GFMC in two ways. Most importantly, it can serve as a guiding
state, $|\phi_G \rangle = |\phi_0 \rangle$, to reduce the variance and
suppress the sign-problem. This guiding state is also used to
calculate so-called mixed estimators for observables. These mixed
estimators also strongly improve with a better guiding state.  The
other aspect, where the DMRG state can help, is in the initial state,
$|\phi \rangle = |\phi_0 \rangle$, but no matter what the quality of
this starting point is, eventually the ground state will be reached.
Especially the quality of the local correlations will increase by this
stochastic process. Without a DMRG state $|\phi_0 \rangle$, the GFMC
would require another guiding state. In practice these are relatively
simple and consequently poor approximations to the ground state, that
are involved and complex to construct.

In this chapter we make the connection between DMRG and GFMC by using
the DMRG ground state as a guiding state, $|\phi_G \rangle = | \phi_0
\rangle$. First we explain the principles of GFMC.  Afterwards the
sign-problem is discussed and a possible cure is described: Fixed-Node
Monte Carlo (FNMC) and the extension to stochastic reconfiguration.
With all that in place we make the connection. In fact the only thing
we need to extract from the DMRG state $|\phi_0\rangle$ is its value
for specific configurations $R$, $\langle R| \phi_0 \rangle$. An
algorithm will be introduced to obtain this value for an arbitrary
configuration.  Naturally no table with an entry for each possible
configuration can be built as it would have a size of ${\cal O}(2^N)$
just like the number of configurations. An extra section is spent on
curing a common problem of the GFMC by switching from discrete
imaginary time intervals, $1-\varepsilon {\cal H}$, to a continuum,
$\exp(-\tau {\cal H})$. This makes the method also more elegant.
Finally, after all these explanatory and introductory sections, the
computations are presented and the results are discussed.

\section{Green Function Monte Carlo}

GFMC has been widely used for at least two decades now
\cite{hetherington84,ceperley80,trivedi90,runge92}. In mathematics it
finds an equivalent in the Markov chain \cite{hetherington84} and in a
broader physical perspective it strongly reminds of diffusion.

The method will be explained along the lines of the frustrated
Heisenberg model, where for the moment we simply ignore the
sign-problem. Following sections will be dedicated to resolving that
complication. The frustrated Heisenberg Hamiltonian is a collection of
spin-pair interactions
\[
\vec{\cal S}_i \cdot \vec{\cal S}_j= \frac{1}{2} ( {\cal S}_i^+ {\cal
  S}_j^- + {\cal S}_i^- {\cal S}_j^+) + {\cal S}_i^z {\cal S}_j^z.
\]
The last term will not alter any of the spins
$\sigma_1,\dots,\sigma_N$ in a state $| R \rangle$ when applied to it,
the first two terms will allow the exchange of an up- and a down-spin.
This limits the number of states $|R' \rangle$ that are connected to
$|R \rangle$ strongly; either they are identical or in the case that
$\sigma_i \neq \sigma_j$, they have spins $i$ and $j$ exchanged,
$\sigma_i'=\sigma_j$ and $\sigma_j'=\sigma_i$. Applying the
Hamiltonian to a configuration reminds of diffusion as it allows the
up-spins to hop from one site to the other.

As mentioned in the introduction we want to project out the ground
state $|\psi_0\rangle$ starting from a ---not yet identified--- $|\phi
\rangle$ by successive applications of ${\cal G} = 1-\varepsilon {\cal
  H}$, equation (\ref{eq:greenproject}). The wave function $\langle R|
\psi_0 \rangle$ cannot be obtained completely because of the size of
the Hilbert space. For most physical systems it is even arguable
whether that is desirable. The physical properties are most
important and GFMC focuses on the determination of these.

There are two categories of observables ${\cal X}$, conserved ones
($[{\cal X},{\cal H}]=0$) and non-conserved ones ($[{\cal X},{\cal H}]
\neq 0$). The conserved observables ${\cal X}$ including the
Hamiltonian itself, can be measured in a fairly simple manner. The
guiding state $|\phi_G \rangle$ will be used to construct a mixed
estimate with exactly the same expectation value as the required
measurement.
\begin{equation}
\langle {\cal X} \rangle_{\rm mixed} \equiv \frac{\langle \phi_G| {\cal X} {\cal G}^{n} | \phi \rangle}{\langle \phi_G| {\cal G}^{n}| \phi \rangle} =
 \frac{\langle \phi_G| {\cal G}^{n/2} {\cal X} {\cal G}^{n/2} | \phi \rangle}{\langle \phi_G| {\cal G}^{n}| \phi \rangle} = \langle \psi_0 | {\cal X} | \psi_0 \rangle. \label{eq:mixedestimator}
\end{equation}
It is essential that the observable ${\cal X}$ commutes with the
Hamiltonian ${\cal H}$ as after the commutations we use $|\psi_0
\rangle \sim {\cal G}^{n/2} | \phi_G \rangle \sim {\cal G}^{n/2} | \phi \rangle$ for $N \gg 1$.

The previous relation, (\ref{eq:mixedestimator}), does not hold if the
observable ${\cal X}$ is not conserved, $[{\cal X}, {\cal H}] \neq 0$.
We will show how the mixed estimate differs from the required one, and
by a simple extension reduce this difference. The guiding state can
always be considered to contain a component along the ground state and
a component orthogonal to it,
\[
|\phi_G \rangle = |\psi_0 \rangle + \delta |\psi_1 \rangle ~~,~~
\langle \psi_0 | \psi_0 \rangle = \langle \psi_1 | \psi_1 \rangle = 1
~~,~~ \langle \psi_1 | \psi_0 \rangle = 0.
\]
A good guiding state should have a small perpendicular component,
$\delta \ll 1$. When the commutations in (\ref{eq:mixedestimator}) are
not allowed, the mixed estimate reads
\[
\langle {\cal X} \rangle_{\rm mixed} = \langle \psi_0 | {\cal X} | \psi_0
\rangle + \delta \langle \psi_1 |{\cal X}| \psi_0 \rangle.
\]
Relating this to the expectation value of the guiding state, $\langle
\phi_G | {\cal X}|\phi_G \rangle$, can reduce the corrections to order
${\cal O}(\delta^2)$;
\begin{eqnarray}
\langle {\cal X} \rangle_{\rm improved} &\equiv&2\frac{\langle \phi_G| {\cal X} {\cal G}^{n} | \phi \rangle}{\langle
  \phi_G| {\cal G}^{n}| \phi \rangle}- \frac{\langle \phi_G | {\cal X}
  | \phi_G \rangle}{\langle \phi_G| \phi_G \rangle} \nonumber \\
& =& \langle \psi_0 |
{\cal X} | \psi_0 \rangle + \delta^2 \left ( \vphantom{\prod} \langle \psi_0 | {\cal X}
  | \psi_0 \rangle - \langle \psi_1 | {\cal X} | \psi_1 \rangle \right
) + {\cal O}(\delta^3) \nonumber \\
& =& \langle \psi_0 | {\cal X} | \psi_0 \rangle +
{\cal O} (\delta^2). \label{eq:improved}
\end{eqnarray}
To remove this ${\cal O}(\delta^2)$ term completely forward walking
schemes \cite{calandra98} are necessary, but it is at present unclear
whether this can be combined with stochastic reconfiguration as there
the weights are frequently changed.  For our purposes only mixed and
improved mixed estimates, $\langle {\cal X} \rangle_{\rm mixed}$ and
$\langle {\cal X} \rangle_{\rm improved}$, are required.

Next is the description of the stochastic nature of the method. In the
introduction, paths $\vec{R}$ through phase space were defined,
equation (\ref{eq:paths}). A selection from all possible paths
$\vec{R}$ has to be made stochastically.  If the path $\vec{R}$ is
selected with a probability $P(\vec{R})$ and assigned a weight
$M(\vec{R})$, the following expectation value has to hold:
\begin{equation}
\langle \phi_G | {\cal X} {\cal G}^n | \phi \rangle =\left \langle
  X(R_n) M(\vec{R}) \right \rangle = \sum_{\vec{R}} X( R_n )
M(\vec{R}) P(\vec{R}), \label{eq:zuivereschatter}
\end{equation}
for all ${\cal X}$ including ${\cal X}\equiv 1$. We use here the local
expectation value
\[
X(R)= \frac{\langle \phi_G | {\cal X} |R \rangle}{\langle \phi_G | R
  \rangle}.
\]
The mixed estimate can then be obtained by choosing a large number of
paths $\{ \vec{R}^\alpha \}$ and calculating
\[
\langle {\cal X} \rangle_{\rm mixed} = \frac{\sum_{\alpha} X(R_n^\alpha )
  M(\vec{R}^\alpha)}{\sum_{\alpha} M(\vec{R}^\alpha)}.
\]
In practice the configurations $R_i$ in a path $\vec{R}$ are selected
successively. The most important advantage of this is that a specific
configuration $R_{i-1}$ connects only to relatively few configurations
$R_i$. Above it is explained that $R_i$ and $R_{i-1}$ can differ at
most in the orientation of two spins for $\langle R_i | {\cal G} |
R_{i-1} \rangle \neq 0$.

The starting configuration, $R_0$, is chosen according to probability
distribution $P_0(R_0)$. Each configuration $R_i$ afterwards is chosen
with probability $P(R_i \leftarrow R_{i-1})$ giving an overall
probability of
\[
P(\vec{R})= \prod_{i=1}^n P(R_i \leftarrow R_{i-1}) P_0(R_0),
\]
directly in line with the theory of Markov chains. The probabilities
$P_0(R)$ and $P(R \leftarrow R')$ have to be normalised
without any negative elements;
\[
\sum_R P_0(R) = \sum_R P(R \leftarrow R')=1 ~~~~,~~~~ P_0(R), P(R
\leftarrow R') \ge 0.
\]
In a similar fashion the weight is also successively constructed from
a starting weight $m_0(R_0)$ and following weight factors $m(R_i)$
combined with 'signs' $s(R_i, R_{i-1})$ \cite{sorella98} and the weight of the path
is  finally rescaled with a factor $m_{\rm fin}(R_n)$;
\[
M(\vec{R})= m_{\rm fin}(R_n) \left [\prod_{i=1}^n s(R_i,R_{i-1}) m(R_i)
\right ] m_0(R_0).
\]
A first approach would be to let the Green function ${\cal G}$ decide;
choosing starting positions according to their quantum mechanical
probability, $ |\langle \phi | R \rangle |^2$ ($\langle \phi| \phi
\rangle =1$), and expressing no favour for any specific path
afterwards:
\begin{equation}
\begin{array}{r@{}c@{}lr@{}c@{}l}
P_0(R_0) &=& |\langle \phi | R_0 \rangle|^2, & m_0(R_0)&=&\displaystyle{\frac{1}{\langle \phi | R_0 \rangle}}, \\
P(R_i \leftarrow R_{i-1}) &=& \displaystyle{\frac{| \langle R_i | {\cal G} | R_{i-1} \rangle |}{\sum_R | \langle R| {\cal G} | R_{i-1} \rangle |}}, & m(R_{i-1}) &=& {\sum_R | \langle R| {\cal G} | R_{i-1} \rangle |}, \\
& & & s(R_i,R_{i-1}) &=&\displaystyle{\frac{ \langle R_i | {\cal G} | R_{i-1} \rangle}{| \langle R_i | {\cal G} | R_{i-1} \rangle |}}, \\
&&&m_{\rm fin}(R_n) &=& \langle \phi_G | R_n \rangle. \label{eq:chancesandweights}
\end{array}
\end{equation}
The equations above also explain why $s(R_i,R_{i-1})$ can be named a
sign. These combinations satisfy the condition
(\ref{eq:zuivereschatter}) as can easily be verified using
\[
s(R_i,R_{i-1}) m(R_{i-1}) P(R_i \leftarrow R_{i-1}) = \langle R_i |
{\cal G} |R_{i-1} \rangle.
\]

In the implementation it is only necessary to store the latest
configuration $R_i$ and the weight up to that moment. Given the form
of the Hamiltonian where up-spins make a random walk through the
system, the name walker become suitable for this latest configuration.  The
walkers are thus combined with the weights to yield the expectation
value (\ref{eq:zuivereschatter}).

In practice too many irrelevant paths are selected with these unbiased
settings. Far better statistics can be achieved using a guiding wave
function $\langle R |\phi_G \rangle$. Indeed this is the same wave
function as was used to complete the mixed estimates
(\ref{eq:mixedestimator}). This wave function helps us to distinguish
important configurations from less important ones and thereby guides
the walkers into the relevant parts of the Hilbert space. The easiest
way to incorporate the guiding state $|\phi_G \rangle$ in our
calculation is by defining an operator $\bar{\cal X}$ associated to
${\cal X}$ by \cite{sorella98}
\[
\bar{\cal X} \equiv \sum_{R,R'} |R \rangle \frac{\langle \phi_G | R
  \rangle}{1}\langle R | {\cal X} | R' \rangle \frac{1}{ \langle
  \phi_G | R' \rangle}\langle R'|.
\]
As this is a similarity transformation, the projector $\bar{\cal G}$
basically remains the same as ${\cal G}$.  The set of equations
(\ref{eq:chancesandweights}) can still be used for the stochastic
process replacing ${\cal G}$ by $\bar{\cal G}$. Only the final and
initial weights have to be altered,
\[
m_{\rm fin}(R)=1 ~~\hbox{and}~~ m_0(R)= \frac{\langle \phi_G| R \rangle}{
  \langle \phi| R \rangle}.
\]
If we choose the guiding wave function as starting position, $|\phi
\rangle = | \phi_G \rangle$, even the initial weight can be dropped,
$m_0(R)=1$.  The transition probabilities $P(R_i \leftarrow R_{i-1})$
are now biased towards the most relevant configurations. This only
reduces the variance, the expectation values are unaltered.

The algorithm thus far prescribes that after $n$ projections a ---mixed---
measurement is made, new walkers are created and the projections
restarts. In practice it is far more efficient to continue using the
same walkers; the existing set is distributed according to $\langle
\phi_G | R \rangle \langle R | \psi_0 \rangle$ while the initial set
is distributed according to $\langle R | \phi_G \rangle$. Relatively
few projections have to be performed to do further measurements that
are both independent of the last ones and representative for the
ground state.

In this process of successive projections the relative weights of the
walkers will spread exponentially. It becomes unwise to continue the
path of certain walkers with negligible weight whereas walkers with
large weight deserve extra attention. Still one does not want to
influence the expectation values.

The technique to perform this task is called branching. Just like the
stochastic process that replaced the projecting, the two requirements
here are that the expectation values are to remain unaltered and the
variance is minimised.

The easiest approach to choose $N$ new walkers out of a set of $N$ old
ones is to draw them from a probability distribution
\[
P_\alpha = \frac{|M_\alpha|}{\sum_{\alpha'} |M_{\alpha'}|}.
\]
In case the walker $\alpha$ is selected, the weight of the new walker
is set to
\begin{equation}
\frac{M_\alpha}{|M_\alpha|} \frac{\sum_{\alpha'} | M_{\alpha'}|}{N}. \label{eq:newweights}
\end{equation}
If one does not want to use correction factors \cite{hetherington84},
the weights can even be set to unity ($M_\alpha \rightarrow 1$).
Despite its elegance the variance of this method is relatively large;
it can happen that $N$ times the same walker is selected.

It is possible to reduce the variance of this branching process
substantially without changing the expectation value. Here we introduce
a small extension of the method introduced by Calandra and Sorella
\cite{calandra98} to reduce the variance of the branching process.
The essential difference with straight selection of $N$ walkers, is
that the scope of the stochastic process is limited as much as
possible. It contains two distinct steps.

The first step is not stochastic in nature. Rescale the weights
\[
\tilde{M}_\alpha = N \frac{M_\alpha}{\sum_{\alpha'} |M_{\alpha'}|}.
\]
This weight $\tilde{M}_{\alpha}$ is truncated to an integer,
$\hbox{int} (\tilde{M}_{\alpha})$. For every $\alpha$ there are
$|\hbox{int} (\tilde{M}_{\alpha})|$ new walkers created, each with the
configuration of walker $\alpha$ and with the weight defined in
(\ref{eq:newweights}). Once this is done for all old walkers, a set of
$N_0$ new walkers is formed. The integer number $N_0$ is always
smaller than $N$. On simple grounds one expects $N_0 \approx
\frac{1}{2} N$.

For the remainder of the weight, $\hat{M}_\alpha = \tilde{M}_\alpha -
\hbox{int} (\tilde{M}_{\alpha})$, the second, stochastic step is performed. We
assign a probability $P_\alpha$ to each old walker,
\[
P_\alpha = \frac{|\hat{M}_\alpha|}{\sum_{\alpha'}
  |\hat{M}_{\alpha'}|}.
\]
The probabilities are put next to each other on the interval $[0,1]$.
In each consecutive interval of length $1/(N-N_0)$ ($[0,1/(N-N_0)]$,
$[1/(N-N_0),2/(N-N_0)]$, $\dots$) one walker is selected by choosing a
random number $\xi$ in that interval and establishing to which
probability interval $P_\alpha$ this number $\xi$ belongs. In this
fashion the remaining $N-N_0$ walkers are selected giving a total of
$N$ new walkers.

This stochastic part is similar to the method of selecting $N$ times
one walker out of a set of $N$ old walkers which we described before.
The essential difference lies in the fact that here one walker is
selected {\it per interval}.  The latter method may be less elegant
than the original proposal, but it reduces the variance of the
branching drastically.

\section{Fixed-Node Monte Carlo}

Unfortunately, the last section does not tell the entire story. The
GFMC is severely ham\-pered by the so-called 'sign-problem' in models
that contain frustration or fermions. The frustrated Heisenberg model
belongs to this class and with it we will exemplify the notion of a
sign-problem.

\begin{figure}
  \centering \epsfxsize=8cm \epsffile{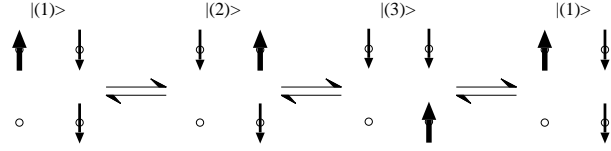}
\caption[]{The possible exchanges that a up-spin
  can make with neighbouring down-spins.  $|(1) \rangle
  \rightleftharpoons |(2) \rangle$ and $|(2) \rangle
  \rightleftharpoons |(3) \rangle$ result from nearest neighbour
  interactions and $|(1) \rangle \rightleftharpoons |(3) \rangle$
  arises through the next nearest neighbour interaction. This figure
  is minimalistic in that only a fraction of the lattice is drawn and
  only the exchanging spins are depicted.}
\label{fig:sign-problem}
\end{figure}

Figure \ref{fig:sign-problem} depicts the relevant situation. The
Hamiltonian contains interactions between $|(1) \rangle$, $|(2)
\rangle$ and $|(3) \rangle$. We know that the matrix element
connecting $|(1) \rangle$ with $|(2) \rangle$ is given by:
\[
\langle (2) | {\cal G} | (1) \rangle = -\varepsilon \langle (2) |
{\cal H} | (1) \rangle = -\varepsilon \frac{J_1}{2}.
\]
In this move the weight $M$ will pick up a minus sign from
$s_{(2),(1)}$. The same holds for the move $|(2) \rangle \rightarrow
|(3) \rangle$. The move from $|(3) \rangle$ back to $|(1)\rangle$ will
also induce a minus sign as $-\varepsilon \langle (1) | {\cal H} | (3)
\rangle = -\varepsilon \frac{J_2}{2}$. When returned to the original
configuration the weight of the walker has thus reversed sign. The
sign that a walker picks up following this loop $|(1) \rangle
\rightarrow |(2) \rangle \rightarrow |(3) \rangle \rightarrow |(1)
\rangle $ cannot be removed by basis transformations or guiding wave
functions.

The foundation of GFMC is that if more and more paths in phase space
are incorporated, the overall weight $\sum_\alpha M_\alpha$ increases
and the average $\sum_\alpha X_\alpha M_\alpha/\sum_\alpha M_\alpha$
improves. Here this line of reasoning does not hold; an extra path can
suppress the previous weights. Given the fact that the underlying
stochastic process is a Markov chain it is easy to show that the
average sign, $\sum_\alpha M_\alpha / \sum_\alpha |M_\alpha|$, will
decrease exponentially in the number $n$ of projections made. Likewise
the signal $\sum_\alpha X_\alpha M_\alpha /\sum_\alpha M_\alpha$ will
become very small with respect to the noise. One can only hope that
before the noises overshadows the measurements, the ground state value
has already been reached. Under normal conditions, this is rare, but
as will be explained in the next sections, one can steer the
calculation towards such a situation.

To complete the argument on the sign-problem, two further assessments
have to be made.  First, in the unfrustrated case, $J_2=0$, such a
loop as described above, does not exist. A basis rotation, ${\cal S}=
\exp(i 2 \pi \sum_{x,y} (x+y) S_{x,y}^z)$ removes the signs all
together from the projector as
\begin{eqnarray}
{\cal S} {\cal G} {\cal S}^\dagger &=& 1 - {\cal S} {\cal H} {\cal
  S}^\dagger = 1 - \varepsilon J_1 \sum_{(i,j)} {\cal S} \vec{\cal
  S}_i \vec{\cal S}_j {\cal S}^{\dagger} \nonumber \\ &=& 1+\varepsilon J_1
\sum_{i,j} \left [ \frac{1}{2} ( {\cal S}_i^+ {\cal S}_j^- + {\cal
    S}_i^- {\cal S}_j^+ ) - {\cal S}_i^z {\cal S}_j^z \right ]. \nonumber
\end{eqnarray}
From this equation, it is clear that the prefactors of the
non-diagonal terms are no longer negative. Once $\varepsilon$ is small
enough this also holds for the diagonal terms.  It is an example of
the fairly general approach of a basis transformation to remove the
signs from the projector ${\cal G}$. GFMC has indeed helped to
establish high quality results for the unfrustrated Heisenberg model
\cite{runge92}.

Marshall \cite{marshall55} has proven that after this rotation the
exact ground state $|\bar{\psi}_0 \rangle = {\cal S} | \psi_0 \rangle$
of the system is free of signs, $\langle R | \bar{\psi}_0 \rangle \ge
0$.  Coming up to our second assessment, this proof can be extended to
the region of small $J_2$ \cite{richter94}: for $J_2 >0$ we just
argued that a sign-problem existed, thus there can even be a
sign-problem when the exact ground state is sign-less!

One of the first successful attempts to overcome this problem
originated in the realm of quantum models with continuous degrees of
freedom \cite{ceperley80}. Later it was extended by van An and van Leeuwen
\cite{an91}, van Bemmel et al. \cite{bemmel94} and ten Haaf et al. \cite{tenhaaf95} to lattice models.
These methods are called Fixed-Node Monte Carlo (FNMC) as both in the
continuous version and in the lattice version the Hamiltonian is
altered by removing the negative projector matrix elements,
\[
\langle R'|\bar{\cal G}| R \rangle < 0 ~~\hbox{or} ~~ \langle R'|
\bar{\cal H} | R \rangle > 0 ~~ \hbox{for} |R' \rangle \neq |R\rangle.
\]
The sign of the out coming wave function $\langle R|{\cal G}^n|\phi
\rangle$ is fixed and the weights $M_\alpha$ are positive definite.
In the continuum version this is all there is to it, but in the
lattice version the projector $\bar{\cal G}$ would be altered so
strongly that no clear connection to the original system remains. An
extra potential has to be introduced to compensate for the restriction
of the hops. In the description of the fixed-node method we follow
Sorella \cite{sorella98} who made a small extension with respect to
the proposal by van Bemmel et al. \cite{bemmel94}.

Define a fixed-node Hamiltonian ${\cal H}^{\rm fn}$ according to the
following rules: if $|R' \rangle \neq | R \rangle$, then
\[
\begin{array}{rclcrcl}
\langle R' |\bar{\cal H}^{\rm fn} | R \rangle &=&\phantom{-\gamma} \langle R' | \bar{\cal H} | R \rangle& \hbox{if}&
 \langle R' | \bar{\cal H} | R \rangle &\le& 0, \\
&=& -\gamma \langle R' | \bar{\cal H} | R \rangle& \hbox{if}&
 \langle R' | \bar{\cal H} | R \rangle &>& 0.
\end{array}
\]
The diagonal element is offset by a sign flip potential,
\begin{eqnarray}
 \langle R|\bar{\cal V}^{\rm sf}|R \rangle &=& \sum_{ \langle R' | \bar{\cal H} | R \rangle>0} \langle R' | \bar{\cal H} | R \rangle, \nonumber \\
\langle R |\bar{\cal H}^{\rm fn} | R \rangle &=&\langle R |\bar{\cal H} | R \rangle + (1+\gamma) \langle R| \bar{\cal V}^{\rm sf} | R \rangle. \nonumber
\end{eqnarray}
For $\gamma=-1$ the original Hamiltonian is completely recovered
including the sign-problem but once $\gamma>0$ the projector
$\bar{\cal G}^{\rm fn}=1-\varepsilon \bar{\cal H}^{\rm fn}$ contains no signs any longer. Van Bemmel et al.
\cite{bemmel94} considered the case $\gamma=0$. Note that the bar over
the sign-flip term, $\bar{\cal V}^{\rm sf}$, is only cosmetic, as it only
appears in the diagonal terms. On the contrary we cannot remove the
bar over ${\cal H}$ in the definition of $\bar{\cal V}^{\rm sf}$ as here
the non-diagonal matrix elements are considered.

It can be proven that this method is variational
\cite{tenhaaf95,sorella98}, i. e.
\[
\langle \phi | {\cal H}^{\rm fn} | \phi \rangle - \langle \phi | {\cal H}
| \phi \rangle = (1 + \gamma) \Delta(\phi,\phi_G) \ge 0 ~~\hbox{for all}~~
|\phi \rangle,
\]
with $\Delta(\phi,\phi_G)$ a well-defined, positive function
independent of $\gamma$. The most important property for this
difference $\Delta(\phi,\phi_G)$ is that it vanishes at $|\phi \rangle
= | \phi_G \rangle$. A direct consequence is that if the ideal guiding
wave would be used, $|\phi_G \rangle = | \psi_0 \rangle$, the
sign-less FNMC would yield the ground state properties exactly
(even without the improved mixed estimator).

Within the framework of FNMC we might state that we start with the
best possible approximation $|\phi_G \rangle$ that can be made prior
to the simulation, and let the (fixed-node) Hamiltonian improve on
that.

This approach can be tested on small systems as there we know both the
ground state wave function $|\psi_0 \rangle$ and its energy $E_0 |
\psi_0 \rangle= {\cal H} | \psi_0 \rangle$. Experimentally we have
found for the frustrated Heisenberg model at $J_2/J_1=0.5$ that a
reasonable guiding wave with an energy
\[
E_G = \frac{\langle \phi_G | {\cal H} | \phi_G \rangle}{\langle \phi_G
  | \phi_G \rangle},
\]
will give rise to an outcome of the FNMC simulation with an error
$\Delta E$ in the energy that is approximately half of the original
error, $\Delta E \approx \frac{1}{2} ( E_G-E_0)$. More often than not
this will not do. Only when a gap in the energy spectrum exist and the
FNMC yields an energy below that of the first excitations, it is clear
that the state has to resemble the true ground state of the system. In
the next section an extension to the FNMC is introduced to get
substantially closer to the ground state.

\section{Stochastic Reconfiguration}

Sorella \cite{sorella98} introduced a method that potentially resolves
the limitations of the FNMC. He named it Green Function Monte Carlo
with Stochastic Reconfiguration (GFMCSR). The new ingredient is the
reconfiguration. The stochastic part refers to branching as defined
before.  It can be interpreted as a sophisticated method to find
repeatedly a suitable starting point for a straight GFMC with
sign-problem.

It was mentioned before that the sign-problem does not need to be a
great obstacle if only a good starting wave function $| \phi \rangle$
could be chosen. The ground state would then be reached before the
noise component in the weights becomes dominant. In this section,
three possible extensions are described, starting with a simple
combination of FNMC and GFMC and finishing with the GFMCSR.

The simplest solution would be to target the ground state of the
fixed-node Hamiltonian ${\cal H}^{\rm fn}$ first through a FNMC and once
that has converged, switch to the projector $\bar{\cal G}$. Ten Haaf
and van Leeuwen \cite{tenhaaf95b} have performed this routine, with
$\gamma=0$ for $\bar{\cal G}^{\rm fn}$, naming it the power method. A
large drawback is that after each measurement the routine has to be
restarted; new starting configurations have to be generated,
distributed according to the $ |\langle \phi_G |R \rangle|^2$

It is actually fairly straightforward to avoid the restart. A FNMC can
be set up with $\gamma > 0$. To each walker two weights are assigned,
$M_\alpha^{\rm fn}$ and $M_\alpha$. The fixed-node weight $M_\alpha^{\rm fn}$
is updated as prescribed before using the projector $\bar{\cal
  G}^{\rm fn}$. The other, normal weight $M_\alpha$ is updated as to
reflect the normal projector $\bar{\cal G}$:
\[
m(R)=m^{\rm fn}(R) ~~\hbox{but} ~~s(R',R)= \frac{\langle R'| \bar{\cal
    G}|R \rangle}{\langle R'| \bar{\cal G}^{\rm fn}|R \rangle}
~~\Leftrightarrow~~ s^{\rm fn}(R',R)=1.
\]
The weights $M_\alpha$ correspond to the projection with $\bar{\cal
  G}$ and will suffer from the sign-problem. After $n$ projections, when the average sign $\sum_\alpha M_\alpha / \sum_\alpha | M_\alpha|$ is not too small,
measurements are made and a new, sign-less starting point is taken by
assigning
\begin{equation}
M_\alpha^{\rm new}=M_\alpha^{\rm fn}. \label{eq:newstart}
\end{equation}
Still is remains unclear whether the weights $M_\alpha$ have converged
enough at the time of measurement and more sophistication is
necessary.

There is information in the walkers that both schemes above do not
use.  In resetting the weights, equation (\ref{eq:newstart}), a lot of
information on the ground state is lost. All kinds of correlation
functions just obtained in the measurements are not used to improve
the starting point. GFMCSR, introduced by Sorella \cite{sorella98},
provides a systematic method to incorporate this information in
the new starting weights.

If we have a set of observables ${\cal X}^i$ with expectation values
\[
X^i=\frac{\sum_\alpha X^i(R^\alpha_n) M_\alpha}{\sum_\alpha M_\alpha},
\]
the new weights $M_\alpha^{\rm new}$ should reflect these,
\begin{equation}
\frac{\sum_\alpha X^i(R^\alpha_n) M_\alpha}{\sum_\alpha M_\alpha}= X_i = \frac{\sum_\alpha X^i(R^\alpha_n) M_\alpha^{\rm new}}{\sum_\alpha M_\alpha^{\rm new}}. \label{eq:improvedstart}
\end{equation}
Moreover the average sign, $\sum_\alpha M^{\rm new}_\alpha / \sum_\alpha
|M^{\rm new}_\alpha|$, should have increased substantially. The solution
to this problem is not unique, but a good handle can be found in the
fixed-node weights $M^{\rm fn}_\alpha$. Start with the expression
\[
M^{\rm new}_\alpha = M^{\rm fn}_\alpha \left ( 1 + \sum_i \beta_i \left
    (X^i(R_n^\alpha) - X_{\rm fn}^i \right) \right ),
\]
with the average
\[
X_{\rm fn}^i = \frac{\sum_\alpha X^i(R_n^\alpha)
  M^{\rm fn}_\alpha}{\sum_\alpha M^{\rm fn}_\alpha},
\]
completely in line with previous definitions. The prefactors
$\beta_i$ are tuned to satisfy equation (\ref{eq:improvedstart})
\cite{sorella98}.

This will yield a starting point with exactly the same properties as
observed in the last measurements. In a longer calculation one can
even consider adjusting the weight $M^{\rm new}_\alpha$ to reflect the
expectation values $X^i$ averaged over several measurements.

With two weights per walker, branching has to be somewhat different
than before. The branching is performed on basis of the normal weights
$M_\alpha$ and afterwards the fixed-node weight $M_\alpha^{\rm fn}$ is
adjusted, $M^{\rm fn}_\alpha = | M_\alpha|$. Usually the branching is
performed just after the reconfiguration.

\section{A guiding wave function from the DMRG}

In the previous sections we have seen that a good guiding wave
function is of tremendous importance for all variants of the GFMC
(straight GFMC, FNMC and GFMCSR). Historically this has been the
bottleneck of the GFMC \cite{goedecker91}; before a calculations could
be performed, a large amount of research time had to be dedicated to
the design of a guiding wave function that would be both similar to
the ground state $|\psi_0\rangle$ and easy to handle in the GFMC.
This latter property means that the inner product between a
configuration $(\sigma_1, \dots, \sigma_N)$ and the guiding wave
function $|\phi_G \rangle$, $\langle \sigma_1 \dots \sigma_N|\phi_G
\rangle$, can rapidly be calculated.

A natural candidate for this guiding wave function is the wave
function $\langle R|\phi_0\rangle$ resulting from a DMRG calculation.
This will also overcome the bottleneck, as the DMRG is based on a
systematic approximation scheme applicable to many different systems.
Still, we have seen in the previous chapter, that especially for
larger width $W \ge 8$ the state $|\phi_0\rangle$ is quite distinct
from the true ground state.  The DMRG state $|\phi_0\rangle$ in
general systematically underestimates the correlations along the
length of the system for relatively wide systems.

It is the distance between the DMRG state $|\phi_0\rangle$ and the
true ground state $|\psi_0\rangle$ that the GFMC has to bridge.  The
obstacles we face implementing the DMRG state $|\phi_0\rangle$ as a
guiding wave function $\langle R|\phi_G\rangle = \langle R | \phi_0
\rangle$ are of a technical nature.  The remainder of this section
will therefore be conceptual straightforward but full of details.

We want to know the value of the wave function $\langle R | \phi_0
\rangle$, but both memory usage and computational effort are an issue.
For a single walker the configurations that are of interest are the
configuration $R$ of the walker itself and those nearby configurations
$\{R'\}$ connected by the Hamiltonian, $\langle R' | {\cal H} | R
\rangle \neq 0$. As these can only differ in the orientation of at
most two spins, an efficient algorithm can find these values $\langle
R' | \phi_0 \rangle $ relatively fast once the value $\langle R |
\phi_0\rangle$ is known. We will first show how to obtain $\langle R |
\phi_0 \rangle$ and afterwards indicate how to use the intermediate
results of this last calculation to obtain $\langle R' | \phi_0
\rangle$ rapidly.

A DMRG calculation provides a state $|\phi_0 \rangle$ on bases of both
the left and the right part of the system. It moreover gives the
necessary transformations to construct these sets. Once we have the
DMRG state $|\phi_0 \rangle$, the representation can be tailored to
suit our purposes. In appendix \ref{app:DMbasis} the technical details
are described. The most important modification is to switch to the
density matrix basis $|\alpha \rangle_l$ for the left $l$ sites and
$|\bar{\alpha} \rangle_l$ for the right $N-l$ sites.  The properties
of our representation can be summarised as follows:
\begin{itemize}
\item For every partition $1\le l < N$ we can represent exactly the
  same state $|\phi_0 \rangle$ as
\[
|\phi_0 \rangle = \sum_\alpha \sqrt{\lambda^l_\alpha} | \alpha
\rangle_l | \bar{\alpha} \rangle_l.
\]
This provides us with $N$ tables of each $m$ values for
$\sqrt{\lambda^l_\alpha}$. It can be seen as an extension of equation
(\ref{eq:exactrepr}). There we have only stated that given a partition
$l$ such a representation can be made. Here we add that for all
partitions the identical state can be representated in this form.
\item All basis transformations $A^l_{\alpha \sigma \alpha'}$ for the
  left and $B^l_{\alpha \sigma \alpha'}$ for the right part are known,
\begin{eqnarray}
|\alpha\rangle_l &=& \sum_{\sigma \alpha'}  A^l_{\alpha \sigma \alpha'} | \sigma \rangle | \alpha'\rangle_{l-1}, \nonumber \\
|\bar{\alpha}\rangle_l &=& \sum_{\sigma \alpha'}  B^l_{\alpha \sigma \alpha'} | \sigma \rangle | \bar{\alpha}'\rangle_{l+1}. \nonumber
\end{eqnarray}
The state $|\sigma \rangle$ contains the spin on site $l$ or $l+1$
respectively. This will yield $N$ matrices $A$ and $B$ of size $2m^2$
each. The estimate of the size is clearly too large as we neglect the
fact the both $A$ and $B$ is very sparse.
\end{itemize}

The wave function $\langle R| \phi_0 \rangle$ can now be evaluate at
the configuration $\langle \sigma_1 \dots \sigma_N | \phi_0 \rangle$
by induction. A reduction to inner products $\langle \sigma_1 \dots
\sigma_l|\alpha\rangle_l$ and $\langle \sigma_{l+1} \dots
\sigma_N|\bar{\alpha}\rangle_l$ is made via
\[
\langle \sigma_1 \dots \sigma_N | \phi_0 \rangle=\sum_{\alpha}
\sqrt{\lambda_\alpha^l} \langle \sigma_1 \dots
\sigma_l|\alpha\rangle_l \langle \sigma_{l+1} \dots
\sigma_N|\bar{\alpha}\rangle_l.
\]
Each of these two inner products can be derived inductively; e.g.
\begin{equation}
\langle \sigma_1 \dots \sigma_l|\alpha\rangle_l = \sum_{\alpha'}
A^l_{\alpha \sigma_l \alpha'} \langle \sigma_1 \dots
\sigma_{l-1}|\alpha'\rangle_{l-1}. \label{eq:iterativeinproduct}
\end{equation}

To optimise the algorithm for the inner products $\langle R' | \phi_0
\rangle$, the intermediate results, $\langle \sigma_1 \dots \sigma_l |
\alpha\rangle_l$ and $\langle \sigma_{l+1} \dots \sigma_N |
\bar{\alpha}\rangle_l$ for all $\alpha$ and $l$, are stored in tables.
We can use these to readily calculate the inner product of a nearby
state $|\sigma_0' \dots \sigma_N' \rangle = ({\cal S}^+_{l_2} {\cal
  S}^-_{l_1} +{\cal S}^-_{l_2} {\cal S}^+_{l_1})$ $ | \sigma_1 \dots
\sigma_N \rangle$ with the state $|\phi_0\rangle$ ($l_2>l_1$). This
new configuration is almost identical to the old one apart from the
exchange of the spins on sites $l_2$ and $l_1$;
\[
|\sigma_0' \dots \sigma_N' \rangle =| \sigma_1 \dots \sigma_{l_2}
\dots \sigma_{l_1} \dots \sigma_N\rangle.
\]
To calculate the inner product, the system can be split up;
\[
\langle \sigma_1' \cdots \sigma_N'| \phi_0 \rangle = \sum_{\alpha}
\sqrt{\lambda_\alpha^{l_2}} \langle \sigma_1 \dots \sigma_{l_2} \dots
\sigma_{l_1} | \alpha \rangle_{l_2} \langle \sigma_{l_2+1} \dots
\sigma_N| \bar{\alpha}\rangle_{l_2}.
\]
The second part of this expression, $\langle \sigma_{l_2+1} \dots
\sigma_N| \bar{\alpha}\rangle_{l_2}$, can be found in the tables.  The
first part can rapidly be built starting from the known, listed inner
product $\langle \sigma_1 \dots \sigma_{l_1-1}| \alpha
\rangle_{l_1-1}$ and iteratively extending this inner product to
location $l_2$ using (\ref{eq:iterativeinproduct}).

Further substantial reductions can be made. The most important one is
to reuse most of the intermediate results for both $\langle R| \phi_0
\rangle$ and $\{\langle R' | \phi_0 \rangle \}$ when a walker moves
from configuration $R$ to a neighbour $R'$. Once a walker has
propagated far enough for the next measurement or reconfiguration and
the next walker will be addressed, all tables are removed. This is
unavoidable as the memory usage has to be limited.

A typical system is of size $L \times L=N$ with open boundary
conditions in both directions. The calculation of the inner products
costs about $2m^2N$ operations for the partial inner products of the
configuration $R$ itself and $N(\sqrt{N} m^2+4 m^2)/2$ for all others.
Here we have again neglected that $A$ and $B$ are very sparse. Still
the calculation duration will scale as $N^{3/2}m^2$. If the tables are
reused, an extra reduction factor of 4 is achieved.

There is one strong restriction in the wave function of the DMRG; when
considering a part of size $l$ the density matrix will select states
that lies in specific ${\cal S}^z$ classes. All other classes, ranging
from ${\cal S}^z=+l/2$ to ${\cal S}^z= -l/2$ will not appear in the
wave function. For the Monte Carlo simulation to relax properly to the
ground state, configurations $|R \rangle$, that are not contained in
the guiding wave function, have to be assigned a fixed and small value
$\beta$;
\[
\langle R| \phi_0 \rangle = 0 ~~\Rightarrow ~~ \hbox{``}\langle R|
\phi_0 \rangle \hbox{''} = \beta.
\]

\section{Continuum imaginary time limit}

The guiding wave function is not perfect and this can lead to
unnecessary large fluctuations in weights of the paths. If a walker
visits a configurations $|R \rangle$ with a low 'probability'
$|\langle \phi_G |R \rangle | \ll 1 $, which neighbours a fairly
likely configuration $|R' \rangle$,
\[
\langle R'|{\cal H}|R \rangle \neq 0 ~~ \hbox{and} ~~ |\langle \phi_G
| R' \rangle | \gg |\langle \phi_G | R\rangle|,
\]
the local estimate of the energy $E(R)$ gets an excessively large
value,
\[
E(R) = \frac{\langle \phi_G| {\cal H} | R \rangle}{\langle \phi_G | R
  \rangle} ~~,~~ |E(R)| \gg 1.
\]
Only for a perfect guiding wave function we could make the replacement
$\langle \phi_G |{\cal H} = E_0 \langle \phi_G |$ and this problem
would disappear.  It will have consequences for the weight factor
$m(R)$ as the projector ${\cal G}$ contains the Hamiltonian ${\cal
  H}$;
\[
m(R)= \frac{\langle \phi_G | {\cal G} | R \rangle}{\langle \phi_G | R
  \rangle} = 1 - \varepsilon E(R).
\]
Naturally the walker will almost certainly leave this configuration
$|R \rangle$ for $|R' \rangle$ the next projection, but the harm has
then already been done. To compensate for this situation, one would
like to send $\varepsilon \rightarrow 0$. Without modifications this
limit leads to the necessity of infinity many projections $n$.

Trivedi and Ceperley \cite{trivedi90} developed an elegant route out
of this trouble. Remember that for $\varepsilon\ll 1$
\begin{equation}
(1-\varepsilon {\cal H})^n = e^{-\varepsilon n {\cal H}}. \label{eq:conttime}
\end{equation}
A continuous time variant can be formulated where only the imaginary
time $\tau = \varepsilon n$ is a relevant parameter. Let us describe
it for a sign free Hamiltonian, like the fixed-node Hamiltonian ${\cal
  H}^{\rm fn}$.

If we start in a configuration $|R \rangle$, the probability to remain
in it for $\Delta n$ steps is given by
\[
P(R \leftarrow R)^{\Delta n}=\left ( \frac{\langle R | \bar{\cal G} |
    R \rangle}{\sum_{R'} \langle R' | \bar{\cal G} | R \rangle} \right
)^{\Delta n}.
\]
Both the numerator and the denominator of the expression can be
simplified using (\ref{eq:conttime});
\begin{eqnarray}
\langle R | \bar{\cal G} | R \rangle &=& \exp(-\varepsilon \langle R| {\cal H} | R \rangle), \nonumber \\
\sum_{R'} \langle R' | \bar{\cal G} | R \rangle &=& 1 - \varepsilon
\frac{\langle \phi_G|{\cal H}|R \rangle}{\langle \phi_G|R\rangle} = \exp(-\varepsilon E(R)). \nonumber
\end{eqnarray}
Thus the probability is given by
\[
P(R \leftarrow R)^{\Delta n}= \exp( - \Delta n \varepsilon(\langle R|
{\cal H} | R \rangle-E(R))).
\]
A random number $\xi$ is chosen to set this time,
\[
\Delta \tau \equiv \Delta n \varepsilon = \frac{\ln
  (\xi)}{E(R)-\langle R| {\cal H} | R \rangle}.
\]
During this time $\Delta \tau$ the weight is multiplied by a factor
\[
m(R)^{\Delta n} = \left(\sum_{R'} \langle R' | \bar{\cal G} | R
  \rangle \right)^{\Delta n} = \exp(-\Delta \tau E(R)).
\]
After this time $\Delta \tau$ a jump to another configuration $|R'
\rangle \neq |R \rangle$ has to be made according to the transition probabilities
\[
P(R' \leftarrow R)= \frac{\langle R'| \bar{\cal G}|R
  \rangle}{\sum_{R'' \neq R} \langle R''| \bar{\cal G} |R \rangle} =
\frac{\langle R'| \bar{\cal H}|R \rangle}{\sum_{R'' \neq R} \langle
  R''| \bar{\cal H} |R \rangle}.
\]
In this new configurations the walker remains for another time
interval. Once the total imaginary time $\tau$ has passed, a
measurement, reconfiguration or branch can be made.

The GFMSR does not differ much from the above prescription. The
fixed-node weight follows it exactly. The normal weight picks up an
extra factor during the stay of the walker at a specific
configuration,
\begin{eqnarray}
\left[\vphantom{\frac{1}{2}}s(R,R) m(R) \right ]^{\Delta n} &=&
\frac{\exp(-\Delta \tau
  \langle R|{\cal H}|R \rangle)}{\exp(-\Delta \tau \langle R| {\cal H}
  + (1 + \gamma) {\cal V}^{\rm sf} |R \rangle)} \exp(-\Delta \tau E(R)) \nonumber \\
&=&
\exp(\Delta \tau (-E(R)+ (1 + \gamma) \langle R|{\cal V}^{\rm sf} |R
\rangle)).\nonumber
\end{eqnarray}
In the hops the factor $s(R,R)=\langle R'| {\cal H}|R \rangle /\langle
R'|{\cal H}^{\rm fn} | R \rangle = 1 ~\hbox{or} -1/\gamma$ is picked up in
the weights.

This approach replaces the discrete imaginary time with a continuum
and resolves the complication of extreme weight factors $m(R)$.

\section{Implementation issues}

In the previous chapter, the spin stiffness $\rho_s$ was studied. It
is possible to obtain the same stiffness in a GFMC simulation by a
trivial extension of the method by Pollock and Ceperley
\cite{pollock87}, but it is as yet not clear whether the same approach
can be combined with the GFMCSR. Instead we will focus on the
correlation functions for various frustrations $J_2$ ranging from
$J_2=0$ to $J_2=1.0$ in steps of $0.1$.

The geometry of the systems are set to $10 \times 10$ with open
boundary conditions in both directions since the correlation functions
do not require periodic boundary conditions as the spin stiffness did.
There are three clear advantages of these open boundary conditions:
first, if a dimer or plaquette phase were to appear, the location of
the dimers or plaquette will be locked by the boundary conditions; the
four corners will always contains such a object and the rest of the
system can then easily be filled in. The second advantage is that DMRG
obtains the highest accuracy in open systems. Finally, the set of
inner products $\langle R' | \phi_0 \rangle$ can be calculated much
faster as no neighbouring states exist with one of the first spins
exchanged with one of the last spins. For all other neighbours the
tables with inner products $\langle \sigma_1 \dots \sigma_l | \alpha
\rangle$ can extensively be used.

\begin{figure}
  \centering \epsfxsize=8cm \epsffile{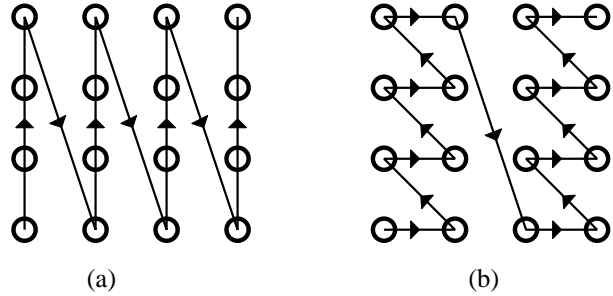}
\caption[]{The sequence in which the DMRG includes the sites in the left
  part of the system. (a) represents the ordinary order. (straight)
  (b) represents a sequence that is more in line with the appearance
  of plaquettes and all members of a plaquette are added to the basis
  successively (meandering).}
\label{fig:straight_meander2}
\end{figure}

The DMRG states are built in two distinct sequences as depicted in
figure \ref{fig:straight_meander2}. Both are based on adding one site
at the time to the basis with $m=75$ basis states. The usual approach,
figure \ref{fig:straight_meander2}(a), is to add column after column,
which we name the straight sequence. For a plaquette order the
meandering sequence of figure \ref{fig:straight_meander2}(b) is
preferable. The individual sites of a plaquette are then added
sequentially allowing strong correlations between them. The energy
$E_{\rm DMRG}$ is systematically lower for the meandering sequence than
for the straight sequence, see table \ref{tab:GFMCSRE0} (Ful
explanation of this table will follow in the next section). Therefore
we use the meandering sequence to build the guiding wave function.  With
increasing frustration, dimer correlations appear in the straight
sequence and plaquette correlations appear in the meandering sequence.

The mixed estimates incorporated in the reconfiguration are the
nearest- and next-nearest-neighbour correlation functions,
\[
\langle \vec{\cal S}_i \cdot \vec{\cal S}_{i+\hat{x}} \rangle_{\rm
  mixed}~,~\langle \vec{\cal S}_i \cdot \vec{\cal S}_{i+\hat{y}}
\rangle_{\rm mixed} ~,~\langle \vec{\cal S}_i \cdot \vec{\cal
  S}_{i+\hat{x}+\hat{y}} \rangle_{\rm mixed} ~,~ \langle \vec{\cal
  S}_i \cdot \vec{\cal S}_{i+\hat{x}-\hat{y}} \rangle_{\rm mixed}.
\]
The guiding states {\it effectively} share two symmetries with the
system geometry: reflections in the lines $y=5\frac{1}{2}$ and
$x=5\frac{1}{2}$. These symmetries are included in the mixed estimates
reducing their number by approximately a factor of four.  No further
geometrical symmetries are included. The reflection through the
diagonal of the system is excluded as the guiding states do not share
this symmetry. Moreover a dimerised state distinguishes itself from a
plaquette state by the lack of this symmetry.

After each reconfiguration a branch is performed. We use 6000 walkers
en set $\gamma=0.5$. Table \ref{tab:GFMCSRE0} lists the imaginary time
intervals $\tau$ between reconfigurations. The times $\tau$ are set to
let the average sign $\sum_\alpha M_\alpha/ \sum_\alpha |M_\alpha|$
decrease from 1 to about 0.8. At the starting of a calculation the
average sign tends to drop to a very small value. At the start of the
computation the configurations are fairly arbitrary in during the
first time intervals $\tau$ they will change frequently. As a
consequence the average sign at the end of one of the initial
intervals will be almost zero and it will only gradually increase to
0.8 over about 50 measurements. During this 'thermalisation period'
none of the calculated mixed estimators can be used for the final
expectation values. These are thus removed when the final averages are
calculated.

At $J_2=0$ there exists a transformation that will remove the
sign-problem altogether. Although we do not perform this
transformation, still the problem becomes almost signless. The length
of the interval $\tau$ is set such that successive measurements are
independent.

Correction factors introduced by Hetherington
\cite{hetherington84,sorella98} are not implemented.  A typical
simulation with a guiding wave function built with $m=75$ states,
takes about 300 hours on a Intel pentium 300 MHz machine.

\section{Results}

\begin{figure}
  
  \epsfig{file=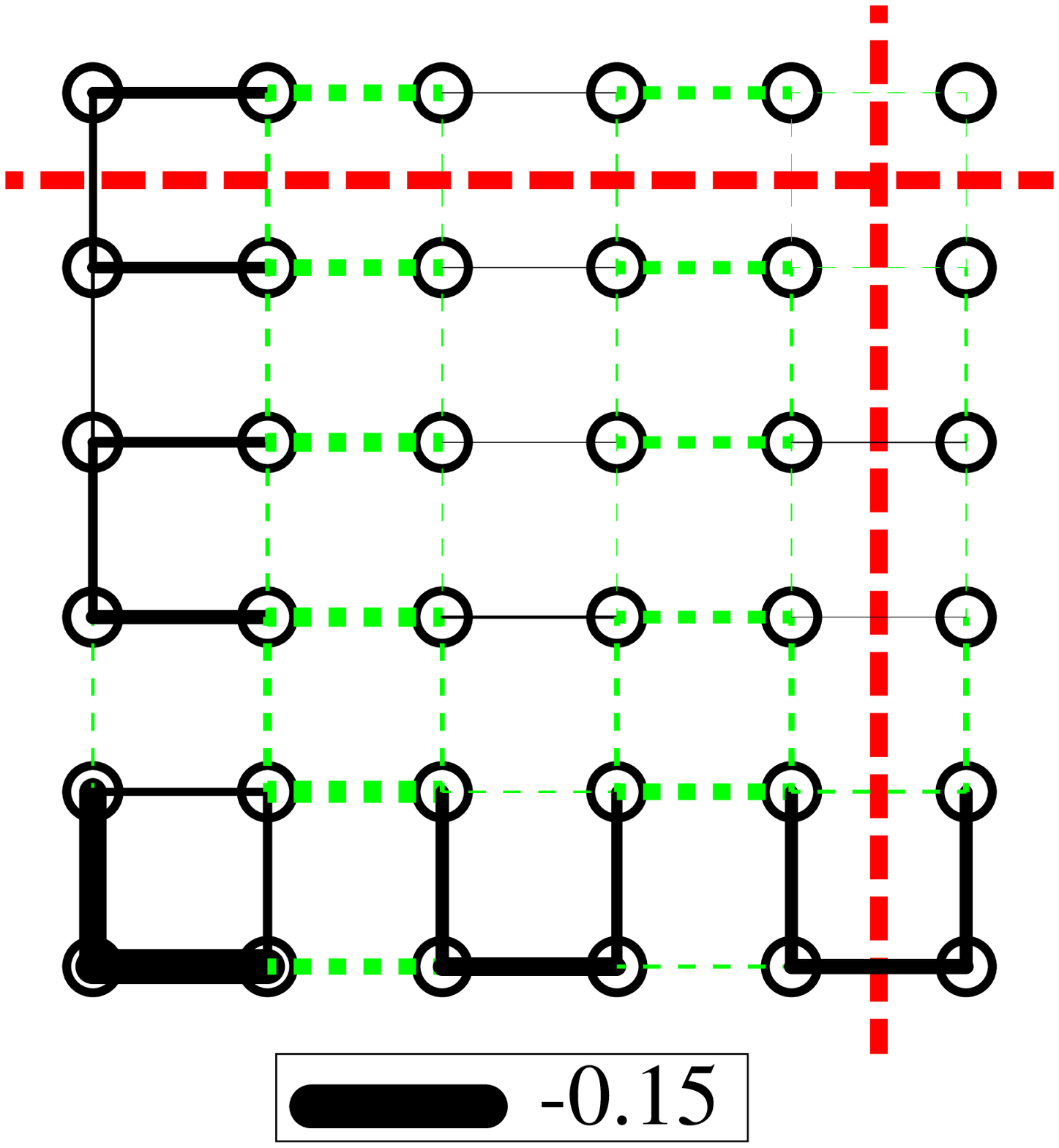,width=6cm}
  \hfill
  \epsfig{file=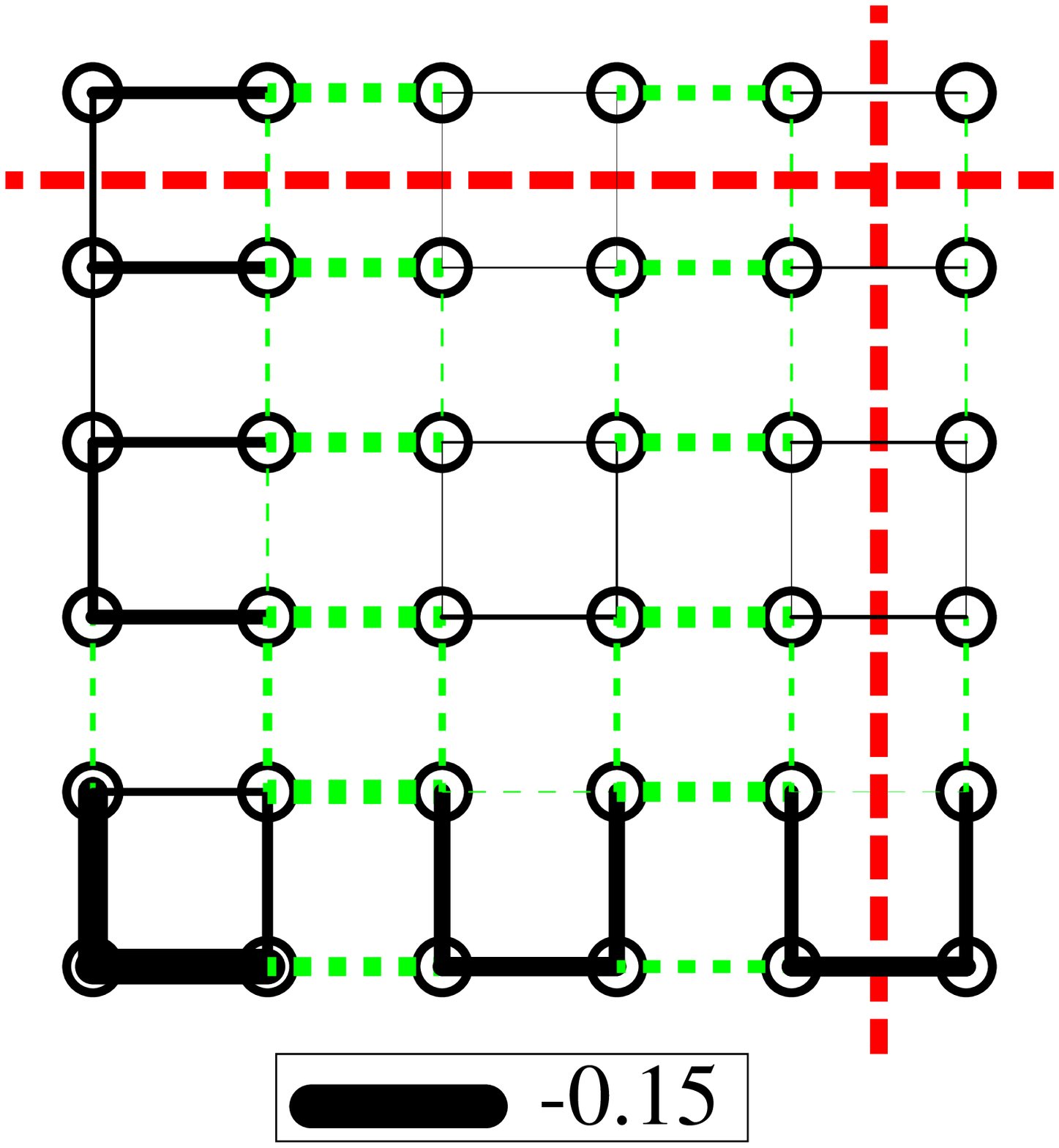,width=6cm} \vspace{0.5cm}\\
  
  \epsfig{file=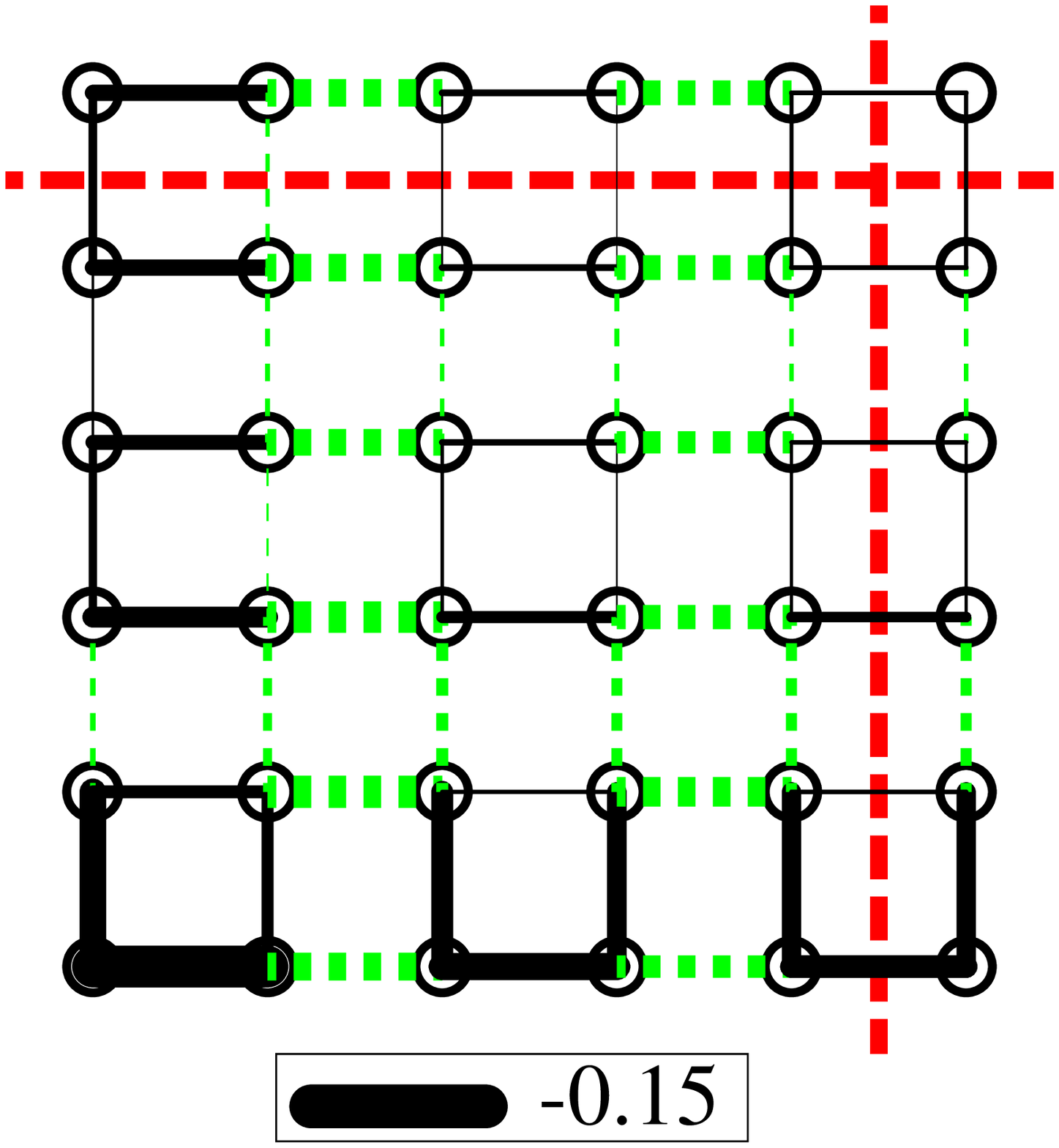,width=6cm}
  \hfill
  \epsfig{file=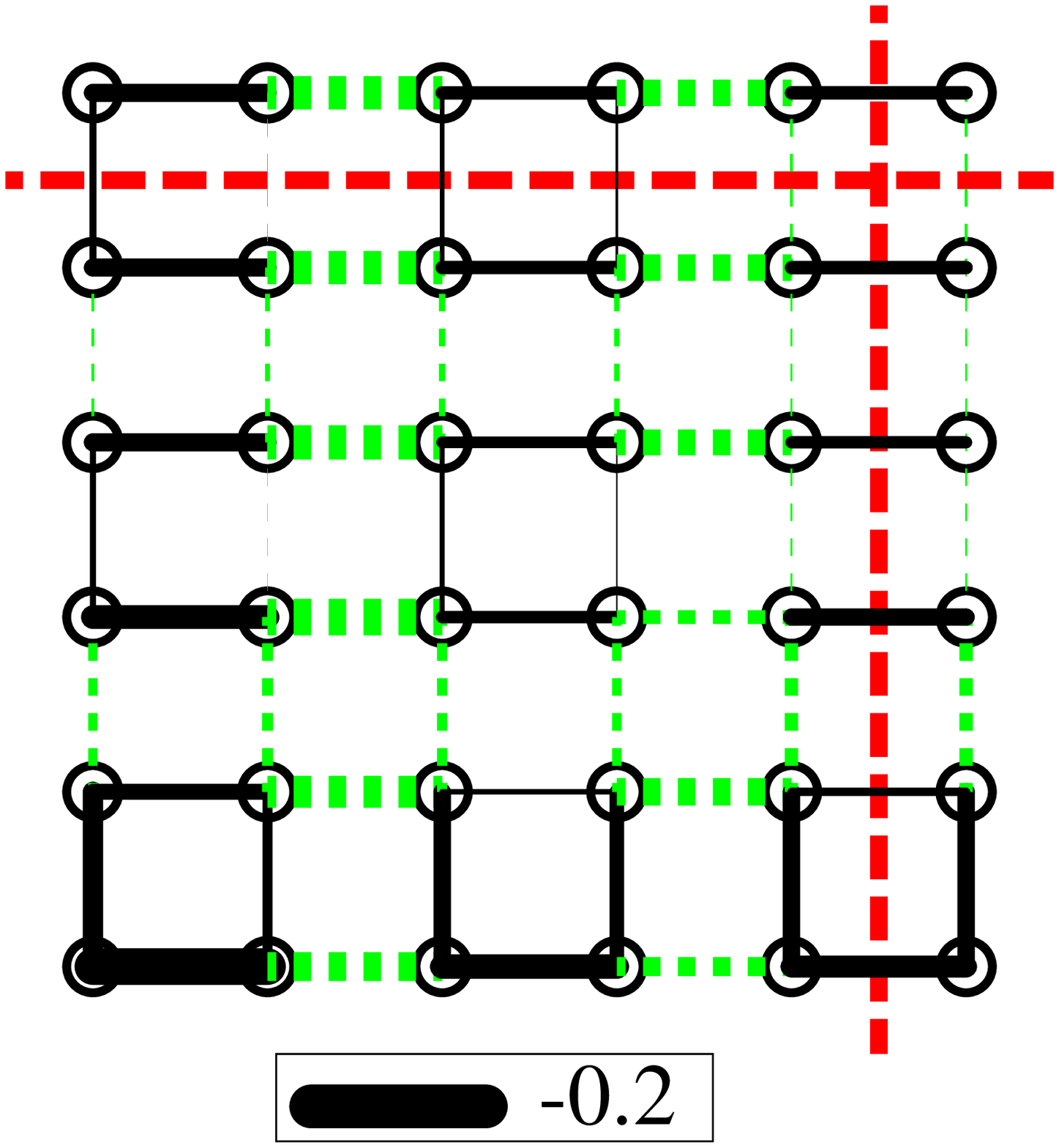,width=6cm}\vspace{0.5cm}\\
  
  \epsfig{file=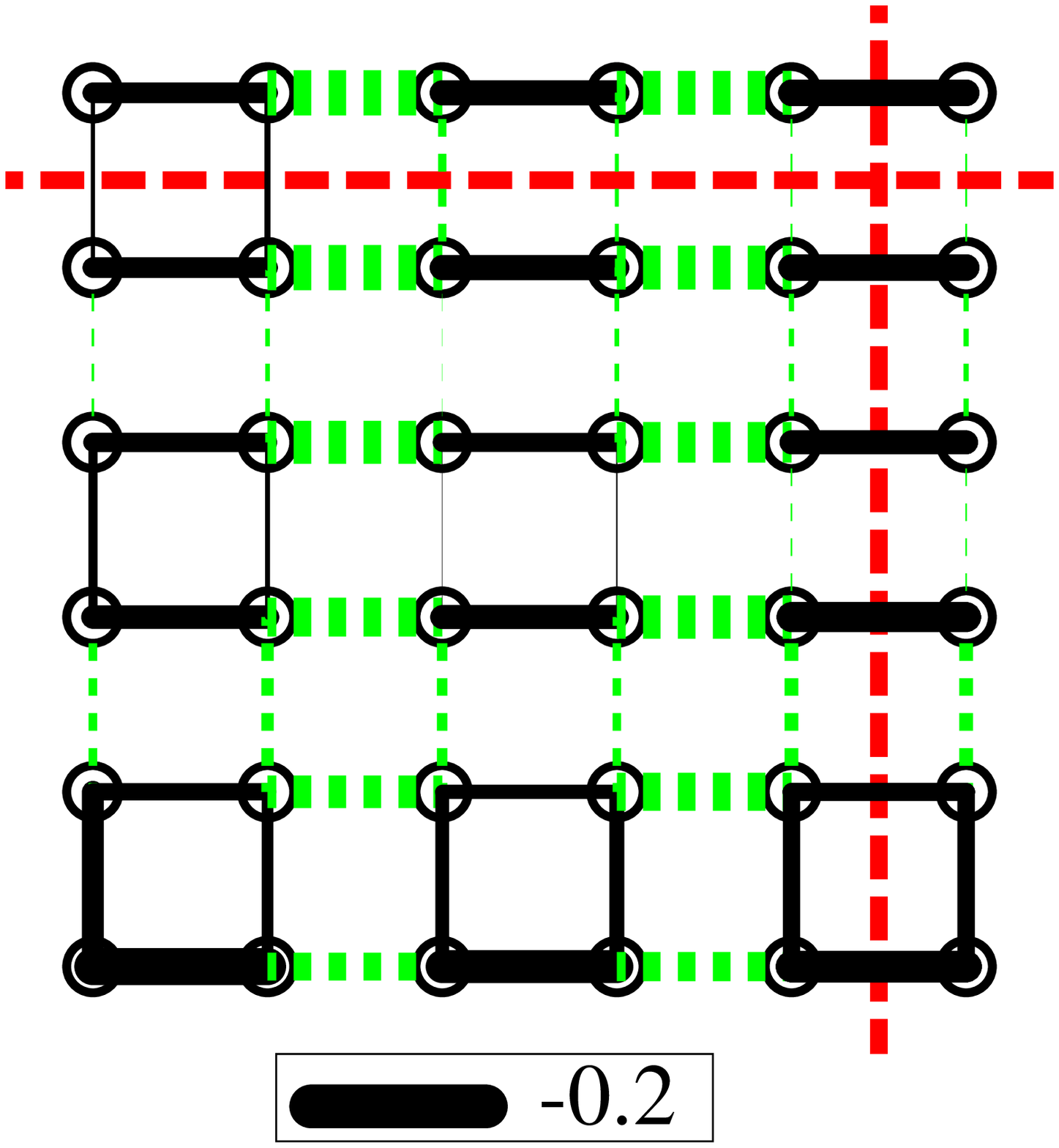,width=6cm}
  \hfill
  \epsfig{file=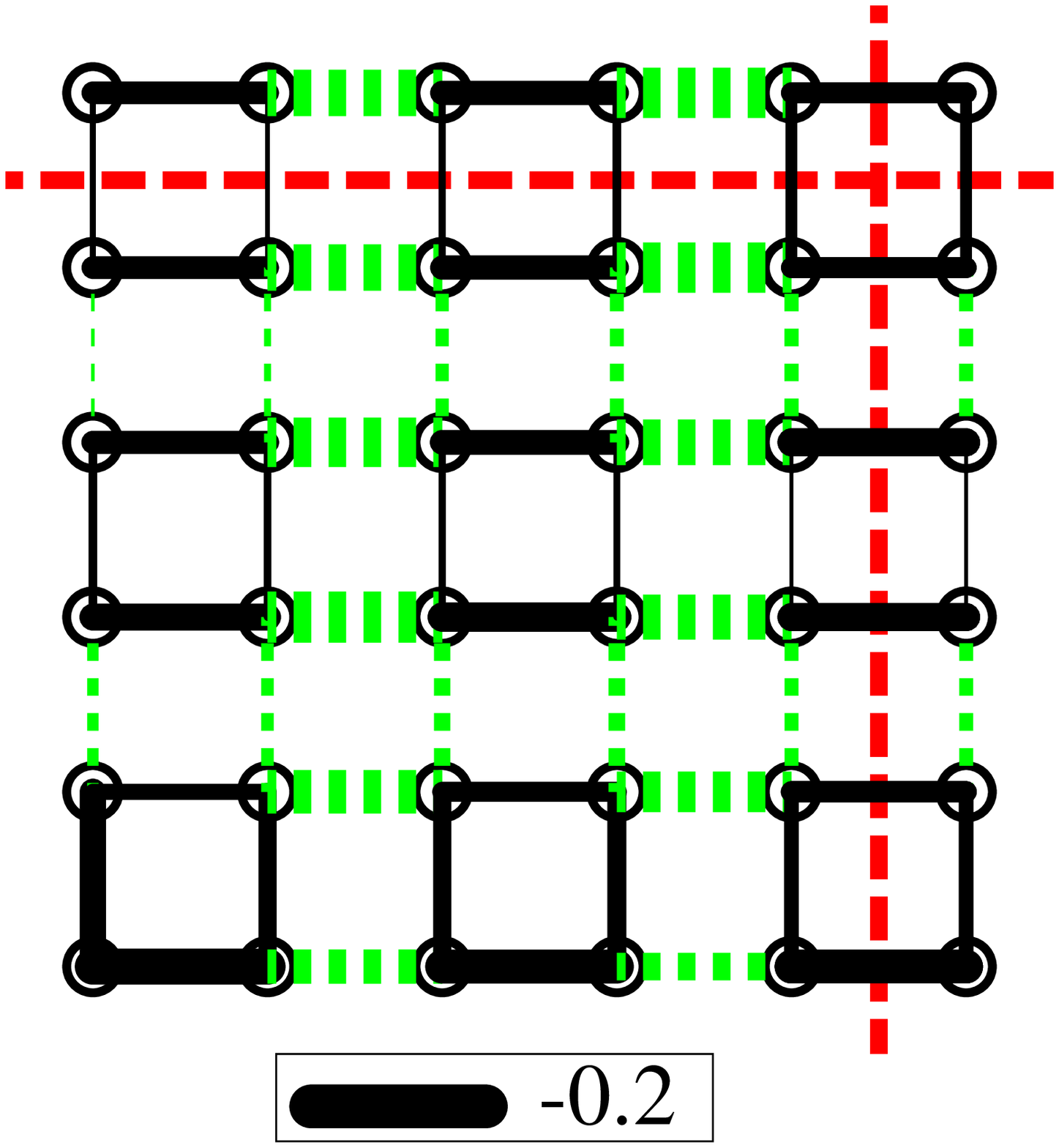,width=6cm}
  \hfill \centering
\caption[]{The relative correlation strengths on $10 \times 10$ lattice.
  All other nearest neighbour correlations can be obtained by
  reflection these picture in the two dashed lines. The DMRG guiding
  state follows the meandering sequence of figure
  \ref{fig:straight_meander2}(b).  More explanation is given in the
  text.  Reading from top left to bottom right, the values for $J_2$
  are $J_2=0,\dots,0.5$ in steps of $0.1$. }
\label{fig:variousmeander2}
\end{figure}

\begin{figure}

  \epsfig{file=result.10x10.m75.e0.014.J20.5.meander2.ss.eps,width=6cm}
  \hfill
  \epsfig{file=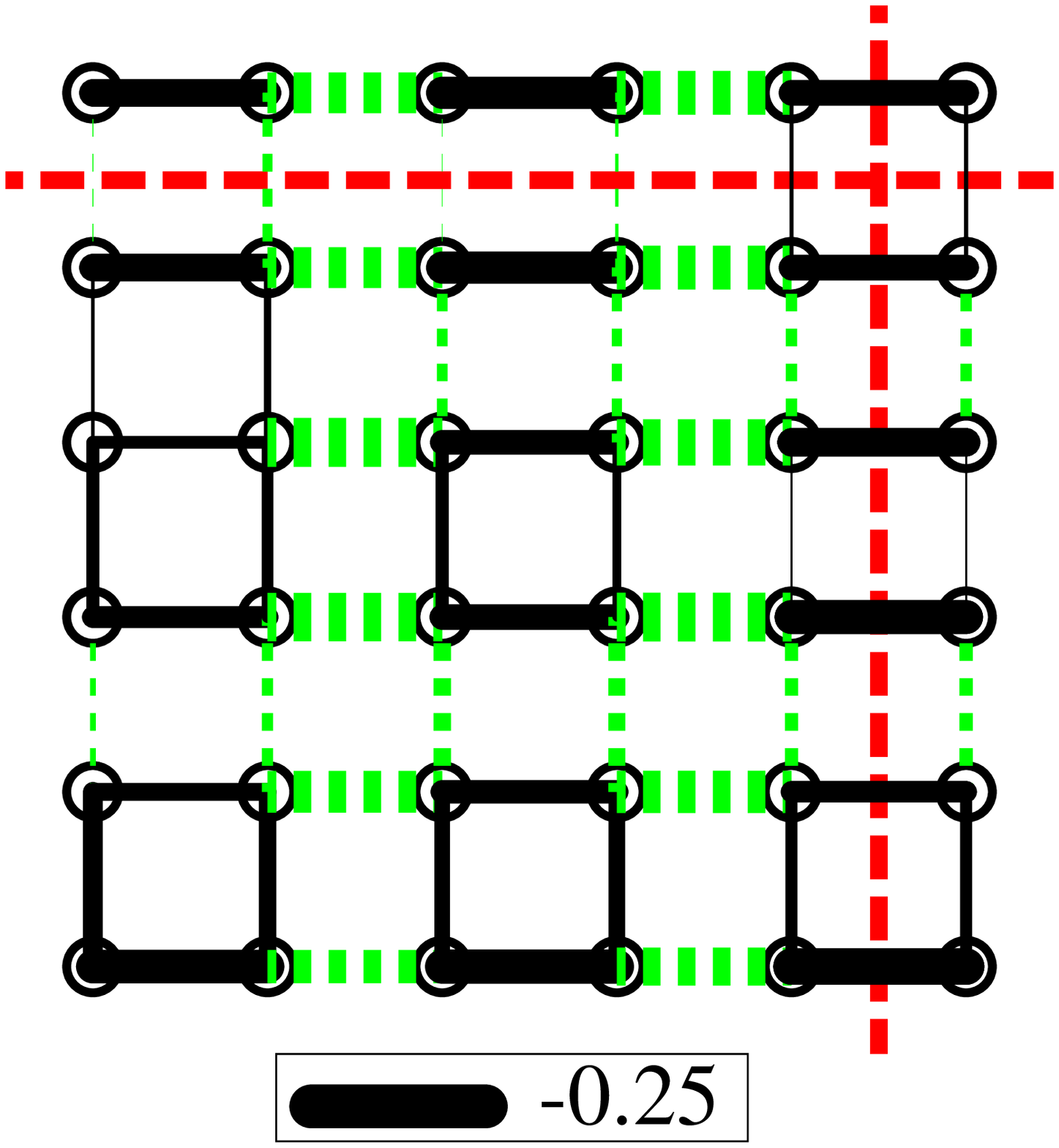,width=6cm} \vspace{0.5cm}\\

  \epsfig{file=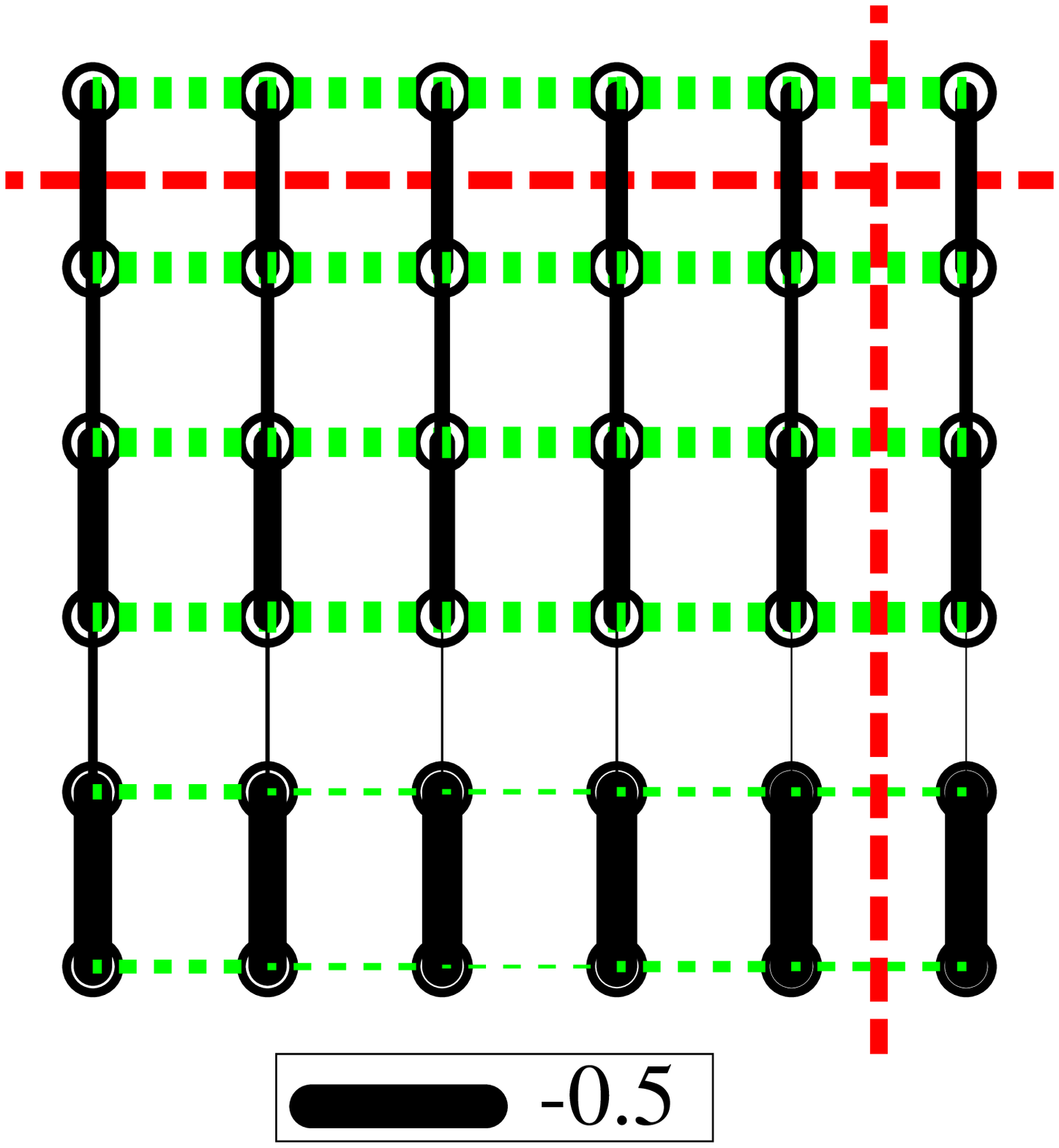,width=6cm}
  \hfill
  \epsfig{file=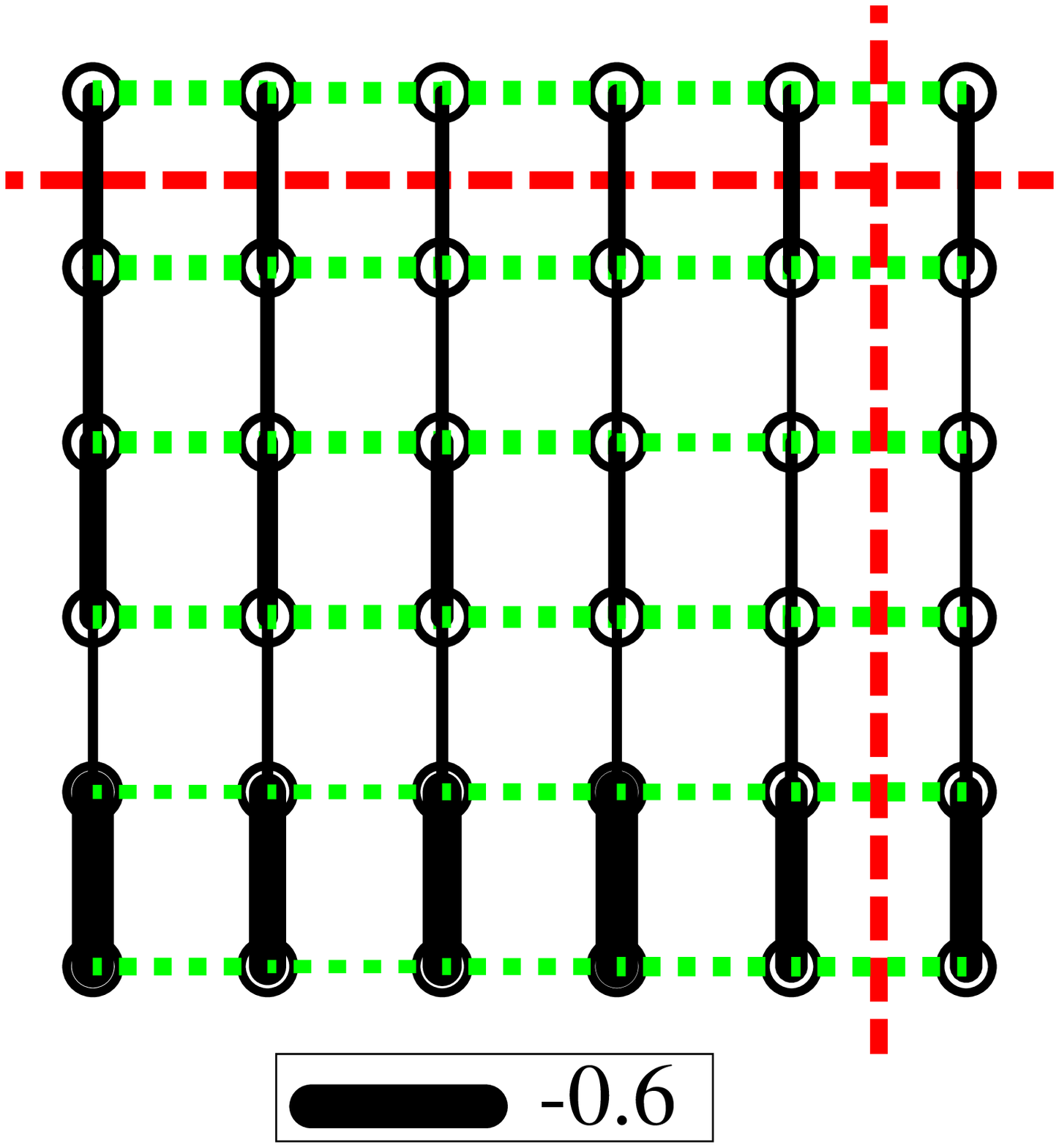,width=6cm}\vspace{0.5cm}\\
  \epsfig{file=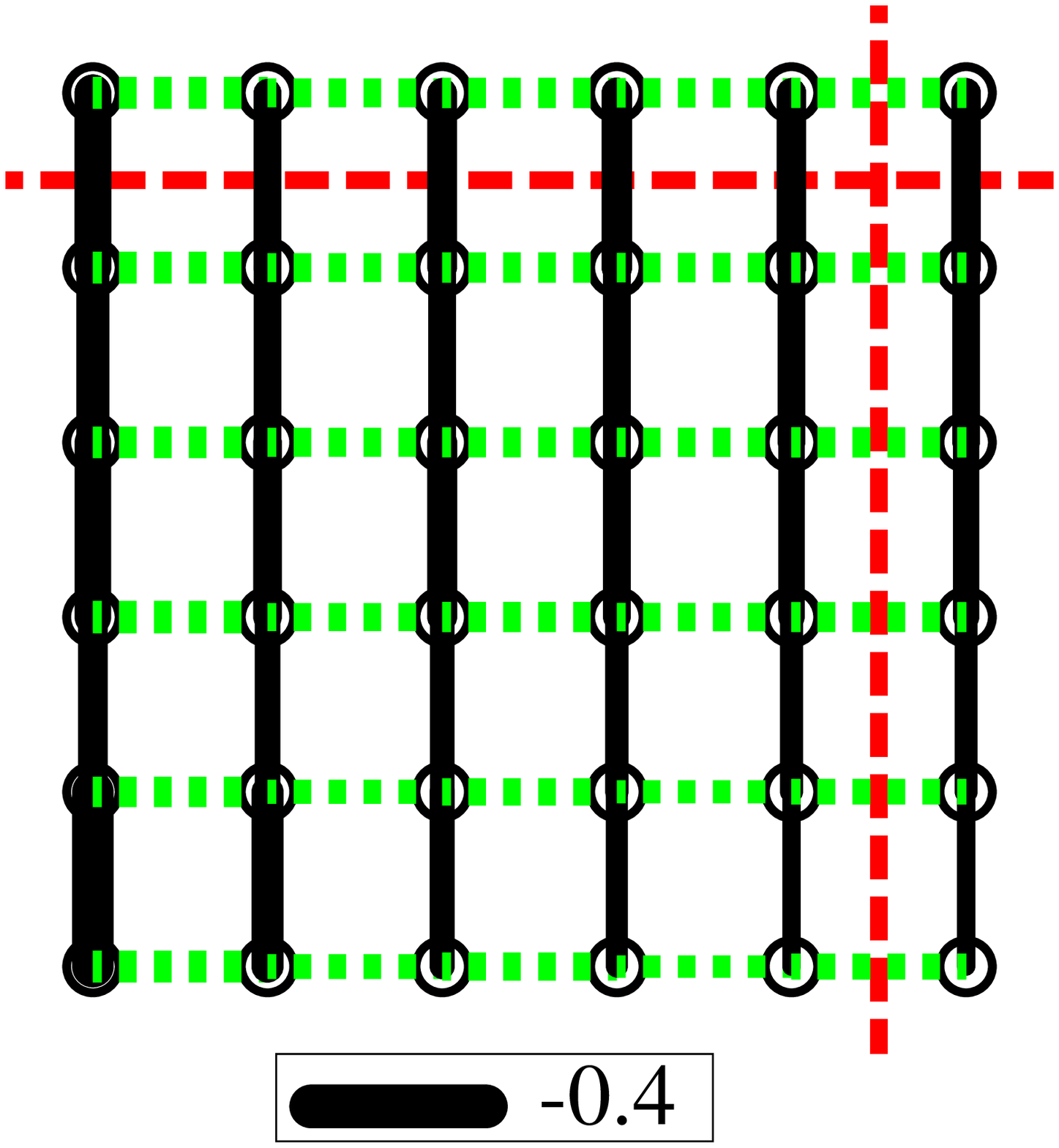,width=6cm}
  \hfill
  \epsfig{file=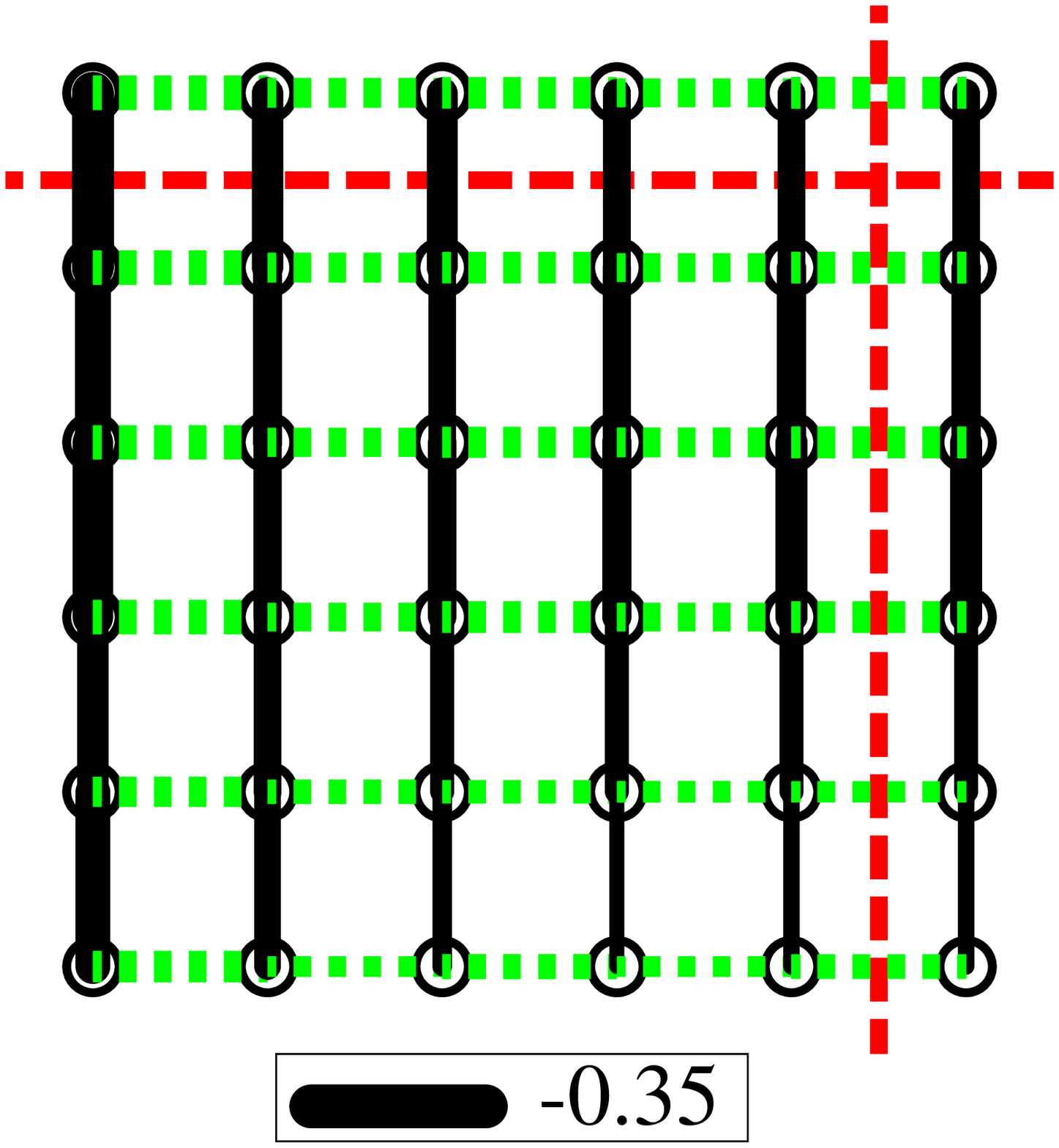,width=6cm}
  \hfill \centering
\caption[]{The continuation of figure \ref{fig:variousmeander2}; the
  relative correlation strengths on $10 \times 10$ lattice.
  $J_2=0.5,\dots,1.0$ in steps of $0.1$. }
\label{fig:variousmeander2b}
\end{figure}

Figures \ref{fig:variousmeander2} and \ref{fig:variousmeander2b}
together with table \ref{tab:GFMCSRE0} contain the results. Let us
first describe the table.

For all values of $J_2$ that we compared, the GFMCSR with a guiding
state that was built meandering through the system, resulted in a
lower final ground state energy $E_0$. The guiding state itself also
has a lower energy that the one obtained from a straight sequence.
(For $J_2=0.7,0.8$ this statement does not hold, but there the values
are close.) This is a clear indication that these GFMCSR calculations
are biased by the guiding state. Future research must determine
whether this dependence can be removed.

The dimerisations (x-dim and y-dim) indicate whether the translational
symmetry is broken in one of the two directions;
\begin{eqnarray}
\hbox{x-dim} & \equiv &  \frac{2}{W(L-2)} \sum_{x=1}^{L/2-1} \sum_{y=1}^W \langle \vec{\cal S}_{2x,y} \cdot \vec{\cal S}_{2x+1,y}  \rangle -\frac{2}{WL} \sum_{x=1}^{L/2} \sum_{y=1}^W \langle \vec{\cal S}_{2x-1,y} \cdot \vec{\cal S}_{2x,y} \rangle , \nonumber\\
\hbox{y-dim} & \equiv & \frac{2}{(W-2)L} \sum_{x=1}^{L} \sum_{y=1}^{W/2-1} \langle \vec{\cal S}_{x,2y} \cdot \vec{\cal S}_{x,2y+1} \rangle -\frac{2}{WL} \sum_{x=1}^{L} \sum_{y=1}^{W/2} \langle \vec{\cal S}_{x,2y-1} \cdot \vec{\cal S}_{x,2y} \rangle . \nonumber
\end{eqnarray}
The expectation values $\langle \vec{\cal S} \cdot \vec{\cal S}
\rangle$ are approximated by the improved mixed estimator.

The abrupt change of these dimerisation indicators from $J_2 =0.6$ to
$J_2=0.7$ already suggest a first order phase transition at that
point. This is in agreement with the results of the SBMF theory and
the numerical results in the last chapter, figure \ref{fig:E0}.

\begin{table}[h]
\small
  \begin{center}
\begin{tabular}{|l|l|l|l|l|l|l|l|l|l|}
 \hline
& & \multicolumn{4}{|c|}{Straight} & \multicolumn{4}{|c|}{Meander} \\
$J_2$ & $\tau$& $E_{\rm DMRG}$ & $E_0$ & x-dim. & y-dim. & $E_{\rm DMRG}$ & $E_0$ & x-dim. & y-dim. \\
\hline
0    & 0.3     & -61.30 & -62.33(8)  & 0.002 & 0.001 & -61.84 & -62.54(4) & 0.012 & 0.002  \\
0.1 & 0.06   & -57.96 &                   &           &           & -58.53 & -59.25(2) & 0.017 & 0.003  \\
0.2 & 0.04   & -54.75 & -56.08(11)& 0.003 & 0.004 & -55.48 & -56.22(4) & 0.022 & 0.004  \\
0.3 & 0.02   & -51.75 & -53.17(4)  & 0.005 & 0.007 & -52.50 & -53.38(3) & 0.034 & 0.006  \\
0.4 & 0.02   & -49.00 & -50.51(8)  & 0.009 & 0.015 & -49.92 & -50.60(5) & 0.035 & 0.004 \\
0.5 & 0.014 & -46.68 & -47.76(6)  & 0.009 & 0.058 & -47.78 & -48.34(4) & 0.063 & 0.021 \\
0.6 & 0.015 & -45.41 &                   &           &           & -46.03 & -46.40(3) & 0.073 & 0.022 \\
0.7 & 0.015 & -45.67 &                   &           &           & -45.60 & -46.00(2) & 0.011 & 0.020\\
0.8 & 0.02   & -49.16 &                   &           &           & -49.13 & -49.60(9) & 0.009 & 0.001\\
0.9 & 0.02   & -53.61 &                   &           &           & -53.70 & -54.52(2) & 0.007 & -0.006\\
1.0 & 0.02   & -58.46 & -59.71(9)  & 0.010 & -0.006 & -58.64 & -59.80(8)& 0.007& -0.005\\
\hline
\end{tabular}
\end{center}
\normalsize
\caption{\label{tab:GFMCSRE0} For each degree of frustration the imaginary time interval
 $\tau$, the energy of the guiding state $E_{\rm DMRG}$ and the properties of the GFMCSR
state are listed. In the text the quantities x-dim and y-dim are explained.}
\end{table}

The figures \ref{fig:variousmeander2} and \ref{fig:variousmeander2b}
give a more qualitative insight in this behaviour. The average
correlation strength between nearest neighbours is negative for each
of these systems, $\frac{1}{(L-1)(W-1)} \sum_{(ij)} \langle \vec{\cal
  S}_i \cdot \vec{\cal S}_j \rangle < 0$. Depicted are the individual
correlation strengths relative to the average. The solid lines
indicate correlations that are more negative and thus stronger than
the average. The dotted lines show which correlations are less
negative than the average. They can even be positive. In all cases the
correlations indicated with the solid lines are also the largest in
absolute terms. To approximate these correlation functions the
improved estimator is used. Especially in figure
\ref{fig:variousmeander2} the ladder structure of the guiding state
still persists in the final result. Forward walking schemes can reduce
this tendency of the guiding state further. Given that this influence
weakens in figure \ref{fig:variousmeander2b}, there is no need for
this extension to get a qualitative picture of the behaviour.

In figure \ref{fig:variousmeander2b} an abrupt change from plaquettes
to dimers appears between $J_2=0.6$ and $J_2=0.7$. The same alteration
is also observed in the DMRG guiding state (not depicted).

\section{Discussion and Conclusion}

In this chapter we have combined the DMRG method with the GFMCSR to
analyse the properties of the frustrated Heisenberg model. Part of the
chapter is methodical in nature and in the remainder the physical
aspects are investigated.

On the methodical side, we find that the combination of the DMRG and
GFMC techniques is successful. A guiding state is generated in a
systematic manner. Previously it was necessary to dedicate a
substantial amount of time to construct an approximation to the ground
state that is both easy to handle and contains the relevant physics.
Here, we present a relatively easy alternative.

The present implementation is not yet state of the art
computationally. Neither have we used the strongest machines currently
available nor is the software fully optimal. A GFMC simulation can
easily be distributed over many processors without communication
between them.  This allows for the cheapest way of upscaling by having
many computers simulate independently.

The Monte Carlo methods themselves could also do with further
improvements. The mixed estimator will not suffice for accurate
expectation values. A forward walking scheme \cite{calandra98} has to
be developed for higher precision, but it is unclear how to combine it
with the stochastic reconfiguration that alters the weights
frequently. We mention that FNMC does not yield high enough accuracy
and an extension is necessary; the GFMCSR is such an extension but
many complications still remain: it is evident that the final result
is biased by the guiding state. This is even the case for conserved
observables like the energy, which should be accurately sampled by the
mixed estimator. Furthermore it remains unclear which and how many
observables should be used for the stochastic reconfiguration. More
investigations are necessary in order to answer all of these
questions.

We select the meandering sequence for the guiding wave function as
that sequence yields the lowest energies $E_{\rm DMRG}$ for the
guiding wave function and $E_0$ for the GFMCSR calculation. The
correlation functions of the GFMCSR are qualitatively the same as
those of the DMRG, although the numerical values lie closer together.
DMRG calculations of the same system with upto $m=512$ support the
same qualitative behaviour although they do not reach the low energy
of the GFMCSR. Therefore there is little doubt that the qualitative
behaviour we find, is correct for the $10 \times 10$ system.

The simulations shed some light on the physical properties of the
frustrated Heisenberg model. A clear change of the spin correlations
presents itself at $J_2 \approx 0.6$. This suggests a first order
phase transition at that location, in line with both the SBMF picture
and the extrapolated DMRG results for the ground state energy $E_0$.
With regard to the spin stiffness (see figure \ref{fig:rhos}), it is
unlikely that a dimer phase exists for stronger coupling then $J_2
\approx 0.6$.  The collinear order must already set in, maybe with
dimer-like tendencies but still conserving long-range order. On the
weak coupling side, $J_2 \ll 1$, plaquette correlations seems to set
in quite soon. The overall picture that appears is that with
increasing next-nearest neighbour coupling plaquette correlations
build up gradually while preserving long-range antiferromagnetic
order. At about $J_2 \approx 0.4$ the N\'eel order disappears in a
second order phase transition and only the local, plaquette order
remains. In a first order phase transition at about $J_2 \approx 0.6$
the system changes into collinear order with dimer-like correlations
which will gradually fade out with increasing $J_2$.

Singh et al. \cite{singh99} recently suggested that the intermediate
phase consists of columnar dimer order where the correlations between
dimers inside a column are stronger than between columns.  
We do not expect that to be correct, although
 figures \ref{fig:variousmeander2} and \ref{fig:variousmeander2b}
show similar behaviour. This behaviour is triggered by 
ladder-like correlations in the DMRG guiding states. We expect that a more
accurate estimator for these correlations will suppress this feature, as the
improved mixed estimator still contains errors of order ${\cal
  O}(\delta^2)$, eq. (\ref{eq:improved}).

%% file: APPENDIX.tex
\appendiceson
\chapter*{Appendices} \addcontentsline{toc}{chapter}{Appendices}

\section{SBMF Approximation for ${\cal U}^\dagger ({\bf q}) {\cal H} {\cal U}({\bf q})$. \label{app:SBMFq}}

In this appendix the derivation of the Schwinger boson mean-field
Hamiltonian for the 'twisted' case, ${\bf q} \neq {\bf 0}$ is
performed. In itself this is mostly a repetition of the untwisted case
combined with a number of small details. In the process the first and
second derivative with respect to ${\bf q}$ are also obtained. These
are used in section \ref{sec:schwingerrho_s} to calculate $\rho_s$.

The Schwinger boson notation is completely equivalent with the spin
representation, therefore ${\cal H}({\bf q})$ -defined in
(\ref{eq:twistH})- can be expressed as
\[
{\cal H}({\bf q}) = \frac{1}{2} \sum_{F} J_{ij} \left ( {\cal
    D}_{ij}^\dagger ({\bf q}) {\cal D}_{ij} ({\bf q}) - \frac{3}{2}
\right) -\frac{1}{2} \sum_{AF} J_{ij} \left ( {\cal B}_{ij}^\dagger
  ({\bf q}) {\cal B}_{ij} ({\bf q}) - \frac{1}{2} \right).
\]
This time we have to distinguish between ferro $F$ and anti
ferromagnetic bonds $AF$ explicitly as they are not in general simple
nearest and next-nearest-neighbours.  where the ${\cal D}_{ij} ({\bf
  q})$ and ${\cal B}_{ij} ({\bf q})$ are given by
\begin{equation}
\begin{array}{ccccc}
  {\cal D}_{ij}({\bf q}) &=& {\cal U}({\bf q}) {\cal D}_{ij} {\cal U}^\dagger ({\bf q}) &=& a_i
  a_j^\dagger e^{\frac{i}{2} {\bf q} \cdot ({\bf r}_i - {\bf r}_j)} +
  b_i b_j^\dagger e^{-\frac{i}{2} {\bf q} \cdot ({\bf r}_i - {\bf
      r}_j)} , \\ {\cal B}_{ij}({\bf q}) &=& {\cal U}({\bf q}) {\cal B}_{ij}
  {\cal U}^\dagger ({\bf q}) &=& a_i b_j e^{\frac{i}{2} {\bf q} \cdot ({\bf
      r}_i - {\bf r}_j)} + b_i a_j e^{-\frac{i}{2} {\bf q} \cdot ({\bf
      r}_i - {\bf r}_j)}.
\end{array}
 \label{eq:dqbq}
\end{equation}
The unitary operator ${\cal U}({\bf q})$ has been defined in
(\ref{eq:unitary}).  In section \ref{sec:spin-stiffness} a paragraph
was dedicated to the correct treatment of the orientation of the
ordering. These conditions correspond to the mean-fields here, which
have to be taken translationally invariant. Without loss of generality
we can take them real;
\begin{eqnarray}
  \kappa_{ij}({\bf q}) &=& \frac{1}{2} \langle {\cal D}_{ij}({\bf q}) \rangle ,
 \label{eq:defkappa}\\ \gamma_{ij}({\bf q}) &=& \frac{1}{2} \langle
  {\cal B}_{ij}({\bf q}) \rangle.  \nonumber 
\end{eqnarray}
The mean-field Hamiltonian ${\cal H}_{\rm MF}$ now becomes
\begin{eqnarray}
  {\cal H}_{\rm MF}({\bf q})&=& \phantom{-}\sum_{F} J_{ij} \kappa_{ij} ({\bf
    q}) \left ({\cal D}_{ij}^\dagger ({\bf q}) + {\cal D}_{ij}({\bf q}) - 2
  \kappa_{ij}({\bf q}) \right ) \nonumber \\ & &- \sum_{AF} J_{ij}
  \gamma_{ij}({\bf q})\left ( {\cal B}_{ij}^\dagger({\bf q})+{\cal B}_{ij}({\bf q})-2
  \gamma_{ij}({\bf q})\right )  \nonumber \\ & & + \lambda \sum_{i}
  (a_i^\dagger a_i + b_i^\dagger b_i - 1) - \sum_{F} J_{ij} \frac{3}{4} +  \sum_{AF} J_{ij} \frac{1}{4} 
\end{eqnarray}
This Hamiltonian is applicable both to the N\'e{e}l and the collinear
ordering.  The spin stiffness has to be derived from the ground state
energy of ${\cal H}_{\rm MF}({\bf q})$ This is very similar to the
situation we had in section \ref{sec:spin-stiffness}; The expression
(\ref{eq:rhofinal}) for $\rho_s$ is recaptured with
\begin{eqnarray}
  \vec{j} &=& \phantom{-}\left .\frac{d}{d{\bf q}} {\cal H}_{\rm MF} ({\bf q}) \right |_{{\bf
      q}=0}, \nonumber \\ \vec{\vec{t}} &=&- \left .\frac{d^2}{d{\bf q}^2} {\cal H}_{\rm MF} ({\bf
    q}) \right |_{{\bf q}=0} , \label{eq:tyjy}
\end{eqnarray}
and ${\cal H}$ replaced by ${\cal H}_{\rm MF}$; The states $|a \rangle$
appearing in (\ref{eq:jfinal}) no longer correspond to the excitations
of the full Hamiltonian but to the excitations of the mean-field
Hamiltonian ${\cal H}_{\rm MF}$.  In order to get explicit expressions for
$j$ and $t$, we will perform some algebra. We define $\vec{\cal
  F}_{ij}$ and $\vec{\cal C}_{ij}$ and use (\ref{eq:dqbq}):
\begin{eqnarray}
  \vec{\cal F}_{ij} &=& \left . \frac{d}{d{\bf q}} {\cal D}_{ij} ({\bf q}) \right |_{{\bf
      q}=0} = \frac{i}{2} ({\bf r}_i - {\bf r}_j)(a_i
  a_j^\dagger - b_i b_j^\dagger) , \nonumber \\ \vec{\cal C}_{ij} &=& \left .
  \frac{d}{d{\bf q}} {\cal B}_{ij} ({\bf q}) \right |_{{\bf q}=0} =
  \frac{i}{2} ({\bf r}_i - {\bf r}_j)(a_i b_j - b_i
  a_j). \nonumber
\end{eqnarray}
The derivatives of neither the mean fields $\kappa_{ij}({\bf q})$,
$\gamma_{ij}({\bf q})$ nor the Lagrange multiplier $\lambda$ will
appear in either $J$ or $T$. It is easy to understand that the
dependence on the last one, $\lambda$, can be neglected. This Lagrange
multiplier is tuned to make $\sum_i a_i^\dagger a_i + b_i^\dagger b_i
- 1 = 0$, so the entire terms drops from the expectation value. The
first two, $\kappa_{ij}({\bf q})$ and $\gamma_{ij}({\bf q})$ do not
appear in $J$, as symmetry considerations yield
\[
\begin{array}{ccccccc}
  \left . \frac{d}{d{\bf q}} \kappa_{ij}({\bf q}) \right |_{{\bf q}=0} &=&
    \left .\frac{1}{2} \langle \frac{d}{d{\bf q}} {\cal D}_{ij} ({\bf q}) \rangle
    \right |_{{\bf q}=0} &=& \frac{1}{2} \langle \vec{\cal F}_{ij} \rangle &=& {\bf 0},
     \\ \left . \frac{d}{d{\bf q}} \gamma_{ij}({\bf q}) \right
    |_{{\bf q}=0} &=& \left .\frac{1}{2} \langle \frac{d}{d{\bf q}} {\cal B}_{ij}
    ({\bf q}) \rangle \right |_{{\bf q}=0} & =& \frac{1}{2} \langle
    \vec{\cal C}_{ij} \rangle &=& {\bf 0}.
\end{array}
\]
Move over in the second derivative these mean field cancel out, e.g.
\begin{eqnarray}
\frac{d^2}{d{\bf q}^2} \left [\kappa_{ij} ({\bf
    q}) \left ({\cal D}_{ij}^\dagger ({\bf q}) + {\cal D}_{ij}({\bf q}) - 2
  \kappa_{ij}({\bf q}) \right ) \right] = \nonumber \\
\left[ \frac{d^2}{d{\bf q}^2} \kappa_{ij}  ({\bf
    q}) \right ]\left ({\cal D}_{ij}^\dagger ({\bf q}) + {\cal D}_{ij}({\bf q}) - 4
  \kappa_{ij}({\bf q}) \right ) + \nonumber \\
\kappa_{ij}  ({\bf
    q}) \left (\frac{d^2}{d{\bf q}^2}\left[{\cal D}_{ij}^\dagger ({\bf q}) + {\cal D}_{ij}({\bf q}) \right ]- 2
  \kappa_{ij}({\bf q}) \right ). \nonumber 
\end{eqnarray}
The expectation value of the first part is zero by equation
(\ref{eq:defkappa}).

Inserting these quantities in (\ref{eq:tyjy}) gives
\begin{eqnarray}
  \vec{j} &=& \sum_{F} J_{ij} \kappa_{ij}(\vec{\cal F}_{ij}^\dagger + \vec{\cal F}_{ij}) -
  \sum_{AF} J_{ij} \gamma_{ij} (\vec{\cal C}_{ij}^\dagger + \vec{\cal C}_{ij}),
  \nonumber \\ \vec{\vec{t}} &=& \frac{1}{2} \sum_{F} J_{ij} \kappa_{ij}
  ({\bf r}_i - {\bf r}_j) ({\bf r}_i - {\bf r}_j)({\cal D}_{ij}^\dagger +
  {\cal D}_{ij}-2 \kappa_{ij}) \nonumber \\ & &- \frac{1}{2} \sum_{AF}
  J_{ij} \gamma_{ij} ({\bf r}_i - {\bf
    r}_j) ({\bf r}_i - {\bf
    r}_j)({\cal B}_{ij}^\dagger + {\cal B}_{ij}-2 \gamma_{ij}). \nonumber
\end{eqnarray} 
These expressions for the current $\vec{j}$ and kinetic term $\vec{\vec{t}}$ can be used to calculate $\rho_s$ as is done in section \ref{sec:schwingerrho_s}.

\section{Transforming the DMRG state to the density matrix basis \label{app:DMbasis}}

In this appendix it will be shown how the same DMRG state $|\phi_0
\rangle$ can be representated in a density matrix basis for all
possible partitions of the system in a left and a right part. The left
part contains the first $l$ sites and the right part the remaining
$N-l$ sites.

We define a basis $\{|i\rangle_l\}$ on the left part of the system
containing the first $l$ spins and let $\{ |j\rangle_l \}$ be a basis
on the remaining $N-l$ spins. The state $|\phi_0\rangle$ can be
represented by
\begin{equation}
|\phi_0 \rangle = \sum_{i,j} \phi^{l_0}_{ij} |i \rangle_{l_0} |j \rangle_{l_0}, \label{eq:trialwave}
\end{equation}
for a certain $l_0$. If we consider the final outcome of a DMRG
calculation, the left part will contain all but one site, $l_0=N-1$.
It is possible to represent the same wave function exactly for all
different partitions $1 \le l < N$. The technique is instructive and
we will elaborate on it: Define the transformations $A^l_{i \sigma
  i'}$ and $B^l_{j \sigma j'}$ by
\begin{eqnarray}
|i\rangle_l &=& \sum_{\sigma i'} A^l_{i\sigma i'} |\sigma \rangle |i' \rangle_{l-1}, \nonumber \\
|j \rangle_l &=& \sum_{\sigma j'} B^l_{j \sigma j'} |\sigma \rangle |j'\rangle_{l+1}. \nonumber
\end{eqnarray}
With the help of these transformations we will construct the split-up
for $l_0+1$. All other split-ups are then trivial iterations of the
same procedure. Insert the transformation of $|j \rangle_{l_0}$ in
equation (\ref{eq:trialwave});
\begin{eqnarray}
|\phi_0 \rangle &=& \sum_{i,j} \phi^{l_0}_{ij} |i \rangle_{l_0} | j \rangle_{l_0} =  \sum_{i,j,\sigma,j'} \phi^{l_0}_{ij} B^{l_0}_{j \sigma j'} |i \rangle_{l_0} | \sigma \rangle | j' \rangle_{l_0+1} \nonumber \\
&\equiv& \sum_{j'} |\beta_{j'} \rangle_{l_0+1} | j' \rangle_{l_0+1}. \nonumber
\end{eqnarray}
where the definition $|\beta_{j'} \rangle = \sum_{i,j,\sigma}
\phi^{l_0}_{ij} B^{l_0}_{j \sigma j'} |i \rangle_{l_0} | \sigma
\rangle$ is used. Orthonormalisation of $|\beta_{j'} \rangle$ will
yield a {\it new} basis $\{ |i \rangle_{l_0+1} \}$ and a {\it new}
transformation $A^{l_0+1}_{i \sigma i'}$. Moreover the prefactors of
the wave function $\phi^{l_0+1}_{ij'}$ can also readily be deduced. It
has to be stressed that both the basis $\{ |i \rangle_{l_0+1} \}$ and
the transformation $A^{l_0+1}_{i \sigma i'}$ follow from this
procedure. If we were to have an expression for the transformation
$A^{l_0+1}_{i \sigma i'}$ already, it is replaced by this new one.
The split-up is changed from $l_0$ to $l_0+1$ and indeed this approach
can trivially be extended to yield for all partitions $1 \le l <N$ the
bases $\{| i \rangle_l \}$, $\{ |j \rangle_l \}$ and the prefactors
$\phi^l_{ij}$.

The next ingredient of our recipe is to switch to the density matrix
basis. The end of section (\ref{sec:DM}) and specifically equation
(\ref{eq:exactrepr}) explain that this can be done by simple basis
rotations on the left and on the right basis yielding a representation
$\{ | \alpha\rangle_l\}$ for the left and $\{ |\bar{\alpha} \rangle_l
\}$ for the right part. These basis rotations give us furthermore:
\begin{itemize}
\item A simple representation of the wave function $|\phi_0\rangle =
  \sum_{\alpha} \sqrt{\lambda_\alpha^l} | \alpha \rangle_l
  |\bar{\alpha} \rangle_l$.
\item Basis transformations $A_{\alpha \sigma \alpha'}^l$ and
  $B_{\alpha \sigma \alpha'}^l$.
\end{itemize}
This is the notation we will use when deriving the value of the
 wavefunction $\langle R| \phi_0 \rangle$.

\appendixoff

%% file: PUBLICATIONS.tex
\chapter*{List of Publications}
\addcontentsline{toc}{chapter}{List of Publications}

\begin{itemize}
\item High-field magnetoresistance oscillations in
  $\alpha$-[bis(ethylenedithio)tetrathiafulvalene] 2KHg(SCN)$_4$: The effects of
  magnetic breakdown, exchange interactions, and Fermi-surface
  reordering, J. Caulfield, S. J. Blundell, M. S. L. du Croo de Jongh,
  P. T. J. Hendriks, J. Singleton, M. Doporto, F. L. Pratt, A. House,
  J. A. A. J. Perenboom, W. Hayes, M. Kurmoo, P. Day , Phys. Rev. B
  {\bf 51}, 8325 (1995)
\item Spin stiffness in the Hubbard model, J. M. J. van Leeuwen, M. S.
  L. du Croo de Jongh, P. J. H. Denteneer, J. Phys. A: Math. Gen. {\bf
    29}, 41 (1996)
\item Spin stiffness in the frustrated Heisenberg antiferromagnet, M.
  S. L. du Croo de Jongh, P. J. H. Denteneer, Phys. Rev. B {\bf 55},
  2713 (1997)
\item Critical behavior of the two-dimensional Ising model in a
  transverse field: A density-matrix renormalisation calculation, M.
  S. L. du Croo de Jongh, J. M. J. van Leeuwen, Phys. Rev. B {\bf 57},
  8494 (1998)
\item Towards Understanding
   Quantum Coherent Dynamics of Molecules: A Simple Scenario for 
   Ultrafast Photoisomerization, D. P. Aalberts, M. S. L. du Croo de Jongh, B. F.
  Gerke, W. van Saarloos, submitted to Phys. Rev. Lett.
\item The Phase diagram of the two-dimensional frustrated Heisenberg model,
M. S. L. du Croo de Jongh, J. M. J. van Leeuwen, W. van Saarloos, to be published.
\end{itemize}

%% file: SAMENVATTING.tex
\chapter*{Samenvatting} \addcontentsline{toc}{chapter}{Samenvatting}

Dit proefschrift is gebouwd op drie zuilen:
\begin{itemize}
\item Numerieke berekeningen,
\item Quantum fase overgangen,
\item Quantum spin systemen.
\end{itemize}

Van deze drie is de eerste, numerieke berekeningen, waarschijnlijk het
meest eenvoudig te doorgronden. Vaak spreekt men ook van numerieke
simulatie om aan te geven dat men de werkelijkheid wil nabootsen. De
dagelijkse weervoorspelling is waarschijnlijk het meest bekende
voorbeeld van een numerieke simulatie. Het laat ook goed zien, dat
deze berekeningen een kunst op zich zijn. Sinds de vorige eeuw weten
we aan welke regels luchtdeeltjes voldoen. Deze regels, ook wel
Navier-Stokes vergelijkingen genaamd, zijn elegant en eenvoudig, maar
de enorme diversiteit aan weersomstandigheden geeft al aan dat het
gedrag van de deeltjes hiermee niet doorzichtig is. Pas recentelijk
kunnen we redelijke voorspellingen maken. Natuurlijk hebben de steeds
snellere computers daaraan bijgedragen, maar computerkracht alleen is
niet opgewassen tegen de complexiteit van dit probleem. We zullen de
computer flink moeten helpen met ons fysisch inzicht. Zo kunnen we
sneeuwval in het weer rond de evenaar uitsluiten. Indien dit soort
vereenvoudigingen niet worden doorgevoerd, zal de computer veel te
veel tijd kwijt zijn om tot bekende conclusies te komen.

Het is ook verstandig om het doel van numerieke berekeningen te
analyseren. Het weerbericht heeft een voorspellend karakter. Er zijn
evenwel meer redenen om simulaties uit te voeren. In de natuurkunde
wordt vaak gepoogd de werkelijkheid te bevatten in een klein aantal
regels en elementen.  Aan de hand hiervan stelt men een model op. Nu
is er vaak onduidelijkheid over de gelijkenis van zo'n model met de
realiteit en een numerieke simulatie van het model kan daar uitkomst
brengen. Men kan zich bijvoorbeeld afvragen of de kleur van het
aardoppervlak van invloed is op het weer.  In de praktijk blijkt dat
zo te zijn, aangezien een donker oppervlak veel meer zonlicht
absorbeert dan een licht oppervlak. De verwachting is dus al dat een
donker oppervlak overdag sneller opwarmt. Is het noodzakelijk om nog
meer kleuren te introduceren, of kunnen we rood en groen afdoen als
half donker, half licht? Met een numerieke simulatie kan de invloed
van die kleuren worden bepaald om vervolgens de nuancering wel of niet
op te nemen in het model.

Verder staan numerieke simulaties ons toe allerlei metingen te doen
die in de {\it echte} wereld niet mogelijk zijn. We kunnen
bijvoorbeeld de tijd even vooruitspoelen om naar het weer van morgen
te kijken, maar we kunnen ook de tijd eenvoudigweg stilzetten.  In dit
proefschrift koelen we ons model af naar het absolute nulpunt, -273,15
$^\circ$ Celcius. Dit is experimenteel niet mogelijk. Toch kan dit
extreme afkoelen enorm helpen om eigenschappen van de natuur te
begrijpen.

Samenvattend zijn numerieke berekeningen nuttig vanwege twee redenen:
Ten eerste, laten ze allerlei metingen toe die experimenteel niet
mogelijk zijn en ten tweede kunnen ze een grote steun zijn bij het
vinden van relevante natuurkundige regels. Ze hoeven evenwel niet
eenvoudig te zijn. Om een goede berekening te doen, dient men inzicht
te hebben in de fysische verschijnselen die men wil analyseren en
verder moet men in staat zijn om de vertaalslag naar een
computerprogramma te maken.

Ook de tweede zuil, quantum fase overgangen, kent een alledaagse
analogie. De fase overgang van water naar ijs door een
temperatuursverandering laat zien dat eigenschappen drastisch kunnen
veranderen bij maar een kleine temperatuursvariatie. Deze overgang is
niet de enige fase overgang in de natuur. Er zijn vele andere fase
overgangen bekend. Opvallend is dat ze vaak niets met een
temperatuursverandering te maken hebben; er zijn vele voorbeelden van
overgangen onder invloed van bijvoorbeeld druk, dichtheid of
elektrisch veld. Indien ijs wordt samengeperst, verandert het weer in
water.  De temperatuur is hierbij niet veranderd. Als tweede voorbeeld
bekijken we de schermen van draagbare computers en horloges. Deze
bevatten polymeren die normaliter ongeordend en transparant zijn.  Zo
gauw een elektrisch veld wordt aangezet, richten zij zich en worden ze
ondoorzichtig. Dit soort overgangen blijken vaak op vergelijkbare
wijze te kunnen worden beschreven als temperatuurafhankelijke
overgangen. Hoe verschillend ze op het eerste gezicht ook mogen zijn;
een duidelijk universeel gedrag wordt waargenomen. De overeenkomsten
zijn veel omvattender dan alleen de gelijke naam 'fase overgang' doet
vermoeden. In dit proefschrift worden twee specifieke gevallen
behandeld, die ieder als voorbeeld kunnen dienen voor gehele klassen
van quantum fase overgangen.

In de eerste twee hoofdstukken bestuderen we een model waar een extern
magneetveld de fase overgang bewerkstelligt. Het is zelf mogelijk te
laten zien dat het magneetveld hier precies dezelfde funktie heeft als
de temperatuur in een gerelateerd model!

De andere fase overgang die in dit proefschrift besproken wordt is in
de praktijk moeizaam te bewerkstelligen. Men moet hierbij denken aan
de variabele samenstelling van een materiaal. Hier kan niet eenvoudig
een vlammetje onder gehouden worden zoals bij ijs om een overgang te
veroorzaken. Men moet vele preparaten, ieder met een net even andere
samenstelling, de revue laten passeren om de fase overgang te kunnen
bestuderen. Slechts \'e\'en preparaat heeft precies de juiste
samenstelling die hoort bij het punt van de fase overgang.  Toch
blijft de fase overgang ook duidelijk invloed uitoefenen op het gedrag
van alle andere preparaten.

De derde zuil, quantum spin systemen, geeft aan dat we fysische
modellen bekijken die uit spins zijn samengesteld.  De
individuele spins doen denken aan magneetjes met een noord- en een
zuidpool. Daarnaast voldoen ze aan nog een aantal andere regels die
diep in de wereld van de quantummechanica thuis horen. Bijvoorbeeld
een spin kan in een toestand verkeren die niet duidelijk
geori\"enteerd is. Pas als men gaat meten zal de spin een unieke
ori\"entatie uitkiezen. Deze spins zijn in ons
geval keurig naast elkaar geplaatst op een vierkant rooster dat lijkt
op een barbeque rooster; elk hokje herbergt \'e\'en spin.

In het proces om fysische systemen zo eenvoudig mogelijk te
beschrijven, worden vaak ---pseudo--- spins ge\"introduceerd. De
feitelijke deeltjes kunnen bijvoorbeeld elektronen zijn, maar alleen
de regels die sterke gelijkenis hebben met de regels voor spins zijn
relevant. Quantum spin systemen hebben dus vaak een voorbeeldfunktie
zoals we al eerder tegenkwamen in de paragrafen over fase overgangen.

Nu de zuilen zijn geplaatst, kunnen we ons richten op het fronton. De
eerste twee- en de laatste drie hoofdstukken vormen ieder een
duidelijk geheel. Het doel van de eerste twee hoofdstukken is om een
nieuwe numerieke methode te doorgronden. Deze methode heet Dichtheids
Matrix Renormalisatie Groep (DMRG). In hoofdstuk drie en vier wordt
deze methode vervolgens toegepast op het gefrustreerde Heisenberg
model.  Dit blijkt maar een matig succes op te leveren en in hoofdstuk
vijf wordt de DMRG met Green Functie Monte Carlo simulaties (GFMC)
gecombineerd om toch de eigenschappen van dit gefrustreerde model te
kunnen analyseren.  De belangrijkste resultaten van dit proefschrift
zijn:
\begin{enumerate}
\item Een goed inzicht in de DMRG is verkregen en gerapporteerd.
\item Door de combinatie van DMRG en GFMC kunnen we GFMC makkelijker
  hanteren voor een hele klasse van problemen.
\item We hebben inzicht gekregen in de fase overgangen van het
  gefrustreerde Heisenberg model.
\end{enumerate}

De eerste twee hoofdstukken zijn dus methodisch van opzet. De DMRG was
al zeer succesvol voor spin ketens en wij willen haar toepassen op
spin roosters. Daarvoor hebben we een zeer bekend model uitgekozen wat
frequent wordt gebruikt als voorbeeld van een quantum fase overgang.
Er is dan ook recentelijk veel numeriek werk aan verricht waaraan wij
houvast hebben bij onze studie. Alleen kleine systeempjes met maximaal 32 rijen van 8 spins elk
kunnen numeriek worden behandeld. Om toch
inzicht te krijgen in grotere systemen wijden we de tweede helft van
hoofdstuk \'e\'en eraan, de specifieke effecten van de kleine
afmetingen te bepalen. Die kunnen vervolgens worden verwijderd om de
eigenschappen van een oneindig groot rooster te verkrijgen. Dit heet
eindige-grootte schaling en wordt al geruime tijd toegepast op dit
soort problemen. De resultaten die we op deze wijze verkrijgen, komen
overeen met de literatuur.

Hoofdstuk twee bespreekt de DMRG methode en bekijkt wat de
mogelijkheden en beperkingen zijn. Het ziet ernaar uit dat de DMRG
alleen goede numerieke kwaliteit kan verkrijgen voor smalle roosters
(strips). Eindige-grootte schaling blijft dus noodzakelijk om de
eigenschappen van grotere roosters te kunnen bepalen. In hoofdstuk
vijf zal evenwel een mogelijke uitweg worden gepresenteerd.

Hoofdstuk drie, vier en vijf richten zich op het gefrustreerde
Heisenberg model. Ook dit model heeft een voorbeeld funktie. Tevens is
recentelijk gesuggereerd dat de chemische verbinding $CaV_4O_9$ er
goed door beschreven zou worden. Indien het model geformuleerd wordt
zonder de quantummechanische aspecten van spins, is het gedrag
volkomen duidelijk. Deze extra quantummechanische regels geven
aanleiding tot nieuwe verschijnselen die maar ten dele begrepen zijn.
De funktie van de temperatuur rond het vriespunt van water wordt hier
bekleed door een frustratie maat $J_2$. Het systeem is maximaal
gefrustreerd rond $J_2=0.5$. Voor grotere waarde van $J_2$ neemt de
frustratie weer af omdat de spins zich dan echt anders gaan
organiseren. Bij weinig frustratie ($J_2 \approx 0$ en $J_2 \approx
1$) begrijpen we goed wat er gebeurt.  In beide gebieden gedraagt het
systeem zich eender als de klassieke variant. Daartussen zit een
onduidelijke gebied ($J_2 \approx 0.5$).  Waarschijnlijk is daar zelfs
een fase die geen klassieke evenknie heeft. Dit geeft aanleiding tot
het veronderstellen van twee quantum fase overgangen.  Tussen de
onbekende fase en de beide zwak gefrustreerde fasen.

Om hier zicht op te verkrijgen gebruiken we in hoofdstuk vier de
spin-stijfheid. Dit is een maat voor het gemak waarmee twee spins,
ieder op een ander uiteinde van het systeem, in verschillende
richtingen geori\"enteerd kunnen worden. Alle andere spins van het
systeem zullen zich aanpassen aan deze beide spins. Indien spins
weinig rekening met elkaar houden is het relatief makkelijk deze twee
spins anders te ori\"enteren; de rest reageert toch niet. Het
tegenovergestelde is het geval indien de spins wel degelijk rekening
houden met elkaar. Deze indicator zou duidelijk andere waarden moeten
aannemen voor de drie verschillende fasen.  Aangezien wederom alleen
kleine systeem numeriek behandeld kunnen worden, is eindige-grootte
schaling noodzakelijk om de eigenschappen van veel grotere systeem te
bepalen. De resultaten zijn redelijk te noemen ofschoon hieruit niet
met zekerheid kan worden geconcludeerd dat deze tussen-fase echt
bestaat.

We kunnen ook bekijken hoe sterk de individuele spins met elkaar
rekening houden. Dit is gedaan in hoofdstuk vijf voor een systeem met
10 rijen van elk 10 spins. In figuur \ref{fig:scdots} op pagina \pageref{fig:scdots} is mooi te zien dat de spins
zich per viertal organiseren. (dikke lijnen geven een sterk verband
aan). Dit resultaat is niet alleen een duidelijke ondersteuning van het
bestaan van de derde fase maar geeft tevens inzicht in de
onderliggende ordening.

Hoofdstuk vijf heeft evenwel nog meer te bieden. In dat hoofdstuk
introduceren we nog een andere numerieke methode, de Green Funktie
Monte Carlo simulatie (GFMC). GFMC omvat systematisch gokwerk om de
eigenschappen van een systeem te bepalen. Dit verklaart tevens het
tweede gedeelte van de naam. Globaal komt het erop neer dat alle spins
een ori\"entatie krijgen opgelegd. De onderlinge relaties worden
vervolgens vastgelegd en de relevantie van deze situatie wordt
bepaald. Nadat vele situaties zijn bekeken, kunnen door combinatie van
de relaties met de relevantie de eigenschappen van het systeem worden
bepaald. DMRG kan hierbij enorm helpen op twee manieren.

Ten eerste kan DMRG bij voorbaat al inzicht geven in welke situaties
het meest relevant zijn. De GFMC bekijkt vervolgens alleen deze. Dit
bespaart buitengewoon veel werk en computertijd.

Het tweede aanknopingspunt is zeer quantummechanisch van aard. Het
blijkt dat in modellen met concurerende wisselwerking (frustratie) of
met elektronen moeilijk een gemiddelde waarde is te bepalen door het
zo geheten teken-probleem. Dit probleem kan verminderd worden door zo
veel mogelijk informatie over het systeem in de berekening te
verwerken. DMRG heeft erg veel informatie te bieden die in de GFMC
methode kan wordt betrokken.

Deze combinatie van DMRG met GFMC kan succesvol worden genoemd. Het
laatste woord is er nog niet over gesproken maar er zijn een heel
aantal fysische problemen bekend waar zij uitkomst zou kunnen bieden.
Deze problemen vari\"eren van supergeleiding tot quantum hall effect.

%% file: CV.tex
\chapter*{Curriculum Vitae} \addcontentsline{toc}{chapter}{Curriculum
Vitae}

Op 27 november 1970 ben ik geboren in Gorssel, een dorp in de buurt
van Deventer. In 1989 haalde ik mijn gymnasium $\beta$ diploma aan het
Geert Groote College te Deventer. Daarna heb ik van 1989 tot en met
1995 natuurkunde en wiskunde gestudeerd aan de Universiteit Leiden.
Tijdens mijn studententijd heb ik nog enkele jaren geroeid op
wedstrijd- en recreatief niveau. In 1993 studeerde ik af in de
wiskunde bij dr. J. Hulshof op een scriptie over parti\"ele
differentiaalvergelijkingen.  Daarna verrichtte ik een jaar lang
experimenteel onderzoek aan de Universiteit van Oxford. In 1995
studeerde ik af in de theoretische natuurkunde bij mijn latere
promotor prof. dr. J. M. J. van Leeuwen. Dit maal omvatte de scriptie
een statistische benadering voor het Hubbard model.

Na een kort verblijf bij een management adviesbureau in Berlijn begon ik in juni
1995 aan mijn promotie. Het onderzoek zoals beschreven in dit
proefschrift concentreerde zich op numerieke methoden om fysische
eigenschappen van quantum rooster modellen te bepalen. Hiervoor heb ik
ook drie maanden prof. dr. S. R. White aan de Universiteit van
California in Irvine bezocht. Uiteraard woonde ik diverse cursussen en
zomerscholen in het buitenland bij.

Naast mijn onderzoekswerk, heb ik nog het werkcollege
'Electromagnetisme' begeleid. Verder heb ik in de zomer van 1996 drie
maanden voor een bank in Londen gewerkt.